%% file: main.tex
\documentclass[sigconf,nonacm=true]{acmart}
\usepackage{booktabs,siunitx}
\usepackage[utf8]{inputenc}
\usepackage{dsfont}
\usepackage{xcolor}
\usepackage{url}
\usepackage{comment}
\usepackage{graphicx}
\usepackage{caption}
\usepackage{subcaption}
\usepackage{amsmath}
\usepackage{multirow}
\usepackage{tikz}

\hypersetup{                    % ...like so
  colorlinks=false
}

\newcommand\blfootnote[1]{%
  \begingroup
  \renewcommand\thefootnote{}\footnote{#1}%
  \addtocounter{footnote}{-1}%
  \endgroup
}

\copyrightyear{2022} 
\acmYear{2022} 
\setcopyright{rightsretained} 
\acmConference[CCS '22]{Proceedings of the 2022 ACM SIGSAC Conference on Computer and Communications Security}{November 7--11, 2022}{Los Angeles, CA, USA}
\acmBooktitle{Proceedings of the 2022 ACM SIGSAC Conference on Computer and Communications Security (CCS '22), November 7--11, 2022, Los Angeles, CA, USA}
\acmDOI{10.1145/3548606.3560663}
\acmISBN{978-1-4503-9450-5/22/11}

\begin{document}

\urlstyle{sf}
\makeatletter
% Inspired by http://anti.teamidiot.de/nei/2009/09/latex_url_slash_spacingkerning/
% but slightly less kern and shorter underscore
\let\UrlSpecialsOld\UrlSpecials
\def\UrlSpecials{\UrlSpecialsOld\do\/{\Url@slash}\do\_{\Url@underscore}}%
\def\Url@slash{\@ifnextchar/{\kern-.05em\mathchar47\kern-.2em}%
    {\kern-.0em\mathchar47\kern-.08em\penalty\UrlBigBreakPenalty}}
\def\Url@underscore{\nfss@text{\leavevmode \kern.06em\vbox{\hrule\@width.3em}}}
\makeatother

\title{Are Attribute Inference Attacks Just Imputation?}

\author{Bargav Jayaraman}
\affiliation{ 
      \institution{University of Virginia}
      %\streetaddress{street}
      \city{Charlottesville} 
      \state{VA} 
      \postcode{22903}
      \country{USA}
    }
\email{bj4nq@virginia.edu}

\author{David Evans}
\affiliation{ 
      \institution{University of Virginia}
      %\streetaddress{street}
      \city{Charlottesville} 
      \state{VA} 
      \postcode{22903}
      \country{USA}
    }
\email{evans@virginia.edu}
%\date{}

\input{preamble}

\begin{abstract}
Models can expose sensitive information about their training data. In an attribute inference attack, an adversary has partial knowledge of some training records and access to a model trained on those records, and infers the unknown values of a sensitive feature of those records. We study a fine-grained variant of attribute inference we call \emph{sensitive value inference}, where the adversary's goal is to identify with high confidence some records from a candidate set where the unknown attribute has a particular sensitive value. We explicitly compare attribute inference with data imputation that captures the training distribution statistics, under various assumptions about the training data available to the adversary. Our main conclusions are: (1) previous attribute inference methods do not reveal more about the training data from the model than can be inferred by an adversary without access to the trained model, but with the same knowledge of the underlying distribution as needed to train the attribute inference attack; (2) black-box attribute inference attacks rarely learn anything that cannot be learned without the model; but (3) white-box attacks, which we introduce and evaluate in the paper, can reliably identify some records with the sensitive value attribute that would not be predicted without having access to the model. Furthermore, we show that proposed defenses such as differentially private training and removing vulnerable records from training do not mitigate this privacy risk.
The code for our experiments is available at \url{https://github.com/bargavj/EvaluatingDPML}.\blfootnote{To appear in the Proceedings of the 2022 ACM SIGSAC Conference on Computer and Communications Security (CCS '22), November 7--11, 2022, Los Angeles, CA, USA.}
\end{abstract}

\maketitle

\input{1_introduction}
\input{2_background}
\input{3_attack_definition}
\input{4_attack_methodology}
\input{5_experiment_setup}
\input{6_blackbox_results}
\input{7_whitebox_results}
\input{8_defenses}

\section{Discussion}\label{sec:discussion}
Despite being a more direct privacy concern than membership inference, attribute inference has received scant attention from the research community. 
Our experiments show that previous works claiming to demonstrate attribute inference~\cite{yeom2018privacy, fredrikson2014privacy, mehnaz2022sensitive} do not seem to learn anything from the model that could not be learned without the model. Both the attribute inference attacks and data imputation depend on some prior knowledge of the training distribution, and it is important to evaluate their effectiveness based on varying assumptions about that prior knowledge. In our experiments, even as we vary the similarly of that distribution to the training distribution, imputation attacks nearly always outperform black-box attribute inference attacks. We introduce a stronger white-box attack that can substantially outperform imputation in some settings. The presence or absence of an individual candidate record in the training dataset has no impact on the vulnerability of that record to the attack, indicating that the attacks reveal what the model has learnt about the training distribution not about specific training records.

This raises this question, \emph{is attribute inference really a privacy threat when it is only revealing statistical properties of the training distribution?} It has been argued that statistical inference is just ``science'' and should not be considered a privacy threat~\cite{mcsherry2016warfarin}. 
Indeed, in cases where the underlying distribution is public and the adversary can make the same inferences using data imputation, there is no additional privacy risk in releasing the model. In many cases, however, the underlying distribution is not public. Then, evidence that an attribute inference attack can outperform imputation raises a legitimate privacy alarm about the risks of releasing the model. Our experiments show that 
white-box attacks trained on a limited data sets and skewed distributions can be surprisingly effective, indicating that the trained model is leaking substantial information about the training distribution. 

To an individual who is harmed by a sensitive value inference, it doesn't matter if the inference is due to distribution inference or dataset inference. As researchers, though, it is essential that we conduct experiments to carefully distinguish between dataset and distribution inference. Separating these two kinds of inference is critical for understanding what kinds of mitigation to explore and how to evaluate them. In cases where the risk is due to dataset inference, an individual aware of these risks may be able to withhold their data. With 
distribution inference, an individual has no hope of preventing the inference by withholding their data, and the only paths to mitigating the privacy risks are limiting model exposure and technical solutions to limit what the model discloses.

\section*{Acknowledgements}
This work was partially supported by a grant from the National Science Foundation (\#1804603). The authors are grateful to the CCS reviewers for providing insights that further improved our paper. We would also like to thank Jerry Su for helping produce the \dataset{Census19} data set.

\bibliographystyle{ACM-Reference-Format}
\bibliography{ref}

\appendix
\input{appendix}

\end{document}

%% file: preamble.tex
%\newtheorem{theorem}{Theorem}[section]
%\newtheorem{corollary}[theorem]{Corollary}
%\newtheorem{lemma}[theorem]{Lemma}
%\newtheorem{proposition}[theorem]{Proposition}
%\theoremstyle{definition}
\newtheorem{defn}{Definition}[section]
\newtheorem{experiment}[defn]{Experiment}
\newtheorem{proc}[defn]{Procedure}
\newtheorem{remark}[theorem]{Remark}
\newcommand{\RR}{\mathbb{R}}
\newcommand{\PP}{\mathbb{P}}
\newcommand{\EE}{\mathbb{E}}

\newcommand{\cA}{\mathcal{A}}
\newcommand{\cD}{\mathcal{D}}
\newcommand{\bD}{\mathbf{D}}
\newcommand{\bX}{\mathbf{X}}
\newcommand{\bY}{\mathbf{Y}}
\newcommand{\bV}{\mathbf{V}}
\newcommand{\cM}{\mathcal{M}}
\newcommand{\bbE}{\mathbb{E}}
\newcommand{\cN}{\mathcal{N}}
\newcommand{\cX}{\mathcal{X}}
\newcommand{\cY}{\mathcal{Y}}
\newcommand{\cV}{\mathcal{V}}
\newcommand{\bS}{\mathbf{S}}
\newcommand{\bU}{\mathbf{U}}
\newcommand{\bC}{\mathbf{C}}
\newcommand{\bB}{\mathbf{B}}
\newcommand{\bT}{\mathbf{T}}
\newcommand{\bz}{\mathbf{z}}
\newcommand{\bt}{\mathbf{t}}
\newcommand{\bv}{\mathbf{v}}
\newcommand{\by}{\mathbf{y}}
\newcommand{\bu}{\mathbf{u}}
\newcommand{\bxi}{\bm{\xi}}
\newcommand{\bI}{\mathbf{I}}
\newcommand{\bgamma}{\bm{\gamma}}

\newcommand{\nbfdataset}[1]{{\small{\textsf{#1}}}}
\newcommand{\bfdataset}[1]{{\small{\textsf{#1}}}}
\newcommand{\dataset}[1]{{\small{\textsf{#1}}}}

\newcommand{\bfattack}[1]{{\small\textbf{\textsf{#1}}}}
\newcommand{\attack}[1]{{\small{\textsf{#1}}}}

\newcommand{\bnote}[1]{\textcolor{purple}{\bf \emph{Bargav: #1}}}

\newcommand{\dnote}[1]{\textcolor{blue}{\bf \emph{Dave: #1}}}

\newcommand{\gnote}[1]{\textcolor{blue}{\bf \emph{Gu: #1}}}

\newcommand{\lnote}[1]{\textcolor{green}{\bf \emph{Lingxiao: #1}}}

\newcommand{\todo}[1]{\textcolor{red}{\bf \emph{TODO: #1}}}

\newcommand\shortsection[1]{\vspace{6pt}{\noindent\bf #1.}}

\newcommand\shortdsection[1]{\vspace{3pt}{\noindent -- #1:}}

%% file: 1_introduction.tex
\section{Introduction}

Models trained using machine learning have been shown to be vulnerable to inference attacks that reveal sensitive information about the model's training data, including membership inference~\cite{shokri2017membership}, attribute inference~\cite{yeom2018privacy}, property inference~\cite{ateniese2015hacking}, and partial memorization~\cite{carlini2019secret}. 
We focus on attribute inference, which assumes an adversary has partial knowledge of some training records and wants to use information obtained from a trained model to infer the value of an unknown sensitive attribute of those records. The goal of attribute inference is the same as that for data imputation, where the task is to fill the unknown parts of the data given the partial data records. The difference is that  an attribute inference adversary additionally exploits access to a model trained on the target record, whereas imputation is done just based on knowledge of a data distribution. We show that most attribute inference attacks are actually doing imputation by using information about the training distribution revealed by the model and that the inference risk is the same for training and non-training records. Thus, the main privacy risk is from what the model reveals about its training distribution, not its training data set. In the research community, such distribution inference attacks are often considered ``science'' and are thus not considered a privacy risk~\cite{mcsherry2016warfarin}, but this is only the case when an adversary already has full knowledge about the distribution.
As noted previously by Graham Cormode~\cite{cormode2011personal}, it should be considered a privacy risk if a ``population inference'' based on the model allows for a more accurate personal information inference about an individual, regardless of whether or not they were included in the training data.
In understanding the risks, however, it is important to distinguish the generally unavoidable risks from data imputation from the potentially avoidable risks due to an individual contributing their own data to a training data set.

Several papers have reported evidence of attribute inference~\cite{fredrikson2014privacy, yeom2018privacy, mehnaz2022sensitive}, but 
as we show in our experiments (summarized in Table~\ref{tab:attribute_inference}), these prior attribute inference attacks do not actually seem to learn more than what could be learnt through imputation without access to the model. Thus, the privacy leakage in such cases is not due to an individual contributing data, but instead is due to the model learning and revealing statistical correlations in the training data distribution.
This is not enough to conclude that there is no attribute inference privacy threat, however, since these attacks may not be extracting all information available from the model. For instance, the attribute inference adversary in Yeom et al.~\cite{yeom2018privacy} uses a membership inference oracle as a black box.
We explore attribute inference attacks that take full advantage of access to the model. Further, previous works do not consider cases where the adversary starts with limited information about the training distribution. We show that in settings where the underlying distribution itself is not fully known to the adversary, even an accurate imputation based on the distributional information leaked by the model can pose serious privacy risks.

\shortsection{Contributions}
To better understand attribute inference risks, we consider threat models where an adversary has limited prior knowledge of the training distribution (\autoref{sec:dataavailable}) and 
study a finer-grained notion of attribute inference that considers the privacy risk of identifying, with high confidence, individuals with a particular sensitive attribute values from a candidate set (\autoref{sec:sensitive_value_inference}). 
We propose a novel white-box attack that identifies neurons in a model that are most correlated with the sensitive value for a target attribute (\autoref{sec:whitebox_attack}).
We perform extensive experimental evaluation sensitive value inference on two large real world data sets with both imputation and black-box attribute inference attacks (\autoref{sec:blackbox_results}) and with our novel white-box attacks (\autoref{sec:whitebox_results}). 

\begin{table}[tb]
    \centering
    \begin{tabular}{l@{\hskip 2em}ccc}
        \toprule
        & \multicolumn{3}{c}{Adversary's Data Set Size} \\ % $\rightarrow$} \\
        & 5\,000 & 500 & 50 \\ \midrule
        Imputation & 0.62 $\pm$ 0.05 & 0.39 $\pm$ 0.03 & 0.24 $\pm$ 0.01 \\
        Black-box Attack & 0.60 $\pm$ 0.04 & 0.42 $\pm$ 0.01 & 0.33 $\pm$ 0.04 \\
        White-box Attack & 0.64 $\pm$ 0.04 & 0.51 $\pm$ 0.09 & 0.50 $\pm$ 0.02 \\ \bottomrule
    \end{tabular}
    \caption{Comparing inference adversaries. \rm Results are the positive predictive value (PPV) over the  top-100 candidate records for predicting \emph{Hispanic} ethnicity from \nbfdataset{Texas-100X} (\autoref{sec:datasets}) with varying amounts of data available to the adversary from the training distribution. Imputation trained on 5\,000 records outperforms both the best black-box and white-box attacks. However, the white-box attacks leak useful information about the training distribution when limited prior information is available.}
    \label{tab:result_summary}
\end{table}

\shortsection{Key findings}
Our experiments show that trained models leak considerable information about the underlying training distribution which can be exploited to infer sensitive attributes about individuals. While  prior attribute inference attacks do not learn anything from the model that could not be learned without it,  our white-box attacks are able to confidently infer sensitive value records, even in the cases where the adversary has limited prior distribution information.
\autoref{tab:result_summary} summarizes some results from our experiments,
which are representative of the results we report on for two data sets and various settings in \autoref{sec:blackbox_results} and \autoref{sec:whitebox_results}. The results consistently show that having access to the model can greatly boost an adversary's inference success rate in the cases where adversary has limited prior knowledge of the training distribution. 

%% file: 2_background.tex
\section{Imputation and Inference} %Background}

This section summarizes prior works on data imputation and attribute inference, and provides experimental results to motivate the need to study sensitive value inference. 

\shortsection{Notation}
We denote $\cD: \bX \times \bY$ as the distribution over data points where $\bX$ is the domain of attributes and $\bY$ is the domain of class labels. Additionally, each data point $z = (v, t)$ consists of a sensitive feature $t$ and non-sensitive features $v$, such that $(v, t) \sim \bX$, and has an associated class label $y \sim \bY$. The support of $t$ is denoted by $\bT$. 
%\dnote{I'm confused by the use of "support" here - is that standard terminology? (from later uses, I see it is just the set of possible values for $t$?} \bnote{yes, it is the set of possible values of $t$, but "support" seems to be a standard terminology in ML -- this is also used by Yeom et al.}
$\psi(z) = v$ and $\pi(z) = t$ are projection functions with domain $\bX$, such that they map to non-sensitive and sensitive attributes of a record $z$ respectively. We denote sampling a data set $\bS$ consisting of $n$ data points from distribution $\cD$ as $\bS \sim \cD^n$.
%\dnote{how are you using bold here? is this following some standard notation? (usually I thought bold is only used for vectors)} \bnote{changed the font to normal}

\subsection{Data Imputation}\label{sec:def_imputation}

Often missing fields or attributes are encountered when dealing with real world data. Data imputation is a longstanding problem of imputing values for missing data, traditionally not considered a privacy issue and studied since the 1970s by the survey research community~\cite{rubin1976inference, rubin1978multiple, rubin1987multiple}. 
%One straightforward way to handle incomplete data is to just remove records that have missing fields. This, however, might drastically reduce the data size when many records have missing fields, and might also introduce biases in the data if missing fields are more common for certain types of data. Another 
A simple strategy is to fill the missing values with mean or median value for the given attribute, or to copy values from a nearby record (as is done by the US census). A more accurate imputation can be performed by taking into consideration factors such as prior probability and correlation between attributes. Machine learning can help find suitable values by using the correlation with known attributes. For instance, expectation maximization algorithms~\cite{little1987statistical, schafer1997analysis} can find the most probable value for the missing attribute given the values for the remaining attributes. Alternatively, k-nearest neighbor~\cite{batista2002study}, linear regression or neural network models~\cite{gupta1996estimating, abdella2005use} can also be employed to predict the missing attribute values. These methods are implemented by automated imputation tools, such as Amelia II~\cite{amelia2008} and MICE~\cite{van2011mice}. 
%While a detailed survey of all the imputation works is beyond the scope of our paper, we refer the readers to the survey papers of 
Gautam and Ravi~\cite{gautam2015data} and Bertsimas et al.~\cite{bertsimas2017predictive} provide detailed literature reviews on the machine learning approaches to data imputation.% including k-nearest neighbours, decision trees and neural networks.

In the context of adversarial attribute inference, the unknown sensitive attributes can be treated as missing values to be imputed. More formally, consider an imputation adversary $\bar{\cA}$ with knowledge of the data distribution $\cD$ who knows the non-sensitive attribute values $\psi(\bC)$ of a set of candidate records $\bC \sim \cD^m$ and wants to infer the sensitive attribute value $t = \pi(z)$ for each candidate record $z$ in $\bC$. Imputation of sensitive attribute $t$ of a record $z$ can be represented in terms of conditional probability $\mathrm{Pr}[t\; | \; \psi(z)]$. Given the support of $t$ is $\bT$, the imputation adversary outputs the value $t_i \in \bT$ that has the maximum conditional probability: %, as given below:
\begin{equation}\label{eq:data_imputation}
    \bar{\cA}(\psi(z), \cD, \bT) = \arg\max_{t_i \in \bT} \mathrm{Pr}[t_i \; | \; \psi(z)]
\end{equation}

While there are many imputation methods that aim to maximize the above conditional probability, neural network based imputation~\cite{sharpe1995dealing, nordbotten1996neural, gupta1996estimating} has been shown to achieve state-of-art performance in many real world settings such as imputing missing values in breast cancer data~\cite{jerez2010missing}, thyroid disease database~\cite{sharpe1995dealing} and census data~\cite{nordbotten1996neural}. Hence, we use this method in our experiments.

\begin{comment}
\begin{table*}[tb]
    \centering
    \begin{tabular}{lcccccccc}
    \toprule
        & \multicolumn{4}{c}{\bfdataset{Census19}} & \multicolumn{4}{c}{\bfdataset{Texas-100X}} \\
        & \multicolumn{2}{c}{Gender} & \multicolumn{2}{c}{Race} & \multicolumn{2}{c}{Gender} & \multicolumn{2}{c}{Ethnicity} \\ 
        & Train & Test & Train & Test & Train & Test & Train & Test \\ \midrule
        Predict Most Common & 0.52 & 0.52 & 0.78 & 0.79 & 0.62 & 0.62 & 0.72 & 0.73 \\
        Imputation (\autoref{eq:data_imputation}) & 0.59 & 0.59 & 0.81 & {\bf 0.80} & 0.66 & 0.65 & 0.83 & {\bf 0.83} \\
        Yeom Attack (\autoref{eq:yeom_ai}) & 0.57 & 0.54 & 0.67 & 0.60 & 0.57 & 0.54 & 0.58 & 0.52 \\
        CSMIA~\cite{mehnaz2022sensitive} (\autoref{eq:csmia}) & 0.63 & 0.55 & 0.10 & 0.07 & 0.59 & 0.51 & 0.60 & 0.55 \\
        CAI (\autoref{eq:cai}) & 0.63 & 0.55 & 0.10 & 0.07 & 0.62 & 0.54 & 0.64 & 0.56 \\
        WCAI (\autoref{eq:wcai}) & {\bf 0.64} & {\bf 0.60} & {\bf 0.82} & {\bf 0.80} & {\bf 0.68} & {\bf 0.66} & {\bf 0.84} & {\bf 0.83} \\
    \bottomrule
    \end{tabular}
    \caption{Comparing prediction accuracy of attribute inference attacks with imputation. The standard deviation is less than 0.01 for all the reported values. The inference attacks do not outperform imputation, and hence the accuracy gap between train and test sets is not a conclusive evidence for the model leaking information about specific training records.}
    \label{tab:attribute_inference}
\end{table*}
\end{comment}

\subsection{Attribute Inference}\label{sec:def_attr_inference}

In an \emph{attribute inference} attack, an adversary $\tilde{\cA}$ has access to the data distribution $\cD$, knows the non-sensitive attribute values $\psi(\bC)$ of a set of candidate records $\bC \subseteq \bS$ and uses the model $\cM_\bS$ trained on $\bS \sim \cD^n$ to infer the sensitive attribute value $t = \pi(z)$ for each candidate record $z$ in $\bC$. The key difference from the imputation adversary is that the attribute inference adversary has access to a model trained on $\bS$, and uses the additional information gained from the model to infer the sensitive attribute value. 

Previous attribute inference works~\cite{fredrikson2014privacy, yeom2018privacy, mehnaz2022sensitive} have only explored the black-box (API) attack setting where the adversary $\tilde{\cA}$ queries the model $\cM_\bS$ and gets either a predicted class label 
%$y'$ 
or a confidence vector.
%$\bV = (V_0, V_1 , \cdots V_l)$ where $V_i \in [0, 1]$ denotes the model's prediction confidence for the $i$-th class label. 
For a given partial record $\psi(z)$ with associated class label $y$, the general approach is to plug in all possible values $t_i \in \bT$ for the sensitive attribute to obtain a complete record, $z_i$, which the adversary queries the model with. The adversary then uses the model's output to infer the underlying sensitive attribute value $t$. We describe previous attribute inference attacks next and compare their effectiveness %with imputation 
in \autoref{tab:attribute_inference} (\autoref{sec:ai_analysis} provides more details on these experiments).

\shortsection{Fredrikson Attack} 
Fredrikson et al.~\cite{fredrikson2014privacy} studied two different inference attacks in their work. In the first attack, called \emph{model inversion}, the adversary queries a face-recognition model with the aim of retrieving the actual face images used in the model training. In the second attack, called \emph{attribute inference}, the adversary has partial information about a training record and aims to infer the unknown sensitive attribute by querying the model. We consider the latter attack here. In this attack, the adversary plugs in different values $t_i \in \bT$ for the sensitive attribute and obtains the query records $z_i$, which are then input to the model $\cM_\bS$ to get the corresponding predicted class labels $y'_i$. The adversary performs multiple queries in this fashion to create a confusion matrix:
\[C[y, y'] = \Pr[\cM_\bS(z) = y'\; | \; y \text{ is the true class label}]\]
The adversary then combines this with the marginal prior of the sensitive attribute $\Pr[t]$ and infers the value for the missing attribute that maximizes the combined value.
\begin{equation}\label{eq:fredrikson_ai}
    \tilde{\cA}(\psi(z), \cD, \cM_\bS, \bT) = \arg\max_{t_i \in \bT} \Pr[t_i] \cdot C[y, y'_i]
\end{equation}
We note that Fredrikson et al.~\cite{fredrikson2014privacy} assume the attributes are mutually independent, which is a very strong assumption. In realistic settings the attributes are often correlated, and in such cases the marginal prior $\Pr[t_i]$ in \autoref{eq:fredrikson_ai} should be replaced with conditional probability $\Pr[t_i \;| \; \psi(z)]$. 

\begin{table}[tb]
    \centering
    \begin{tabular}{lcccc}
    \toprule
        & \multicolumn{2}{c}{\bfdataset{Census19}} & \multicolumn{2}{c}{\bfdataset{Texas-100X}} \\
        & Gender & Race & Gender & Ethnicity \\ \midrule
        Predict Most Common & 0.52 & 0.78 & 0.62 & 0.72 \\
        Imputation (\autoref{eq:data_imputation}) & 0.59 & 0.82 & 0.66 & 0.72 \\
        Yeom Attack (\autoref{eq:yeom_ai}) & 0.57 & 0.65 & 0.57 & 0.58 \\
        CAI (\autoref{eq:cai}) & 0.63 & 0.06 & 0.62 & 0.64 \\
        WCAI (\autoref{eq:wcai}) & {\bf 0.64} & {\bf 0.83} & {\bf 0.68} & {\bf 0.74} \\
        CSMIA~\cite{mehnaz2022sensitive} (\autoref{eq:csmia}) & 0.63 & 0.06 & 0.59 & 0.60 \\
    \bottomrule
    \end{tabular}
    \caption{Comparing prediction accuracy of attribute inference attacks. \rm Results reported are average of five trials. The standard deviation is less than 0.01 for all the reported values. Note that in many cases, the attribute inference attacks do worse than just na\"ively predicting the most common attribute value.}
    \label{tab:attribute_inference}
\end{table}

\shortsection{Yeom Attack}
Yeom et al.~\cite{yeom2018privacy} propose an attribute inference attack that relies on black-box access to a membership inference oracle. In their attack, the adversary tries all possible values $t_i \in \bT$ for the unknown sensitive attribute and queries the membership inference oracle $\mathcal{O}_{MI}$ with the corresponding record $z_i$. The oracle outputs a binary membership decision indicating whether the record $z_i$ is part of the model training set. The adversary then chooses the value $t_i$ with the highest prior probability among the ones that pass the membership test. The attack can be described by this equation:
\begin{equation}\label{eq:yeom_ai}
    \tilde{\cA}(\psi(z), \cD, \cM_\bS, \bT) = \arg\max_{t_i \in \bT} \Pr[t_i] \cdot \mathcal{O}_{MI}(\cM_\bS, z_i)
\end{equation}

While the above formulation can support any membership inference attack as an oracle, the oracle $\mathcal{O}_{MI}$ of Yeom et al. takes the model confidence vector $\bV = (V_0, V_1 , \cdots V_l)$ and uses a threshold on the model confidence for the correct class label $V_y$ to predict the membership. Similar to the Fredrikson et al.~\cite{fredrikson2014privacy} attack, Yeom et al.'s attack also embeds a strong assumption that the attributes are mutually independent.

%\dnote{are these in some other work, or are they our extension? (I think new here, but paper doesn't make this clear} \bnote{we extend their attack, I have clarified this in the prose}
We improve Yeom et al.'s attack by removing the dependence on the oracle, and instead directly using the model confidence for the correct class label $V_y$ for attribute inference. We call the resulting black-box attack \emph{confidence-based attribute inference} (CAI), and describe it using this equation:
\begin{equation}\label{eq:cai}
    \tilde{\cA}(\psi(z), \cD, \cM_\bS, \bT) = \arg\max_{t_i \in \bT} V_y
\end{equation}

Combining the above attack with conditional probability of the sensitive attribute yields the  \emph{weighted CAI} (WCAI) attack:
\begin{equation}\label{eq:wcai}
    \tilde{\cA}(\psi(z), \cD, \cM_\bS, \bT) = \arg\max_{t_i \in \bT} \Pr[t_i | \psi(z)] \cdot V_y
\end{equation}

These modified versions of Yeom et al.'s attack outperform the original attack in \autoref{eq:yeom_ai} (as shown in \autoref{tab:attribute_inference}), and are later used for our black-box sensitive value inference attacks.% \dnote{can this be shown in Table 1?} \bnote{done}

\shortsection{Mehnaz Attack} Mehnaz et al.~\cite{mehnaz2022sensitive} recently proposed a \emph{confidence score-based model inversion attack} (CSMIA) and empirically showed it to surpass the attack of Fredrikson et al.~\cite{fredrikson2014privacy}. In their attack, the adversary queries the model $\cM_\bS$ with records $z_i$ having different values for the sensitive attribute $t_i \in \bT$, and obtains the corresponding predicted class labels $y_i'$ and model confidence for the predicted label $V_{y_i'}$. The adversary then uses this information to predict the sensitive value as follows:
\begin{equation}\label{eq:csmia}
    \tilde{\cA}(\psi(z), \cD, \cM_\bS, \bT) = 
    \begin{cases}
        t_i \in \bT | y_i' = y, & \text{if } \forall t_j \neq t_i, y_j' \neq y \\
        \arg\min_{t_i \in \bT} V_{y_i'}, & \text{if } \forall t_i \in \bT, y_i' \neq y \\
        \arg\max_{t_i \in \bT} V_{y_i'}, & \forall t_i \in \bT, y_i' = y
    \end{cases}
\end{equation}

We note that the above attack does not consider the prior marginal or conditional probability of sensitive attribute, and hence is similar to the CAI attack from \autoref{eq:cai}.

\subsection{Empirical Analysis}\label{sec:ai_analysis}

Previous experimental results on attribute inference (including Mehanaz et al.~\cite{mehnaz2022sensitive}) do not differentiate whether a successful inference is due to the model leakage or due to the correlations existing between the attributes which the inference captures. 
\autoref{tab:attribute_inference} summarizes the results of our experiments comparing the prediction accuracy of existing state-of-art attribute inference attacks with imputation on the \bfdataset{Census19} and \bfdataset{Texas-100X} data sets (\autoref{sec:datasets} provides details on these data sets, which we created as expanded versions of standard benchmarks to enable more extensive experiments).
We select the \emph{gender} and \emph{race} as sensitive attributes for \bfdataset{Census19} data set, and for \bfdataset{Texas-100X} data set we chose the \emph{gender} and \emph{ethnicity}. 
For our comparison, we include the membership oracle attack of Yeom et al.~\cite{yeom2018privacy} (Yeom Attack) and its modified versions (CAI and WCAI) that directly use the model confidence instead of relying on the membership oracle. We also include the CSMIA attack of Mehnaz et al.~\cite{mehnaz2022sensitive} that uses the model confidence and is shown to outperform the Fredrikson et al.~\cite{fredrikson2014privacy} attack.\footnote{The authors of CSMIA did not release a reference implementation of their attack. Hence, we implemented our own version of the attack based on the description in their paper~\cite{mehnaz2022sensitive}, and confirmed that it performed similarly to their reported results.}
For reference, we also include the prediction accuracy of a na\"{i}ve baseline that always predicts the most common value for the sensitive attribute. Note that this silly baseline already outperforms the three of the attribute inference attacks (Yeom, CAI, and CSMIA) in three of four settings, and is only consistently outperformed by WCAI. Imputation outperforms Yeom attack, CAI and CSMIA, and the accuracy gap between CAI and WCAI indicates that the inference success is mainly due to the imputation. Furthermore, the accuracy of WCAI is similar to that of imputation. Thus, the black-box attacks do not appear to pose a significant privacy risk beyond imputation on these metrics in these settings. This motivates us to develop more meaningful metrics for evaluating inference attacks (which we introduce in Section~\ref{sec:sensitive_value_inference} and use in our later experiments), and to explore white-box attacks that extract more meaningful information from the models (Section~\ref{sec:whitebox_attack}). 

\subsection{Theoretical Results}
While the above three works empirically explore attribute inference attacks, other works have theoretically explored the effectiveness of attribute inference. Yeom et al.~\cite{yeom2018privacy} reduce the attribute inference problem to querying a membership inference oracle, and prove that the success of attribute inference attacks is limited by the membership inference success. Zhao et al.~\cite{zhao2021onthe} also conclude that attribute inference fails when membership inference fails. Thus, the general view of these works is that attribute inference attacks do not pose a threat against differentially private models that are invulnerable to membership inference attacks. 
This conclusion is based on the \emph{attribute advantage metric} proposed by Yeom et al.~\cite{yeom2018privacy} that bounds the adversary's inference accuracy between training and non-training inputs, and is proven to be bounded by differential privacy. We evaluate an adversary's success in inferring an attribute by measuring the attack precision for a subset of records (see \autoref{sec:defenses}), regardless of whether they occur in the training set. Our advantage metric thus corresponds to the gap between the accuracy that can be achieved with (``attribute inference'') and without (``imputation'') access to the trained model. The attack success is due to the model revealing information about the distribution which is not mitigated by differential privacy where the individual data record is the unit of privacy. Our results corroborate the claim of Wu et al.~\cite{wu2016methodology} that differential privacy does not mitigate the ability of an adversary to infer an attribute, even if it can bound the gap between accuracy on training and non-training records.

%% file: 3_attack_definition.tex
\section{Threat Model}\label{sec:threatmodel}

The prediction accuracy experiments in the previous section which correspond to prior notions of attribute inference assume that the adversary has access to a large number of records sampled from the same distribution at the training dataset, and evaluate the uniform privacy risk for all values of the targeted sensitive attribute, averaged across all records in a test set. Neither of these assumptions holds for realistic scenarios where attribute inference matters, so we develop an extended threat model that considers different possibilities for the data available to the adversary (\autoref{sec:dataavailable}), and a more realistic attack goal where the adversary seeks to identify individuals with sensitive attributes with high confidence (\autoref{sec:sensitive_value_inference}).

\input{threat_model_fig}

\subsection{Data Availability}\label{sec:dataavailable}

Access to data is a critical component of any attribute inference or imputation attack. In our threat model, we consider a range of adversaries based on the auxiliary information available to the adversary summarized in \autoref{tab:threat_model}. This information is composed of two parts: (i) access to a model trained on the sensitive data, and (ii) knowledge of data. We use $M_{\mathrm{aux}}$ to denote the adversary's access to the trained model, $\cM_\bS$, which varies from highest to lowest, as: (a) having white-box access to $\cM_\bS$, (b) having black-box access to $\cM_\bS$, and (c) having no access to $\cM_\bS$. The last case corresponds to the situation where the model need not exist, and makes the inference adversary ($\cA$) equivalent to an imputation adversary ($\bar{\cA}$). 

The adversary's knowledge of data is denoted with $D_{\mathrm{aux}}$ and can be characterized according to two dimensions: the distribution the adversary has access to, and how many records they are able to sample. In the best case for the adversary, they have access to the same distribution (but no overlapping records) as was used to train the model ($D_{\mathrm{aux}} \sim \cD$); in more realistic cases, they have access to a different distribution ($D_{\mathrm{aux}} \sim \cD^* (\not\approx \cD)$) which may be more or less similar to the training distribution. All the prior attribute inference works assume that the adversary knows the training distribution. 

The data available to the adversary, $D_{\mathrm{aux}}$ can be further classified based on the adversary's sample set size ($|D_{\mathrm{aux}}|$) sampled from the distribution known to the adversary. This determines how much information the adversary has about the distribution $\cD$ (or $\cD^*$), independent of the model. Larger $|D_{\mathrm{aux}}|$ indicates the adversary has more accurate information about the underlying distribution. 

The importance of considering different levels of data available to the adversary will become clear when we compare the effectiveness of imputation and attribute inference attacks across various settings. In cases where the adversary has enough information to accurately model the training distribution, we find that an imputation adversary is accurate enough that little additional information can be learnt from access to a trained model. The serious privacy risks from attribute inference are limited to scenarios where the training distribution itself is not publicly available, so an adversary is able to make better predictions with access to a model trained on that distribution than they can by imputation alone from the limited or skewed data available.

\subsection{Sensitive Value Inference}\label{sec:sensitive_value_inference}

The uniform average accuracy metric does not capture well the risk of attribute inference in realistic scenarios, where some values of an attribute are more common than others and
the records with minority value for a sensitive attribute are likely to have the highest associated privacy risk. What matters most for privacy risk is not the average accuracy of the inferences across the entire candidate set, but whether it is possible to predict a minority value with high confidence for some records in a candidate set. To more realistically capture the risks of attribute inference, we propose an alternative definition which we call \emph{sensitive value inference}. Our definition captures both the asymmetric nature of the risks of inferring a sensitive value and emphasizes the threat of an adversary being able to make high confidence predictions for some candidate records.

Similar to the attribute inference threat model, the sensitive value inference adversary knows the non-sensitive attribute values $\tilde{\bC} = \psi(\bC)$ of a set of candidate records $\bC \subseteq \bS$. The adversary’s goal is to infer a subset of candidate records from $\bC$ for which $t = t^*$, where $t^*$ is the targeted sensitive value. We formalize the sensitive value inference attack using the following adversarial game inspired by the attribute inference game of Yeom et al.~\cite{yeom2018privacy}. 

\begin{experiment}[Sensitive Value Inference]~\label{exp:svi} \rm
Let $\cD$ be the distribution over training data points $(z, y)$, where $(v, t) = z \sim \bX$ are the attributes and $y \sim \bY$ is the class label. Let $\cA(D_{\mathrm{aux}}, M_{\mathrm{aux}}, t^*)$ be a sensitive value inference adversary that wants to infer the sensitive value $t^*$ for attribute $t$, and has auxiliary information about data $D_{\mathrm{aux}}$ and model access $M_{\mathrm{aux}}$. The sensitive value inference experiment proceeds as follows:
\begin{enumerate}
    \item Sample a training set $\bS$ from the distribution $\cD$ and train a model $\cM_\bS$.
    \item Sample a subset of candidates $\bC$ from $\bS$.
    \item Output $\mathds{1} [\cA(\psi(z), D_{aux}, M_{aux}, t^*) = \pi_{t^*}(z)], \forall  z \in \bC$.
\end{enumerate}
where $\mathds{1}$ is the indicator function and $\pi_{t^*}(z)$ is the projection function that outputs 1 if the record $z$ has the sensitive attribute value $t = t^*$, and outputs 0 otherwise.
\end{experiment}

The sensitive value inference adversary $\cA(D_{\mathrm{aux}}, M_{\mathrm{aux}}, t^*)$ takes partial candidate records $\psi(\bC)$ and outputs a binary decision for each candidate record $z \in \bC$, denoting whether $z$ has the sensitive value $t^*$. Next, we provide a straightforward method to obtain a sensitive value inference attack from any score-based attribute inference attack, such as the attacks discussed in \autoref{sec:def_attr_inference}.

\shortsection{Converting AI Attacks to Sensitive Value Inference}\label{ssec:ai_to_svi}
Consider a score-based attribute inference attack $\tilde{\cA}$ that infers the sensitive attribute value $t$ of a query record $z$ by selecting the value that maximizes the $\textit{score}$ as given below: \[\tilde{\cA}(\psi(z), D_{aux}, M_{aux}, \bT) = \arg\max_{t_i \in \bT} \textit{score}(z_i = (\psi(z), t_i))\]

The prior attacks discussed in \autoref{sec:def_attr_inference} assumed access to the training distribution (i.e. $D_{aux} \sim \cD$) and black-box access to model $\cM_\bS$. Hence these attacks were denoted by $\tilde{\cA}(\psi(z), \cD, \cM_\bS, \bT)$. Here we keep a more general notation $\tilde{\cA}(\psi(z), D_{aux}, M_{aux}, \bT)$ to denote a broader class of attribute inference attacks, and use this notation consistently for the rest of the paper. We can convert the above attribute inference adversary $\tilde{\cA}$ into a sensitive value inference adversary $\cA$ by setting a threshold score as:

\begin{equation}\label{eq:ai_to_svi}
    \cA(\psi(z), D_{aux}, M_{aux}, t^*) = 
    \begin{cases}
    1, & \text{if } \textit{score}(z^* = (\psi(z), t^*)) \ge \varphi \\
    0, & \textit{otherwise}
    \end{cases}
\end{equation}
Here, $t^*$ is the sensitive attribute value for the attribute $t$, and $\varphi$ is the threshold on the $\textit{score}$ metric. As described in Experiment~\ref{exp:svi}, the adversary $\cA$ receives a set of partial candidate records $\psi(\bC)$ and runs the sensitive value inference (\autoref{eq:ai_to_svi}) independently for each partial record $\psi(z)$ in the candidate set. 

Note that the attack success depends on the threshold value $\varphi$. To find a suitable threshold $\varphi$ on the $\textit{score}$ metric, the adversary can sort the candidate records based on their $\textit{score}$ and  select the threshold value that maximizes the positive predictive value (i.e., attack precision) of the sensitive value inference attack. Another way to think about this, which we use in our visualizations, is that by sorting the candidate records by $\textit{score}$ the adversary can select the $k$ candidate records most likely to have attribute value $t^{*}$ for any value of $k$. 
In the following sections, we show how the above procedure can be used to construct black-box and white-box sensitive value inference attacks. 

%% file: threat_model_fig.tex
\begin{table}[tb]
    \centering
    \setlength{\tabcolsep}{4pt}
    \begin{tabular}{clcc@{\hskip 1em}cc}
        \toprule
        \multicolumn{2}{c}{Similarity of Distribution} & \multicolumn{2}{c}{$D_{\mathrm{aux}} \sim \cD$} & \multicolumn{2}{c}{$D_{\mathrm{aux}} \sim \cD^*$} \\
        \multicolumn{2}{c}{Size of Dataset ($|D_{\mathrm{aux}}|$)} & \multicolumn{1}{c}{Large} & Small & Large & Small \\ \midrule
        \parbox[t]{2mm}{\multirow{3}{*}{\rotatebox[origin=c]{90}{$M_{\mathrm{aux}}$}}} & No Access & \multicolumn{4}{c}{Imputation} \\ \cmidrule(r){2-6}
        & Black-box & $\bullet$ & & & $\bigtriangleup$ \\
        & White-box & & $\bigtriangleup$ & $\bigtriangleup$ & $\bigtriangleup$ \\ \bottomrule
    \end{tabular}
    \caption{Threat models considered in this paper. \rm Rows describe the adversary $\cA$'s access to model ($M_{\mathrm{aux}}$), and columns describe $\cA$'s knowledge of data ($D_{\mathrm{aux}}$), which may be sampled from the same distribution as the training data ($\cD$) or a different distribution ($\cD^* \not\approx \cD)$). Cases where we have observed inference attacks that do significantly better than imputation are denoted with $\bigtriangleup$. Previous attribute inference works \cite{fredrikson2014privacy, yeom2018privacy, mehnaz2022sensitive} consider only the adversary with access to a large dataset from the training distribution, and black-box access to the model ($\bullet$).}
    \label{tab:threat_model}
\end{table}

%% file: 4_attack_methodology.tex
\section{Black-Box Attacks}\label{sec:blackbox_attack}
In a black-box attack, the adversary has unlimited API access to the model, and is able to submit queries to obtain output prediction confidence vectors for any inputs it wants.  We consider both pure black-box inference attacks that only use a released model, and combined attacks that combine imputation based on knowledge of an underlying distribution with black-box inference from a model. While we can use any black-box attribute inference attack discussed in \autoref{sec:def_attr_inference}, we choose the modified versions of Yeom et al.~\cite{yeom2018privacy} attack, CAI (\autoref{eq:cai}) and WCAI (\autoref{eq:wcai}), for our experiments as these perform the best among the known black-box attacks. 

\shortsection{Model Confidence Attack (\bfattack{BB})} 
As mentioned in \autoref{eq:cai}, the CAI attack uses the model's confidence for correct class label as the score metric to infer the sensitive attribute value. We can plug in this score metric into \autoref{eq:ai_to_svi} to obtain the corresponding sensitive value inference attack. For a given partial candidate record $\psi(z)$ with corresponding class label $y$, the sensitive value inference adversary $\cA$ queries the model $\cM_\bS$ with record $z^* = (\psi(z), t^*)$ and obtains the model confidence vector $\bV = (V_0, V_1 , \cdots V_l)$. The adversary then uses $V_y$ to infer if the record $z$ has the sensitive value $t^*$:
\begin{equation}\label{eq:bb}
    \cA(\psi(z), D_{aux}, M_{aux}, t^*) = 
    \begin{cases}
    1, & \text{if } V_y \ge \varphi \\
    0, & \textit{otherwise}
    \end{cases}
\end{equation}
We denote the above black-box attack as \attack{BB} in our experiments. For brevity, we refer to $V_y$ as the \emph{model confidence} for the rest of the paper, unless specified otherwise.

\shortsection{Weighted Model Confidence Attack (\bfattack{BB$\cdot$IP})}
Similar to the \attack{BB} attack above, we convert the WCAI attack from \autoref{eq:wcai} to the corresponding sensitive value inference attack as follows:
\begin{equation}\label{eq:bb_ip}
    \cA(\psi(z), D_{aux}, M_{aux}, t^*) = 
    \begin{cases}
    1, & \text{if } \Pr[t^* \; | \; \psi(z)] \cdot V_y \ge \varphi \\
    0, & \textit{otherwise}
    \end{cases}
\end{equation}
We name the above attack \attack{BB$\cdot$IP} since it combines the \attack{BB} attack with the imputation by multiplying the model confidence $V_y$ with conditional probability of sensitive value $\Pr[t^* \;| \; \psi(z)]$.

\shortsection{Decision Tree Model Confidence Attack (\bfattack{BB$\mathbf{\diamondsuit}$IP})}
While \attack{BB$\cdot$IP} multiplies the model confidence and the imputation output, there are other ways to combine the information from these two approaches. One effective way is to train a machine learning model that takes both model confidence and imputation output and outputs a combined confidence score that reflects the likelihood of the query record having the sensitive value $t^*$. To obtain such a model, the adversary $\cA$ first obtains the known set $\bU$ of complete records based on the auxiliary data knowledge $D_{aux}$. The adversary then gets the model confidence $V_y$ and imputation output $Pr[t^* | \psi(z)]$ for each record $z$ in $\bU$. These are then used to train a  model $f$ to output a confidence score $f(Pr[t^* | \psi(z)], V_y)$ between 0 and 1, such that the score is high for records that have the sensitive value $t = t^*$, and the score is low for the remaining records in $\bU$.
We use decision tree for this because it is a non-linear model that is easy to interpret.
In the testing phase, $\cA$ uses $f$ to infer the sensitive value of candidate records. We call this attack \attack{BB$\diamondsuit$IP}, and define it as:
\begin{equation}\label{eq:bb_ip_f}
    \cA(\psi(z), D_{aux}, M_{aux}, t^*) = 
    \begin{cases}
    1, & \text{if } f(\Pr[t^* | \psi(z)], V_y) \ge \varphi \\
    0, & \textit{otherwise}
    \end{cases}
\end{equation}

\section{White-Box Attacks}\label{sec:whitebox_attack}

As discussed in \autoref{sec:def_attr_inference}, previous research on attribute inferences has focused on black-box attacks~\cite{fredrikson2014privacy, yeom2018privacy, mehnaz2022sensitive}. There has been no work on developing practical white-box attribute inference attacks that exploit the model weights, even though work on other inference attacks~\cite{carlini2019secret, carlini2020extracting, hisamoto2020membership, lehman2021does} has shown that the deep neural networks may leak additional information to attacks that consider their internal parameters and activations. While there have been a few white-box model inversion attacks~\cite{nguyen2016synthesizing, zhang2020secret} that use an image-recognition model's weights to infer the face images similar to the ones in the training data, they target a different problem setting and are not applicable to attribute inference.
We propose a novel white-box attack that takes advantage of the neuron activation values in neural network models. The intuition for this attack is that a subset of neurons in the neural network tend to be correlated with the different attribute values of the input records. Hence, a white-box adversary can benefit from identifying the neurons that are correlated to the sensitive value. We test this hypothesis and find that we can identify neurons that have higher activation values for inputs matching the sensitive attribute value.

\begin{figure}[tb]
    \centering
    \includegraphics[width=0.45\textwidth]{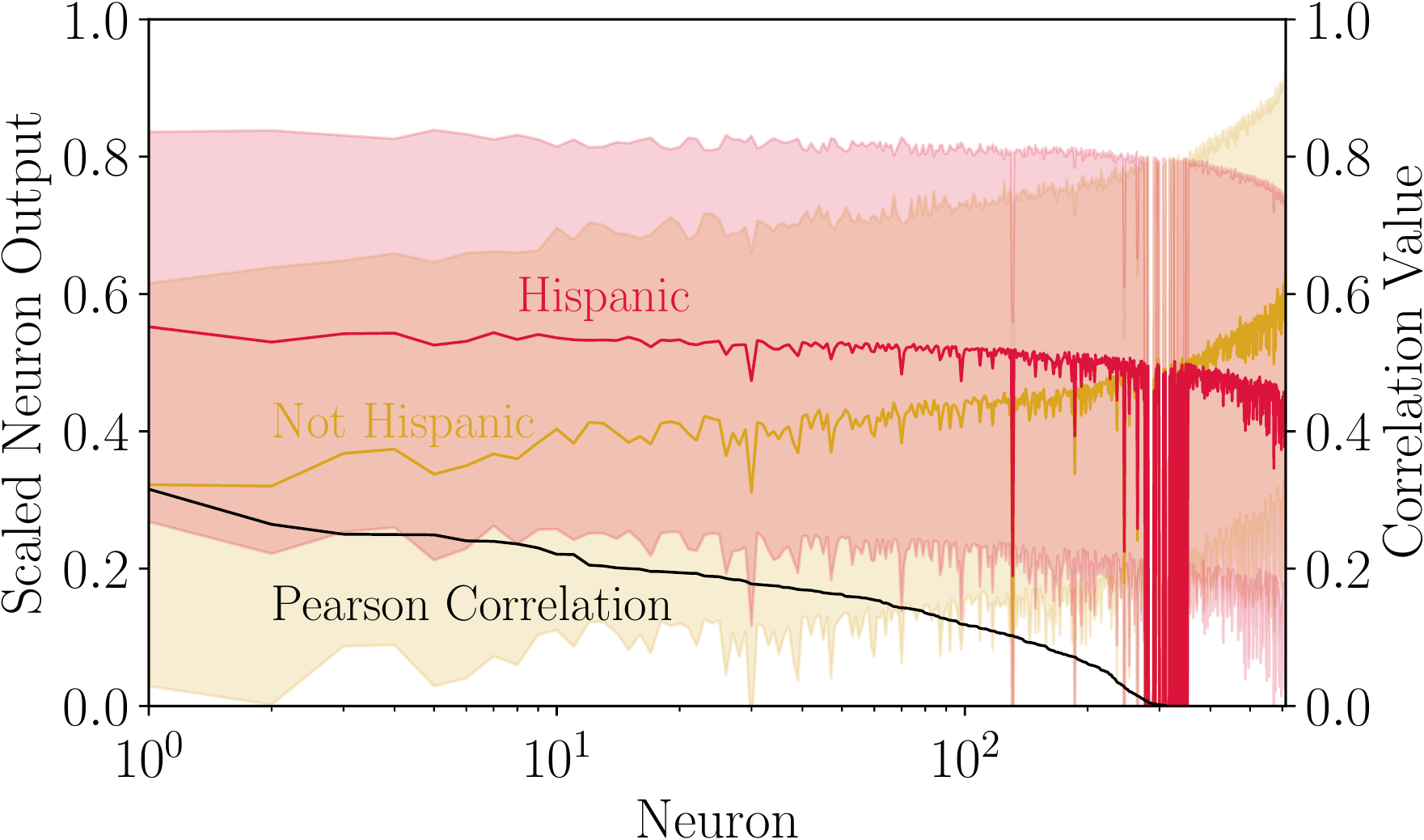}
    \caption{Correlation of neuron activations with attributes. \rm The graph shows the scaled activation of neurons (mean $\pm$ std) of neural network model correlated to Hispanic ethnicity in \bfdataset{Texas-100X}, sorted in decreasing order of correlation value.}% \dnote{need a more extensive caption to understand this figure}}
    \label{fig:texas_ethnicity_low_neuron_output}
\end{figure}

\shortsection{Neuron Output Attack (\bfattack{WB})}
In this attack, the adversary $\cA$ has a training set $\bU$ of records for which they know the sensitive attribute value. This could be obtained by the adversary based on the auxiliary knowledge $D_{aux}$. For each record $z_i \in \bU$ where $t_i \neq t^*$, $\cA$ flips its value $t_i$ to $t^*$ and keeps the other records the same. $\cA$ then identifies neurons that have higher activation value on an average for the records with sensitive value $t^*$, and have lower activation value for the records that originally had $t \neq t^*$. We do this by calculating the Pearson correlation coefficient for each neuron with respect to the sensitive value $t^*$ and sorting them in decreasing order. The Pearson correlation coefficient lies between -1 and 1, where 1 denotes strong positive correlation.
As an example, \autoref{fig:texas_ethnicity_low_neuron_output} shows the activation of neurons (uniformly scaled to the 0--1 range using the quantiles information) correlated to the the records with Hispanic ethnicity in \bfdataset{Texas-100X} (described in Section~\ref{sec:datasets}), sorted in decreasing order of correlation value. 
The most correlated neuron has $0.39 \pm 0.04$ Pearson correlation coefficient across five runs, and the correlation value slowly decreases to 0. 
We find the top 10 neurons have more than $0.27 \pm 0.01$ correlation value on average. We find similar correlations with other data sets and sensitive attributes. 

For our experiments with this attack, we use these top-10 most correlated neurons. Though this choice seems heuristic and there is more information that might be gleaned from additional neurons, we find that these top 10 neurons give a strong signal for our white-box attack. Indeed in our initial experiments across two data sets we vary the top-$k$ neurons between 1, 2, 5, 10 and 100, and find top-10 to give the best results. 
We obtain the aggregate neuron output $\textit{op}$ by taking a weighted average of the scaled activation values of top-10 neurons using the neuron correlation values as weights. The $\textit{op}$ value is between 0 and 1 and can be considered as the confidence of the white-box attack in predicting the sensitive value. We call this attack \attack{WB}, and define it as:
\begin{equation}\label{eq:wb}
    \cA(\psi(z), D_{aux}, M_{aux}, t^*) = 
    \begin{cases}
    1, & \text{if } \textit{op} \ge \varphi \\
    0, & otherwise
    \end{cases}
\end{equation}

\shortsection{Weighted Neuron Output Attack (\bfattack{WB$\cdot$IP})}
Analogous to the \attack{BB$\cdot$IP} attack, the \attack{WB$\cdot$IP} attack multiplies the aggregate neuron output $\textit{op}$ with the conditional probability $\Pr[t^*\;|\; \psi(z)]$:

\begin{equation}\label{eq:wb_ip}
    \cA(\psi(z), D_{aux}, M_{aux}, t^*) = 
    \begin{cases}
    1, & \text{if } \Pr[t^* | \psi(z)] \cdot \textit{op} \ge \varphi \\
    0, & \textit{otherwise}
    \end{cases}
\end{equation}

\shortsection{Tree Based Neuron Output Attack (\bfattack{WB$\diamondsuit$IP})}
We also evaluate a decision tree based combination of aggregate neuron output and imputation confidence, similar to the \attack{BB$\diamondsuit$IP} attack:
\begin{equation}\label{eq:wb_ip_f}
    \cA(\psi(z), D_{aux}, M_{aux}, t^*) = 
    \begin{cases}
    1, & \text{if } f(\Pr[t^* | \psi(z)], \textit{op}) \ge \varphi \\
    0, & otherwise
    \end{cases}
\end{equation}

\begin{comment}
\begin{defn}[($k$, $\theta$)-Attribute Inference]
Given a model $\cM_\bS$ trained on data set $\bS \sim \cD$, an adversary $\cA$ achieves $\theta$-attribute inference if there exists a non-empty candidate set $\bC \subseteq \bS$ of size $k$, such that the attribute experiment 
\[ Exp(\cA, \cM, \cD) \ge \theta \]
\end{defn}

\bnote{The above is a bound on the metric, not an AI defn.. check the pufferfish defn of attribute privacy: https://arxiv.org/pdf/2009.04013.pdf (seems like property inference). We can provide a bound on the metric along with a privacy definition/theorem linking the bound with DP.}
\end{comment}

%% file: 5_experiment_setup.tex
\section{Experimental Design}\label{sec:exp_setup}
In this section, we introduce the data sets used and describe the model training procedure and evaluation method for our sensitive value inference attack experiments. 

\subsection{Data Sets}\label{sec:datasets}

Realistic attribute inference attack experiments require tabular data sets that are proxy for sensitive user information, such as patient health records or census data which include sensitive attributes like gender, medical condition, race and ethnicity. Since the datasets used in previous inference experiments are either image datasets or small-scale toy data sets, we created two large data sets and have made them publicly available.\footnote{\url{https://github.com/bargavj/Texas-100X}; \url{https://github.com/JerrySu11/CensusData}}

\shortsection{\bfdataset{Texas-100X}}
The \dataset{Texas-100X} data set is an extended version of the \dataset{Texas-100} hospital data set from Shokri et al.~\cite{shokri2017membership}. Each record consists of patient's demographic information, such as age, gender, race and ethnicity, and medical information such as duration of hospitalization, type of admission, source of admission, admitting diagnosis, patient status, medical charges and principal surgical procedure. The task is to predict one of the 100 surgical procedures based on the patient's health record. The original \dataset{Texas-100} data set consists of 60\,000 records with 6\,000 anonymized binary attributes and hence is not suitable for our problem setting. We curate a new data set from the same public source files\footnote{Texas Hospital Inpatient Discharge Public Use Data File, [Quarters 1, 2, 3 and 4 from year 2006]. Texas Department of State Health Services, Austin, Texas.} used to create \dataset{Texas-100}. Our new data set, \dataset{Texas-100X}, has 925\,128 patient records collected from 441 hospitals, and retains the original 10 demographic and medical attributes in non-anonymized form. This allows us to target sensitive features, such as ethnicity, in evaluating our attacks. We find that the race attribute is highly correlated with ethnicity and hence we drop race from model training when inferring ethnicity.

\shortsection{\bfdataset{Census19}}
The \dataset{Census19} data set is curated from the 2019 US Census Bureau Database\footnote{American Community Survey (ACS) Public Use Microdata Sample (PUMS). 
\url{https://www.census.gov/programs-surveys/acs/microdata/access.html}.} and is similar to the widely used \dataset{Adult} census data set~\cite{adult} (derived from 1994 Census data). The \dataset{Adult} data set consists of around 48\,000 records with 14 features, some of which are duplicates or highly correlated to other features. We derived a similar census data set from the public use microdata sample (PUMS) files for 2019. The records are geographically grouped based on the public use microdata areas (PUMAs) which are unique non-overlapping regions within the states each with at least 100\,000 people. The data set we curate has records from 2\,351 PUMAs, although these PUMA identifiers are only for sampling data and are not used as attributes for model training or inference. 
The resulting \dataset{Census19} data set consists of 1\,676\,013 records with 12 features denoting the personal and demographic information about individuals in the United States. These features include age, gender, race, marital status, education, occupation, work hours and native country, as well as attributes related to cognitive, ambulatory, vision and hearing disability. The classification task is to predict whether an individual earns more than \$90\,000 annually, similar to traditional task for the \dataset{Adult} data set (which used \$50\,000 for the income threshold, which we have inflation-adjusted from 1994 to 2019).

\subsection{Model Training}~\label{sec:model_training}

For our experiments with both \dataset{Texas-100X} and \dataset{Census19}, we randomly select 50,000 records to form the training set and use it to train a two-layer neural network model. We also randomly sample 25,000 additional records from the remaining data to form the test set such that the training and test sets are mutually exclusive. The neural network consists of two hidden layers, each having 256 neurons with ReLU activation function. The output layer is a softmax layer consisting of one neuron for each output class. This is a standard neural network architecture used in prior inference works~\cite{shokri2017membership, jayaraman2020revisiting}, and we use the training hyperparameter settings of Jayaraman et al.~\cite{jayaraman2020revisiting} to obtain models with reasonable utility. For the \dataset{Texas-100X} data set with a 100-class classification task, the model achieves 61\% training accuracy and 46\% test accuracy. 
For the \dataset{Census19} data set with binary classification task, the neural network model achieves 88\% training accuracy and 86\% test accuracy. These results are similar to what this model architecture and training process achieves on the \dataset{Adult} data set~\cite{shokri2017membership}.

\subsection{Attack Evaluation}\label{sec:attack_evaluation}

For each of our experiments, we randomly select 10,000 candidate records from the training set. The goal of the adversary is to infer a specific-size subset of those candidates that have the sensitive attribute value. For the \dataset{Texas-100X} data set, we choose the ethnicity as the target attribute and the sensitive value is Hispanic. Ethnicity is a binary attribute with around 28\% records having the value that corresponds to Hispanic ethnicity. For the \bfdataset{Census19} data set, we select the race attribute which has seven values. %: White, Black, Asian, Pacific Islander, Native American, Some Other Race, and Mixed Race. 
The adversary's goal is to identify the candidate records that have attribute value Asian, which accounts for around 6\% records. While the choice of these target attributes is somewhat arbitrary, they have been selected as representative proxies for real-world sensitive attributes---race and ethnicity can be sensitive (and have been considered by prior attribute inference works), and we selected sensitive values that are minority values for these attributes. We evaluate the attack success using the positive predictive value (PPV) metric that tells what fraction of the sensitive attribute value predictions made by the adversary are correct. A PPV of 1 would mean the attack always predicts correctly.

We explore various threat model settings (\autoref{sec:dataavailable}) in our experiments by varying the adversary's knowledge of data, $D_{aux}$, and access to the model, $M_{\mathrm{aux}}$. The adversary's access to the trained model varies between white-box access, black-box access and no access. The adversary's knowledge of data is broadly categorized into two groups: %\dnote{include the notations from Table 1}
(b) adversary knows the training distribution $\cD$ and can sample records similar to the training records from that distribution, and (b) adversary does not know the training distribution and instead relies on a different data distribution $\cD^*$ to sample records. We consider different ways of skewing the distribution, discussed below. For both these groups, $D_{\mathrm{aux}}$ is further varied based on the number of sample records available to the adversary. We consider $|D_{\mathrm{aux}}|$ as 50, 500, 5\,000, and 50\,000 records.

\shortsection{Distribution Skewing}
As described in \autoref{sec:model_training}, the training set is uniformly randomly sampled from the whole data set for both \dataset{Texas-100X} and \dataset{Census19}. For \dataset{Texas-100X}, this corresponds to uniformly sampling without replacement from 441 Texas hospitals, and for \dataset{Census19}, the training data is uniformly sampled without replacement from 2\,351 public use microdata areas (PUMAs). To simulate the threat setting where the adversary has access to the training distribution $\cD$, we uniformly sample the adversary's candidate set $\bU$ from the respective data sets such that there is no overlap with the training set $\bS$. For the threat setting where the adversary does not know the training distribution $\cD$ and instead has access to a different distribution $\cD^*$, we skew the data distribution available to the adversary to be different from the training distribution. For \dataset{Texas-100X}, we consider two skewed distributions: $\cD_{LP}$, consisting of set of 266 hospitals that have the lowest population of patients, and $\cD_{HP}$, consisting of set of 7 hospitals that have the highest population of patients. The proportion of records that are Hispanic in $\cD_{LP}$ is 19\%, while $\cD_{HP}$ is 30\% Hispanic (we speculate that this reflects the demographics of Texas where larger urban areas with large hospitals have higher concentrations of Hispanic population than rural areas with small hospitals). Similarly for \dataset{Census19}, we consider two skewed distributions: $\cD_{LP}$, consisting of set of 210 PUMAs that have the lowest population, and $\cD_{HP}$, consisting of set of 58 PUMAs that have the highest population. For this data set, $\cD_{LP}$ has 4.3\% Asian records and $\cD_{HP}$ has 3.7\% Asian records. 

%% file: 6_blackbox_results.tex
\section{Imputation and Black-Box Attacks}\label{sec:blackbox_results}

\begin{table*}[tb]
    \centering
    \begin{tabular}{clccccccccc}
        \toprule
        & & \multicolumn{3}{c}{$\cA$ knows train distribution $\cD$} & \multicolumn{3}{c}{$\cA$ knows different distribution $\cD_{HP}$} & \multicolumn{3}{c}{$\cA$ knows different distribution $\cD_{LP}$} \\ 
        & $|D_{aux}| $ & 5\,000 & 500 & 50 & 5\,000 & 500 & 50 & 5\,000 & 500 & 50 \\ \midrule
        \parbox[t]{2mm}{\multirow{7}{*}{\rotatebox[origin=c]{90}{\dataset{Texas-100X}}}} & \attack{IP} & 0.62 $\pm$ 0.05 & 0.39 $\pm$ 0.03 & 0.24 $\pm$ 0.01 & 0.63 $\pm$ 0.03 & 0.39 $\pm$ 0.03 & 0.40 $\pm$ 0.03 & 0.44 $\pm$ 0.02 & 0.41 $\pm$ 0.05 & 0.37 $\pm$ 0.05 \\ 
        & \attack{WB} & 0.49 $\pm$ 0.02 & {\color{red} 0.52 $\pm$ 0.03} & {\color{red} 0.47 $\pm$ 0.05} & 0.49 $\pm$ 0.03 & 0.50 $\pm$ 0.08 & 0.45 $\pm$ 0.04 & {\color{red} 0.49 $\pm$ 0.02} & 0.49 $\pm$ 0.04 & {\color{red} 0.52 $\pm$ 0.07} \\
        & \attack{WB$\cdot$IP} & 0.59 $\pm$ 0.04 & {\color{red} 0.62 $\pm$ 0.08} & {\color{red} 0.42 $\pm$ 0.03} & 0.58 $\pm$ 0.02 & {\color{red} 0.51 $\pm$ 0.04} & 0.48 $\pm$ 0.07 & {\color{red} 0.52 $\pm$ 0.02} & 0.49 $\pm$ 0.05 & {\color{red} 0.54 $\pm$ 0.04} \\
        & \attack{WB$\diamondsuit$IP} & {0.64 $\pm$ 0.04} & 0.51 $\pm$ 0.09 & {\color{red} 0.50 $\pm$ 0.02} & 0.59 $\pm$ 0.07 & {\color{red} 0.54 $\pm$ 0.06} & {\color{red} 0.50 $\pm$ 0.04} & {\color{red} 0.50 $\pm$ 0.03} & 0.47 $\pm$ 0.05 & 0.48 $\pm$ 0.08 \\
        & \attack{BB} & 0.28 $\pm$ 0.04 & 0.28 $\pm$ 0.04 & 0.28 $\pm$ 0.04 & 0.28 $\pm$ 0.04 & 0.28 $\pm$ 0.04 & 0.28 $\pm$ 0.04 & 0.28 $\pm$ 0.04 & 0.28 $\pm$ 0.04 & 0.28 $\pm$ 0.04 \\
        & \attack{BB$\cdot$IP} & 0.58 $\pm$ 0.05 & {\color{red} 0.47 $\pm$ 0.02} & {\color{red} 0.35 $\pm$ 0.02} & 0.60 $\pm$ 0.05 & 0.42 $\pm$ 0.02 & 0.43 $\pm$ 0.03 & 0.49 $\pm$ 0.04 & 0.36 $\pm$ 0.02 & 0.34 $\pm$ 0.04 \\
        & \attack{BB$\diamondsuit$IP} & 0.60 $\pm$ 0.04 & 0.42 $\pm$ 0.01 & {\color{red} 0.33 $\pm$ 0.04} & 0.60 $\pm$ 0.07 & 0.39 $\pm$ 0.03 & 0.43 $\pm$ 0.05 & {\color{red} 0.48 $\pm$ 0.01} & 0.39 $\pm$ 0.04 & 0.38 $\pm$ 0.03 \\ \midrule
        
        \parbox[t]{2mm}{\multirow{7}{*}{\rotatebox[origin=c]{90}{\dataset{Census19}}}} & \attack{IP} & 0.91 $\pm$ 0.03 & 0.25 $\pm$ 0.02 & 0.55 $\pm$ 0.06 & 0.89 $\pm$ 0.02 & 0.11 $\pm$ 0.04 & 0.36 $\pm$ 0.06 & 0.90 $\pm$ 0.03 & 0.46 $\pm$ 0.04 & 0.42 $\pm$ 0.04 \\
        & \attack{WB} & 0.85 $\pm$ 0.03 & {\color{red} 0.82 $\pm$ 0.05} & 0.67 $\pm$ 0.06 & 0.85 $\pm$ 0.01 & {\color{red} 0.82 $\pm$ 0.05} & {\color{red} 0.64 $\pm$ 0.09} & 0.86 $\pm$ 0.05 & {\color{red} 0.85 $\pm$ 0.04} & {\color{red} 0.65 $\pm$ 0.12} \\
        & \attack{WB$\cdot$IP} & 0.88 $\pm$ 0.04 & {\color{red} 0.80 $\pm$ 0.05} & {\color{red} 0.71 $\pm$ 0.05} & 0.87 $\pm$ 0.03 & {\color{red} 0.77 $\pm$ 0.06} & {\color{red} 0.65 $\pm$ 0.10} & 0.87 $\pm$ 0.02 & {\color{red} 0.84 $\pm$ 0.04} & {\color{red} 0.74 $\pm$ 0.07} \\
        & \attack{WB$\diamondsuit$IP} & 0.87 $\pm$ 0.04 &{\color{red}  0.85 $\pm$ 0.03} & {\color{red} 0.73 $\pm$ 0.04} & 0.87 $\pm$ 0.04 & {\color{red} 0.83 $\pm$ 0.02} & {\color{red} 0.70 $\pm$ 0.08} & 0.82 $\pm$ 0.02 & {\color{red} 0.84 $\pm$ 0.05} & {\color{red} 0.78 $\pm$ 0.07} \\
        & \attack{BB} & 0.05 $\pm$ 0.02  & 0.05 $\pm$ 0.02 & 0.05 $\pm$ 0.02 & 0.05 $\pm$ 0.02 & 0.05 $\pm$ 0.02 & 0.05 $\pm$ 0.02 & 0.05 $\pm$ 0.02 & 0.05 $\pm$ 0.02 & 0.05 $\pm$ 0.02 \\
        & \attack{BB$\cdot$IP} & 0.86 $\pm$ 0.03 & 0.24 $\pm$ 0.03 & 0.50 $\pm$ 0.06 & 0.90 $\pm$ 0.03 & 0.11 $\pm$ 0.05 & 0.25 $\pm$ 0.03 & 0.88 $\pm$ 0.02 & 0.45 $\pm$ 0.04 & 0.34 $\pm$ 0.06 \\
        & \attack{BB$\diamondsuit$IP} & 0.89 $\pm$ 0.02 & 0.27 $\pm$ 0.04 & 0.60 $\pm$ 0.07 & 0.88 $\pm$ 0.03 & {\color{red} 0.28 $\pm$ 0.09} & {\color{red} 0.55 $\pm$ 0.04} & 0.84 $\pm$ 0.03 & {\color{red} 0.56 $\pm$ 0.03} & {\color{red} 0.63 $\pm$ 0.05} \\ \bottomrule
    \end{tabular}
    \caption{Average PPV of inference attacks for predicting the sensitive value of top-100 records with varying adversarial knowledge about data and model access, and different attack methods. \rm Reported results are for predicting \emph{Hispanic} ethnicity in \bfdataset{Texas-100X} and \emph{Asian} race in \bfdataset{Census19}. Cases in which the inference attack PPV (mean - std) is significantly greater than the imputation PPV (mean + std) are highlighted in red. These are the cases where our experiments show an attribute inference adversary benefiting from having access to the trained model over an imputation adversary with the same training data but no access to the model.}
    \label{tab:ppv_top_100}
\end{table*}

\begin{comment}
\begin{table*}[tb]
    \centering
    \begin{tabular}{lr!{\;}ccccccc}
        \toprule
    \multicolumn{1}{c}{Data Set} & \multicolumn{1}{c}{$k$} & \attack{IP} & \attack{BB} & \attack{BB$\cdot$IP} & \attack{BB$\diamondsuit$IP} & \attack{WB} & \attack{WB$\cdot$IP} & \attack{WB$\diamondsuit$IP} \\ \midrule
        \bfdataset{Texas-100X} & 10 & \textbf{0.92 $\pm$ 0.07} & 0.34 $\pm$ 0.15 & 0.90 $\pm$ 0.09 & 0.80 $\pm$ 0.15 & 0.86 $\pm$ 0.08 & 0.90 $\pm$ 0.00 & \textbf{0.92 $\pm$ 0.04} \\
        \bfdataset{Texas-100X} & 100 & 0.79 $\pm$ 0.06 & 0.35 $\pm$ 0.01 & 0.75 $\pm$ 0.04 & 0.81 $\pm$ 0.01 & 0.86 $\pm$ 0.03 & \textbf{0.87 $\pm$ 0.01} & \textbf{0.87 $\pm$ 0.04} \\[0.5ex]
        \bfdataset{Census19} & 10 & 0.92 $\pm$ 0.07 & 0.00 $\pm$ 0.00 & 0.86 $\pm$ 0.08 & 0.90 $\pm$ 0.09 & 0.90 $\pm$ 0.06 & \textbf{0.96 $\pm$ 0.05} & 0.92 $\pm$ 0.10 \\
        \bfdataset{Census19} & 100 & 0.86 $\pm$ 0.02 & 0.03 $\pm$ 0.01 & 0.88 $\pm$ 0.02 & 0.83 $\pm$ 0.02 & 0.85 $\pm$ 0.04 & \textbf{0.89 $\pm$ 0.04} & 0.85 $\pm$ 0.03 \\ \bottomrule
    \end{tabular}
    \caption{Comparing the average PPV of inference attacks for predicting the sensitive value of top-$k$ records at high adversarial knowledge. Reported results are for predicting \emph{Hispanic} ethnicity in \bfdataset{Texas-100X} and \emph{Asian} race in \bfdataset{Census19}.}
    \label{tab:ppv_top_k}
\end{table*}
\end{comment}

Prior black-box attribute inference approaches~\cite{yeom2018privacy, mehnaz2022sensitive} do not evaluate whether the success of the inference attack is due to the model or due to the imputation, as noted in Section~\ref{sec:def_attr_inference}. We make this difference explicit in our experiments. Our results suggest that the black-box attack success is nearly all due to the imputation, and hence these are not dataset inference attacks per se, but reveal that the trained model leaks statistical information about its training distribution.
\autoref{tab:ppv_top_100} compares the average PPV of the black-box attacks to results from imputation without any access to the model in predicting the sensitive value of top-100 records for \dataset{Texas-100X} and \dataset{Census19} across different threat models. Across all experiments, the black-box attack (\attack{BB}) has very low PPV (never exceeding 0.5 for the \dataset{Texas-100X}, and close to 0.0 for \dataset{Census19}), and is always significantly worse than imputation (\attack{IP}). When combined with imputation (\attack{BB$\cdot$IP} and \attack{BB$\diamondsuit$IP}), the combined attack only occasionally outperforms imputation alone (\attack{IP}) but in most of such cases, the improvement is within the error margin.

For these experiments, we randomly choose 10\,000 candidate records from the training set and run the attacks to identify $k$ records with sensitive value. We report the PPV of the top $k$ records as ranked by the scoring mechanism of the attack. Around 28\% of the records are Hispanic in \dataset{Texas-100X}, and around 6\% of the records are Asian in \dataset{Census19}. Hence a random guess would have PPV of 0.28 and 0.06 for the respective scenarios. A perfect inference attack would achieve PPV of 1.0 for an identification subset up to size $k$ = 2\,800 for \dataset{Texas-100X}. We vary the $D_{\mathrm{aux}}$ available to the adversary in terms of what distribution the adversary has access to (i.e. train distribution $\cD$ or other skewed distributions $\cD_{HP}$ or $\cD_{LP}$) and how many data samples the adversary can obtain from the distribution (50, 500, or 5\,000). Higher sample size $|D_{\mathrm{aux}}|$ allows the adversary to train stronger attacks and hence infer sensitive value records with higher confidence.

The imputation adversary (\attack{IP}) achieves close to 0.62 PPV for predicting the Hispanic ethnicity in top-100 records in \dataset{Texas-100X} when it is trained on 5\,000 records from the training distribution $\cD$. 
However, we see a steep drop in PPV from 0.62 to 0.39 when \attack{IP} is only trained on 500 records. This further drops to 0.24 PPV when the sample set has 50 records, at which point \attack{IP} does worse than random guessing of 0.28 PPV. We a observe similar trend when the adversary has access to other distributions $\cD_{HP}$ (hospitals with high patient population) and $\cD_{LP}$ (hospitals with low patient population). %\dnote{remind what this is for Texas}. 
While \attack{IP} trained on 5\,000 records from $\cD_{HP}$ also achieves similar PPV as when it is trained on the train distribution. This may be due to a smaller gap between the two distributions: $\cD$ has around 28\% Hispanic records and $\cD_{HP}$ has around 30\% Hispanic records. Though we note that \attack{IP} trained on 5\,000 records from $\cD_{LP}$ only achieves 0.44 PPV. This could be attributed to the larger gap between $\cD_{LP}$ and $\cD$, as $\cD_{LP}$ only has 20\% Hispanic records. 

\begin{figure*}[tb]
     \centering
     \begin{subfigure}[b]{0.42\textwidth}
         \centering
         \includegraphics[width=\textwidth]{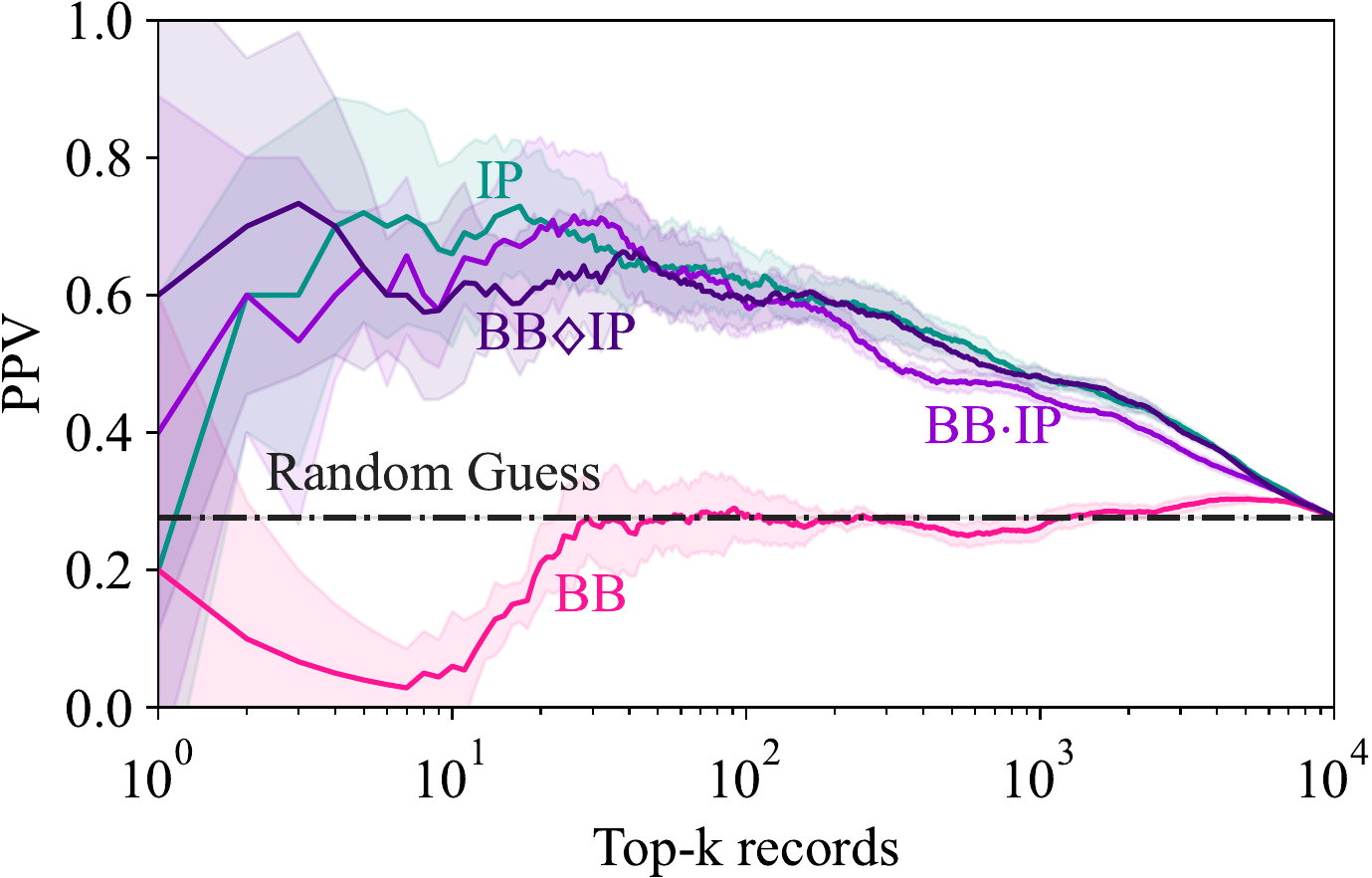}
         \caption{Black-box Attacks}
         \label{fig:ppv_texas_med_bb_main}
     \end{subfigure}\qquad\qquad
     \begin{subfigure}[b]{0.42\textwidth}
         \centering
         \includegraphics[width=\textwidth]{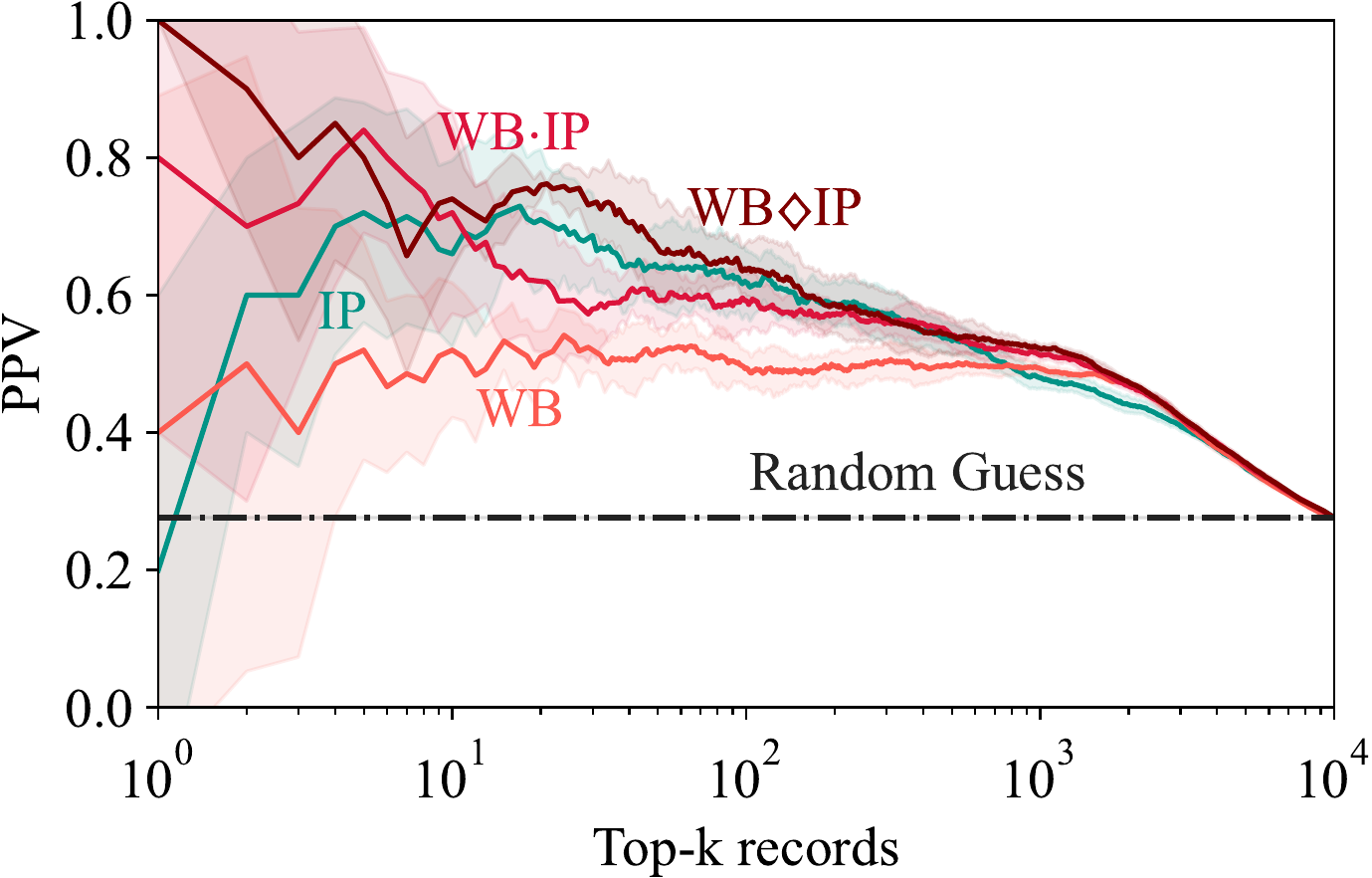}
         \caption{White-box Attacks}
         \label{fig:ppv_texas_med_wb_main}
     \end{subfigure}
    \caption{Comparing the PPV of attacks against imputation on predicting Hispanic ethnicity among 10\,000 candidate training records in \dataset{Texas-100X}. \rm Imputation is trained on 5\,000 records sampled from the training distribution. Results are averaged over five runs. Black-box attacks perform worse than imputation, and combining imputation with black-box attacks does not significantly improve upon imputation. White-box attacks do not outperform an imputation attack trained on large amount of data. Although combining with imputation does improve the white-box attacks.}
    \label{fig:ppv_texas_med_main}
\end{figure*}

The black-box attack (\attack{BB}) by itself does no better than random guessing in any of our experiments. Neither of the methods for combining \attack{BB} with the imputation (\attack{BB$\cdot$IP}, \attack{BB$\diamondsuit$IP}) significantly improves upon the performance of imputation alone when \attack{IP} is trained on 5\,000 records. This is also corroborated in \autoref{fig:ppv_texas_med_bb_main},
which shows the average positive predictive value (PPV) of black-box attacks in inferring candidate records with Hispanic ethnicity in \dataset{Texas-100X} across five runs in the threat setting where the adversary has access to 5,000 records from the training distribution. 
Neither \attack{BB$\cdot$IP} nor \attack{BB$\diamondsuit$IP} significantly outperforms the imputation \attack{IP} across varying top-$k$ settings. However, this trend changes when the imputation has less data to train. As shown in \autoref{tab:ppv_top_100}, the combined black-box attacks \attack{BB$\cdot$IP} and \attack{BB$\diamondsuit$IP} achieve higher PPV than \attack{IP} when the adversary has only 50 or 500 records, although this difference is within the error margin for most cases. This indicates that the model leaks information about the training distribution when the adversary has low distributional information to begin with. 

For \dataset{Census19}, \attack{IP} trained on 5\,000 records achieves around 90\% PPV on average for top-100 records across all distribution settings as shown in \autoref{tab:ppv_top_100}. We note that \attack{IP} achieves high PPV even when it does not have access to the training distribution. This is because the training distribution $\cD$ has 6\% Asian records whereas the skewed distributions $\cD_{HP}$ and $\cD_{LP}$ have around 3.7\%  and 4.3\% Asian records respectively. Recall that $\cD_{HP}$ for \dataset{Census19} corresponds to adversary sampling records from the most populous regions (PUMAs) within the United States, and $\cD_{LP}$ corresponds to sampling records from least populous PUMAs. Since the gap between $\cD$, $\cD_{HP}$ and $\cD_{LP}$ is not large, \attack{IP} manages to achieve similar PPV when trained on large amounts of data from either of these distributions. As with the \dataset{Texas-100X} case, we observe that the black-box attacks do not outperform the imputation attack trained on 5\,000 records even when combined with \attack{IP} (\attack{BB$\cdot$IP} and \attack{BB$\diamondsuit$IP}) for inferring candidate records with Asian race in \dataset{Census19}. However, when the adversary has smaller data set to train imputation, combined black-box attacks \attack{BB$\cdot$IP} and \attack{BB$\diamondsuit$IP} achieve higher PPV than \attack{IP} alone. The gap is within the error margin for most of the cases, however, similar to what we observed for \dataset{Texas-100X}. These results corroborate that when the adversary has limited data and hence cannot train a strong imputation attack, the model tends to leak significant information about the training distribution. In \autoref{sec:whitebox_results}, we show how our white-box attacks take advantage of this leakage.

Detailed results for \dataset{Texas-100X} and \dataset{Census19} across different threat models are in \autoref{fig:ppv_texas_ethnicity_bb} and \autoref{fig:ppv_census_race_bb} in the appendix. These experiments corroborate our previous findings that the predictions made by combining imputation with black-box attacks are mainly due to the imputation. We explore this in more depth next, analyzing the inferences for each individual record. 

\begin{figure*}
     \centering
     \begin{subfigure}[b]{0.32\textwidth}
         \centering
         \includegraphics[width=\textwidth]{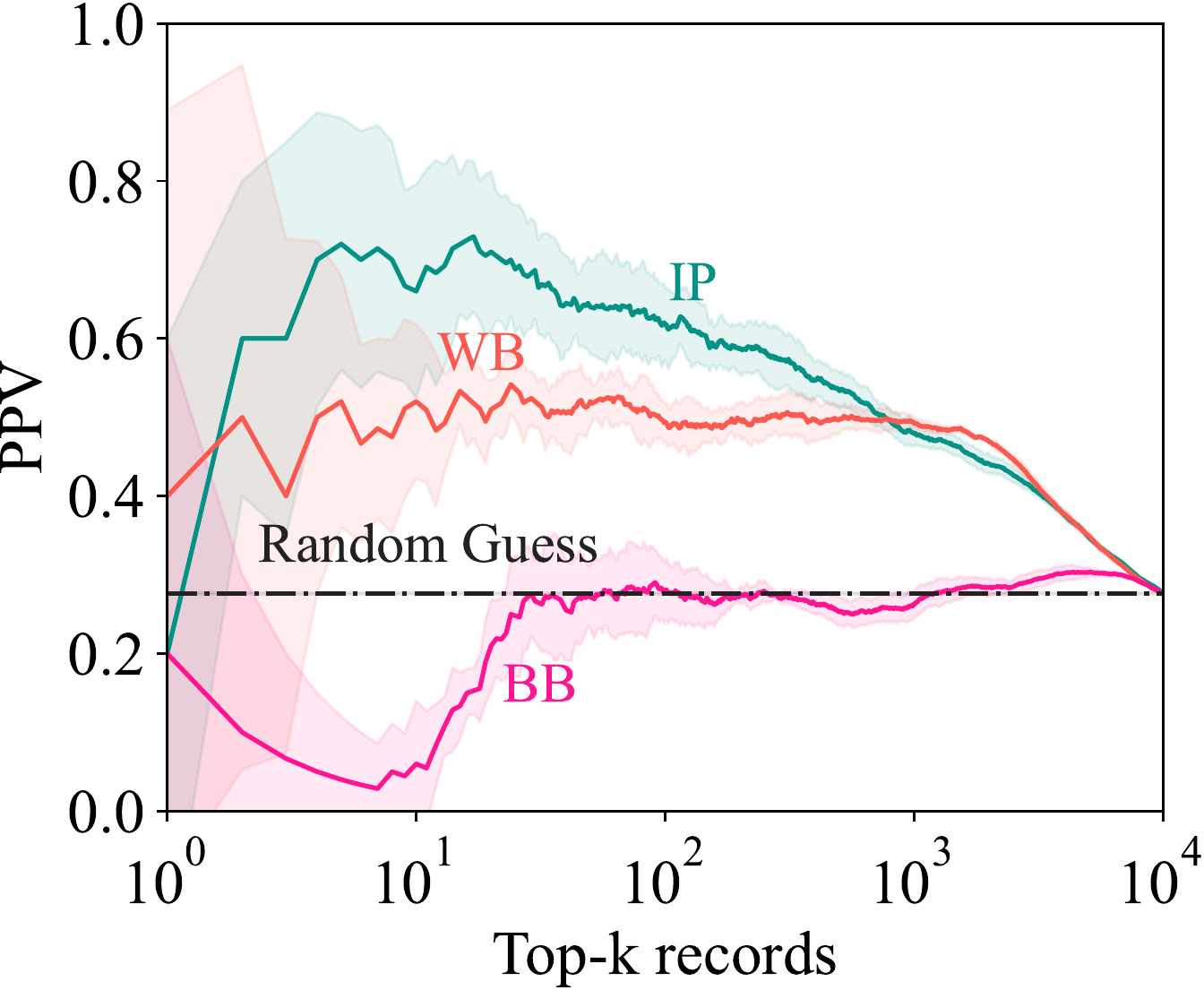}
         \caption{$|D_{aux}|$ = 5\,000}
         \label{fig:ppv_texas_med_5000}
     \end{subfigure}
     \begin{subfigure}[b]{0.32\textwidth}
         \centering
         \includegraphics[width=\textwidth]{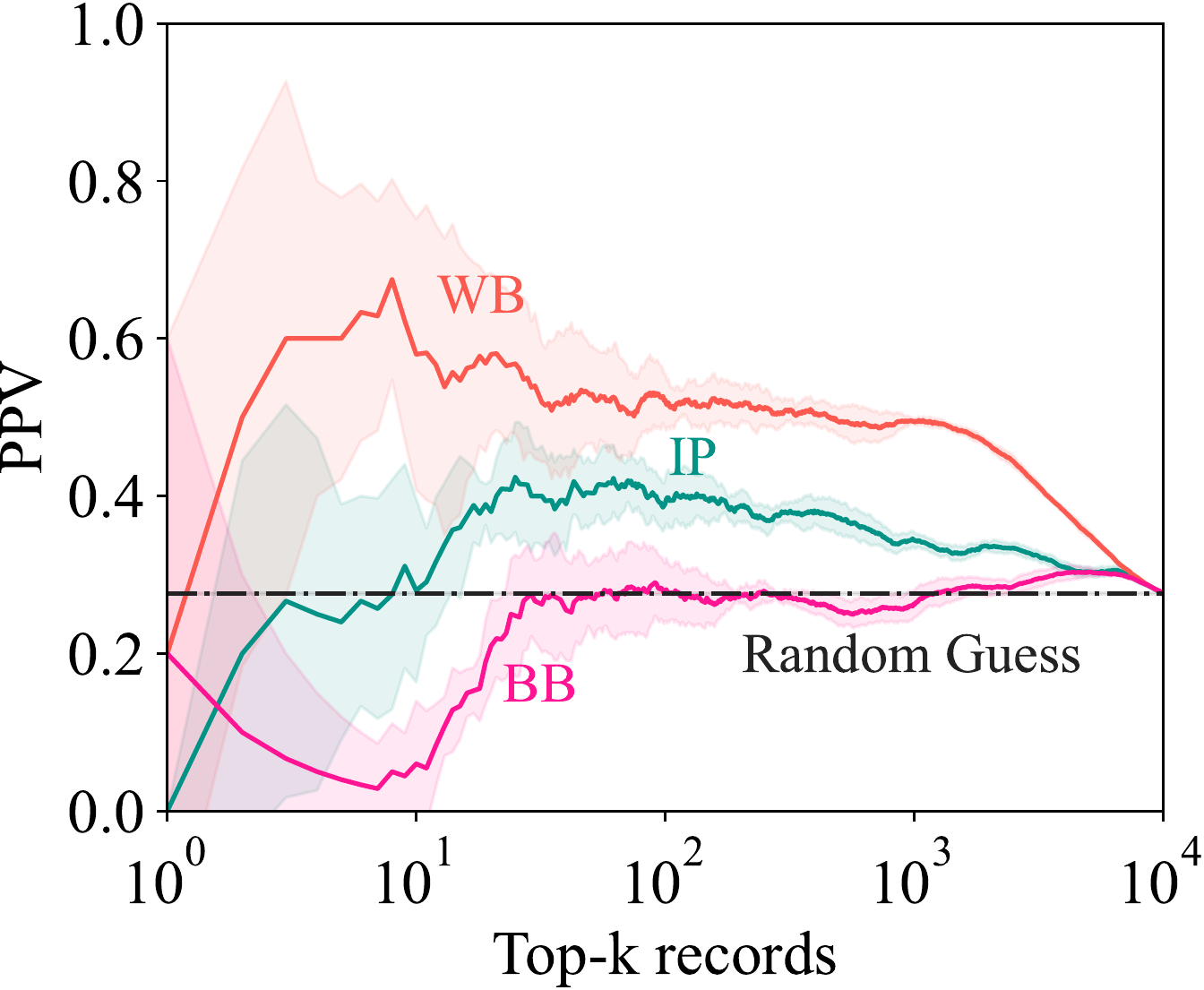}
         \caption{$|D_{aux}|$ = 500}
         \label{fig:ppv_texas_med_500}
     \end{subfigure}
     \begin{subfigure}[b]{0.32\textwidth}
         \centering
         \includegraphics[width=\textwidth]{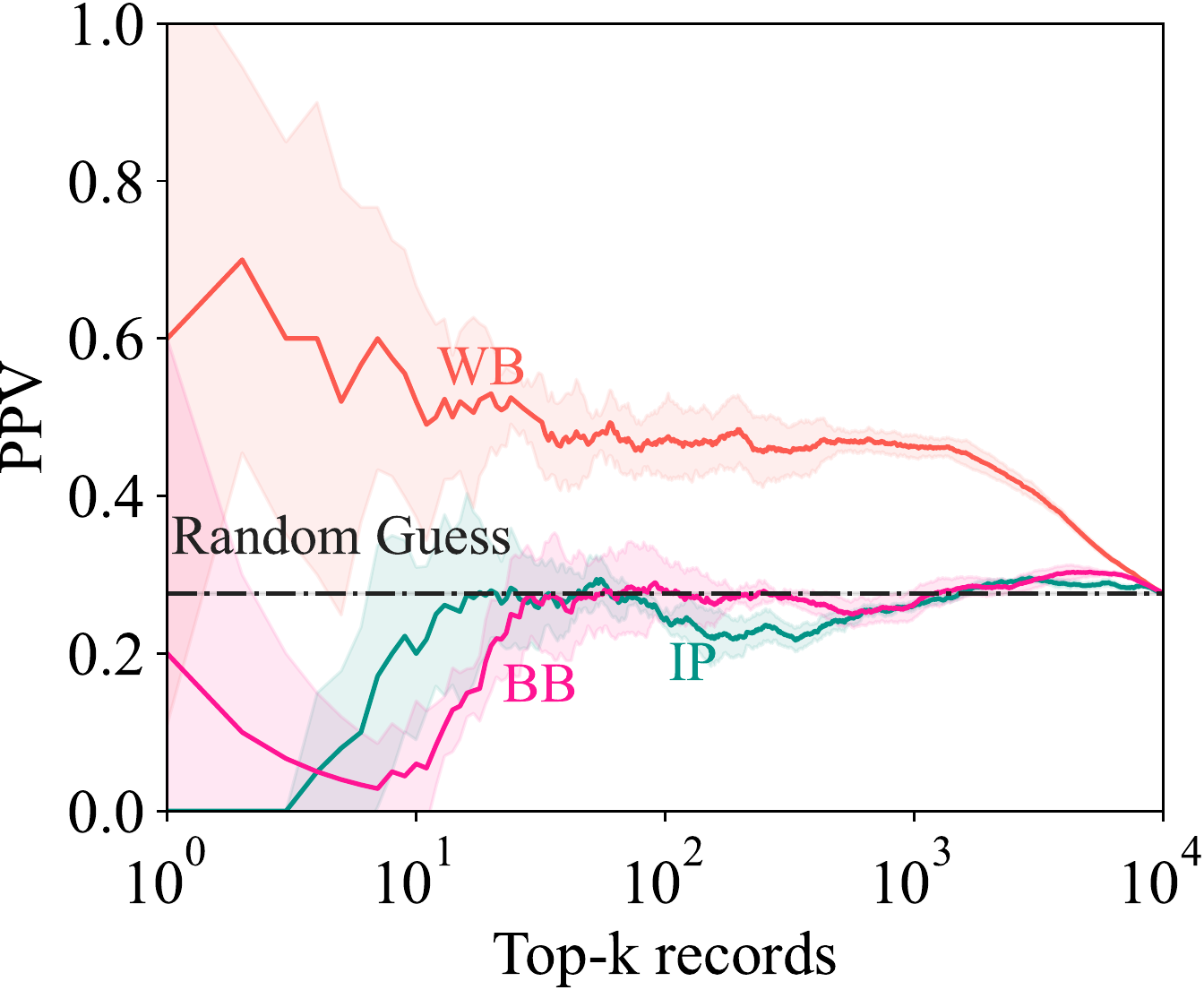}
         \caption{$|D_{aux}|$ = 50}
         \label{fig:ppv_texas_med_50}
     \end{subfigure}
    \caption{Comparing the PPV of attacks against imputation across different data knowledge ($|D_{aux}|$) settings. \rm Results are for predicting the Hispanic ethnicity among 10,000 candidate training records in \bfdataset{Texas-100X}, averaged over five runs. Imputation attack gets weaker as the adversary's data set size decreases, but white-box attack continues to pose privacy threat.}
    \label{fig:ppv_texas_med}
\end{figure*}

\begin{figure*}[tb]
    \centering
    \begin{subfigure}{0.24\textwidth}
        \includegraphics[width=0.9\textwidth]{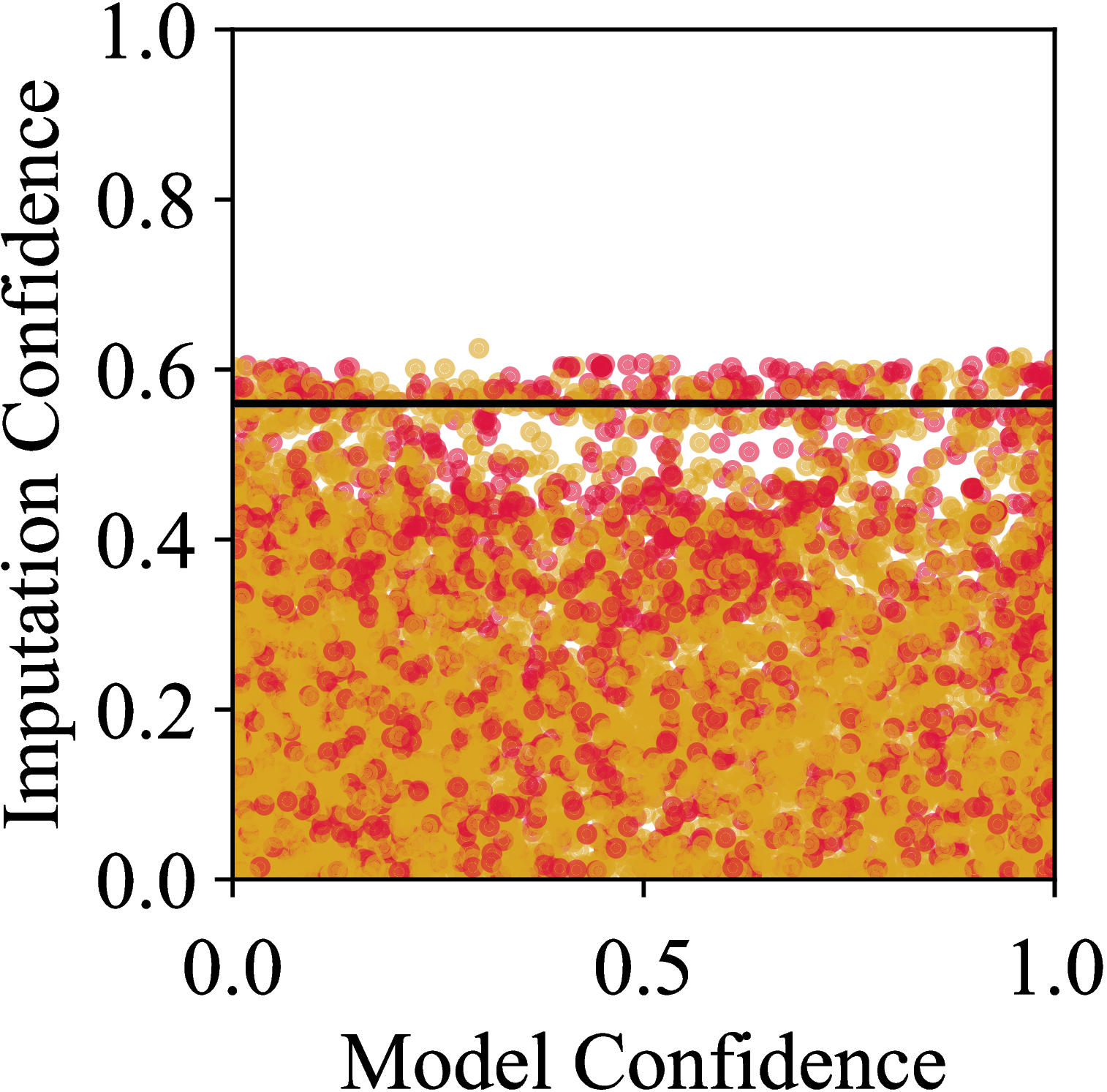}
        \caption{BB vs IP ($|D_{aux}|$ = 5,000)}
        \label{fig:scatter_texas_med_5000_bb}
    \end{subfigure}
    \begin{subfigure}{0.24\textwidth}
        \includegraphics[width=0.9\textwidth]{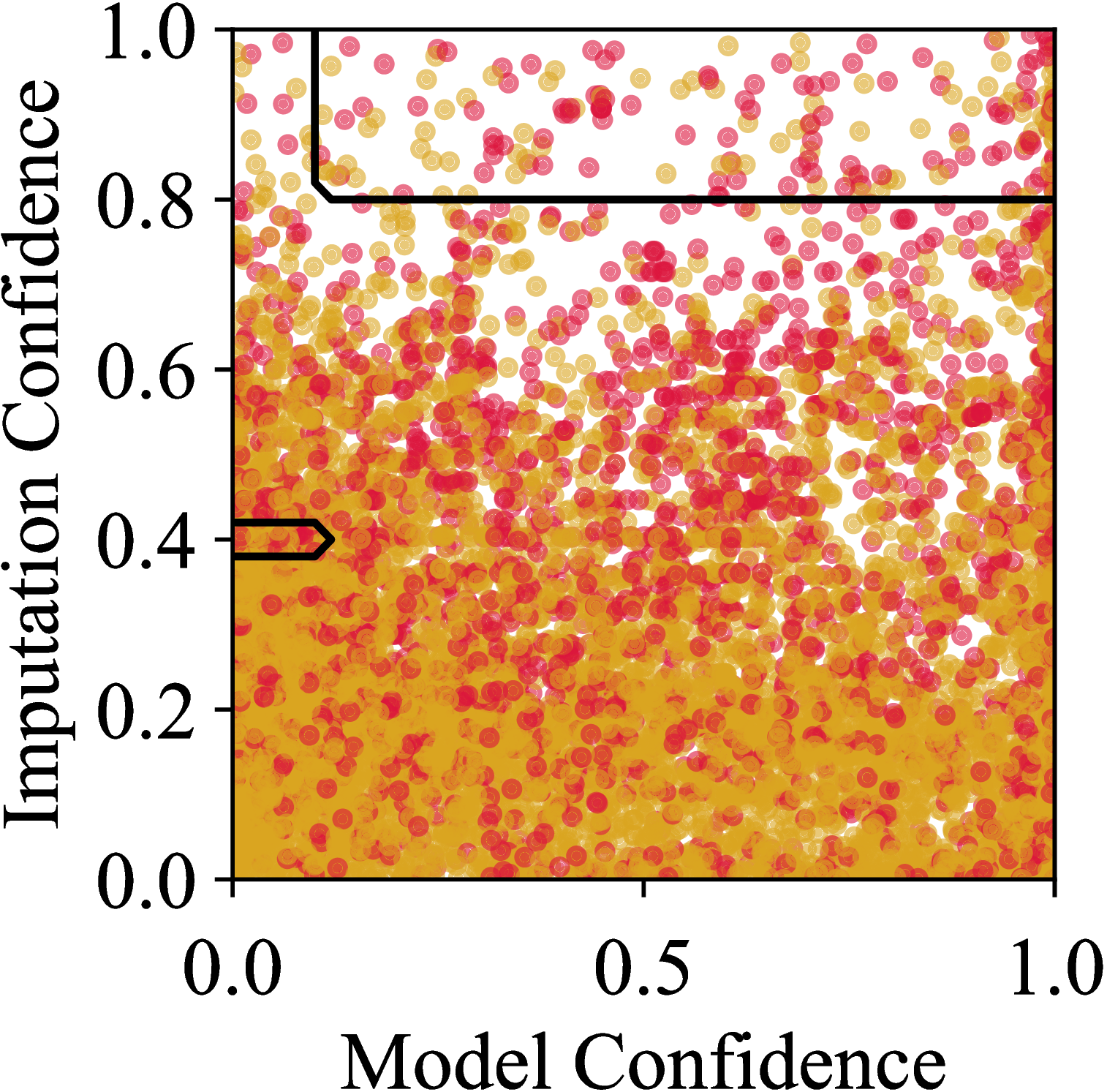}
        \caption{BB vs IP ($|D_{aux}|$ = 50,000)}
        \label{fig:scatter_texas_med_50000_bb}
    \end{subfigure}
    \begin{subfigure}{0.24\textwidth}
        \includegraphics[width=0.9\textwidth]{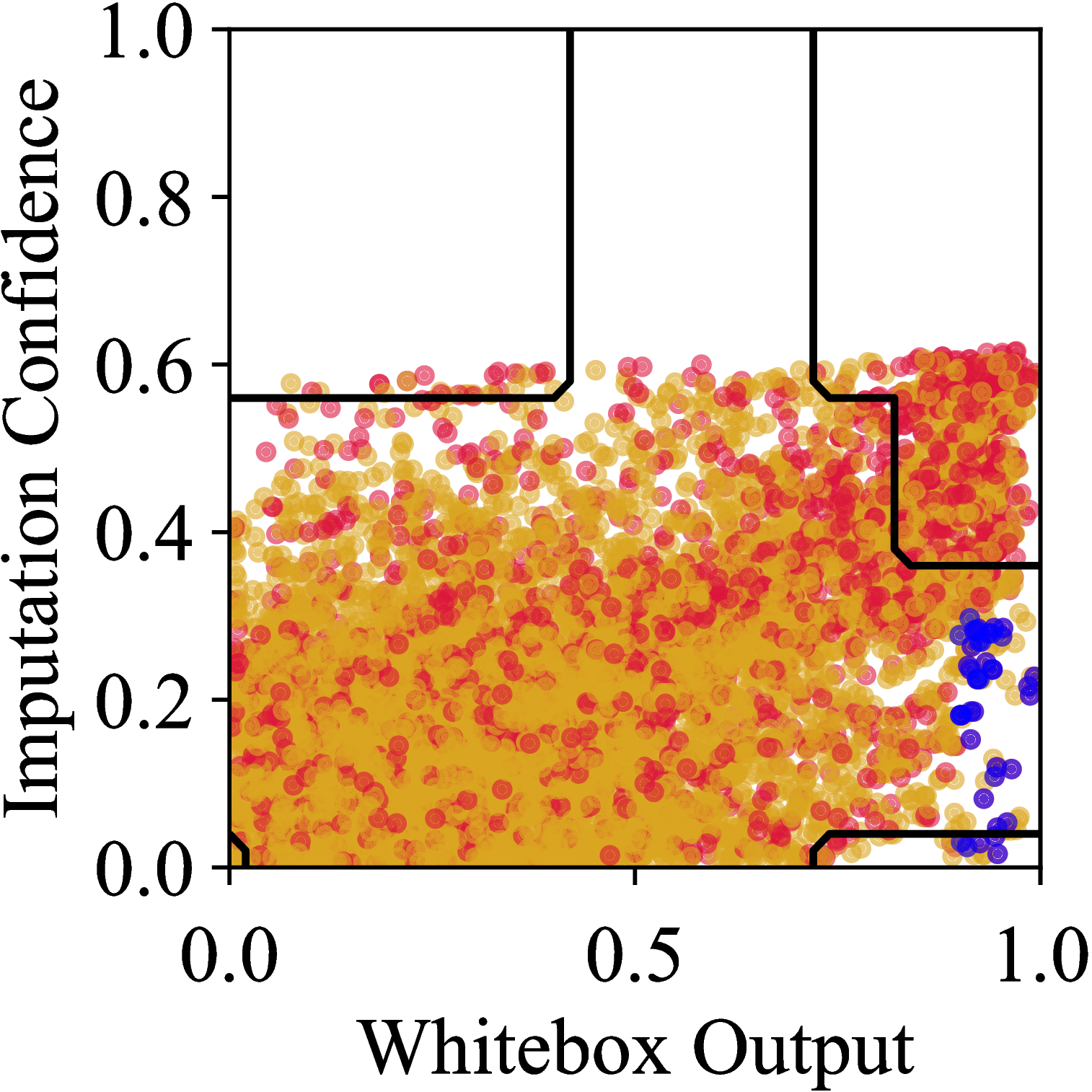}
        \caption{WB vs IP ($|D_{aux}|$ = 5\,000)}
        \label{fig:scatter_texas_med_5000_wb}
    \end{subfigure}
    \begin{subfigure}{0.24\textwidth}
        \includegraphics[width=0.9\textwidth]{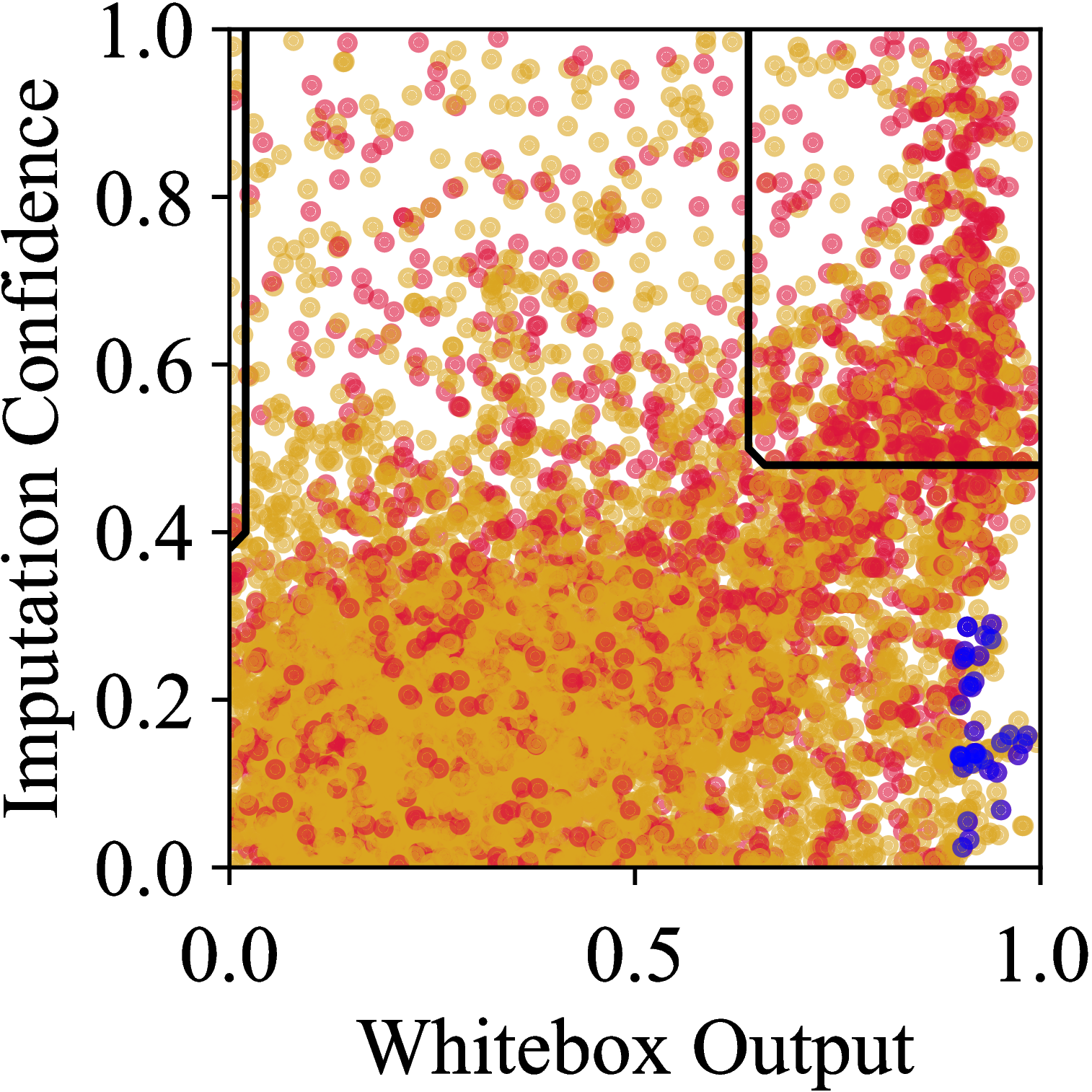}
        \caption{WB vs IP ($|D_{aux}|$ = 50\,000)}
        \label{fig:scatter_texas_med_50000_wb}
    \end{subfigure}
    \caption{Comparing attack output with imputation confidence on \dataset{Texas-100X}. \rm Red dots are Hispanic records, yellow dots are the Non-Hispanic records and blue dots are the vulnerable Hispanic records that are misclassified by imputation, but are identified correctly by white-box attack. Line denotes the decision boundary of decision tree used by \attack{BB$\diamondsuit$IP} (a, b) and \attack{WB$\diamondsuit$IP} (c, d).}
    \label{fig:scatter_texas_med}
\end{figure*}

\shortsection{Understanding Inferences on Individual Records}
\autoref{fig:scatter_texas_med_5000_bb} shows both the model confidence and imputation confidence outputs for all the candidate records for one run for \dataset{Texas-100X} in the threat setting where the adversary has 5\,000 records from the training distribution. The red points denote the candidates with Hispanic ethnicity and the yellow points denote the remaining candidates. The sensitive value is only slightly correlated with the model confidence, which explains the low performance of black-box adversary (\attack{BB}). \attack{BB$\cdot$IP} multiplies the model confidence with imputation confidence, which helps the attack to eliminate most of the non-Hispanic records (yellow dots) for which model confidence is high but imputation confidence is low. However, this attack also misclassifies many Hispanic records (red dots) that have low model confidence but high imputation confidence. We note that the imputation confidence in this setting never exceeds 0.6 since very few (28\% of 5\,000) records belong to the positive class, and hence the model has less information to learn about the Hispanic records. When we train the imputation over a larger data set of 50\,000 records, the maximum prediction confidence of \attack{IP} increases to 1.0 (as shown in \autoref{fig:scatter_texas_med_50000_bb}). However, the low absolute confidence scores do not affect our experiments as we obtain top-$k$ predictions by sorting all the candidate records with respect to the imputation confidence regardless of the absolute value of the imputation confidence.

\autoref{fig:scatter_texas_med_5000_bb} also shows the decision boundary of the decision tree used by \attack{BB$\diamondsuit$IP}. The decision tree predicts Hispanic ethnicity with less than 50\% confidence below the decision line, and more than 50\% confidence above it. This corresponds to predicting top-186 records for \dataset{Texas-100X}. As visible in \autoref{fig:scatter_texas_med_5000_bb}, the decision boundary is horizontal close to 0.55 imputation confidence. We observe minor variations in the boundary across multiple runs, which could due to the noise in the training process, but the trend remains the same. Hence, the decision tree learns to only consider imputation confidence when combining both imputation confidence and model confidence except for some minor cases. We observe a similar outcome for the \dataset{Census19} data set.

%% file: 7_whitebox_results.tex
\section{Results for White-Box Attacks}\label{sec:whitebox_results}

The results from the previous section indicate that the black-box attack does not appear to learn much (if anything) from the model that wouldn't be learned by imputation alone from the same available auxiliary data. 
In this section we evaluate the white-box attacks introduced in \autoref{sec:whitebox_attack}. These attacks exploits the individual neurons of the trained model, and are able, in some cases, to make sensitive value attribute inferences that cannot be made using imputation. 

Table~\ref{tab:ppv_top_100} summarizes the PPV of white-box attacks in predicting the sensitive value of top-100 records. The results show the effectiveness of the white-box attack (\attack{WB}) compared to imputation and the black-box attacks. While imputation outperforms all the inference attacks when trained on considerable amount of data as expected, the white-box attack and its combinations with \attack{IP} (\attack{WB$\cdot$IP} and \attack{WB$\diamondsuit$IP}) consistently outperform the imputation attack when the adversary has limited information about the training distribution.
For the \dataset{Texas-100X} data set, the imputation \attack{IP} trained on 500 records from the training distribution achieves around 0.39 PPV, while our white-box attack \attack{WB} is able to achieve close to 0.52 PPV for the same setting. Furthermore, with only 500 records available, \attack{WB$\cdot$IP} is able to achieve around 0.62 PPV which is the same as that obtained by \attack{IP} when trained on 5\,000 records. This gap between our white-box attacks and imputation further widens when the adversary has fewer training records.
\autoref{fig:ppv_texas_med_wb_main} shows the PPV of white-box attacks for inferring candidate records with Hispanic ethnicity in \dataset{Texas-100X} when the adversary has access to 5\,000 records from the training distribution. The imputation attack trained on 5\,000 records achieves higher PPV that the white-box attack across varying top-$k$ records. Combining white-box with imputation (\attack{WB$\cdot$IP} and 
\attack{WB$\diamondsuit$IP}) improves the attack, but it still does not improve upon the strong imputation attack that has considerable knowledge of the training distribution. 

As the imputation has less and less data available, it has less prior knowledge about the training distribution, and this is where the white-box attacks provide considerable information about the training distribution. As shown in \autoref{fig:ppv_texas_med}, the overall PPV of \attack{IP} decreases across varying top-$k$ records as we decrease the imputation training size from 5\,000 to 50, however the white-box attack continues to achieve PPV in the same range.
Thus, the white-box attack demonstrates that an adversary can obtain substantial sensitive information about the underlying training distribution from the model. Note that this does not necessarily provide evidence for training dataset inference, however, since the model reveals information about the training distribution (not about individual training records). We discuss more on this in \autoref{sec:discussion}. 

The privacy risk is further amplified when the adversary only has access to a skewed data distribution. For \dataset{Texas-100X} where the adversary has access to the distribution $\cD_{LP}$ (with only 20\% Hispanic records), \attack{IP} achieves 0.44 PPV for top-100 predictions even when trained on 5\,000 records,  while \attack{WB} achieves 0.49 PPV. The PPV gap further widens when the adversary has smaller data set. Thus having white-box access to a trained model compensates for the lack of adversary's knowledge about the training data distribution, indicating the compounded risk of releasing the model. We observe similar results for the \dataset{Census19} data set as shown in \autoref{tab:ppv_top_100}. \attack{IP} achieves around 0.46 PPV in inferring top-100 records with Asian race when trained on 500 records from $\cD_{LP}$, while \attack{WB} achieves around 0.85 PPV for the same setting. %\dnote{to what? don't make readers flip back 4 pages to find a table}.

\begin{table}[tb]
    \centering
    \begin{tabular}{lcc}
        \toprule
        & \dataset{Texas-100X} & \dataset{Census19} \\ \midrule
        \# Total Records & 75.80 $\pm$ 5.04 & 3.80 $\pm$ 2.04 \\
        \# Vulnerable Records & 38.60 $\pm$ 4.41 & 3.00 $\pm$ 1.67 \\
        Average PPV & 0.51 $\pm$ 0.06 & 0.82 $\pm$ 0.17 \\ \bottomrule
    \end{tabular}
    \caption{Candidate records in the vulnerable region that are labelled non-sensitive by imputation but considered sensitive by the white-box attack. \rm Vulnerable records are the ones with sensitive value. Results are averaged over five runs.}
    \label{tab:vulnerable_records}
\end{table}

Detailed results showing white-box attack across different settings are in \autoref{fig:ppv_texas_ethnicity_wb} and \autoref{fig:ppv_census_race_wb} in the appendix. The results show that white-box attacks consistently outperform imputation when the adversary has limited knowledge of the underlying distribution.

\shortsection{Vulnerability of Individual Records}
\autoref{fig:scatter_texas_med_5000_wb} plots the imputation confidence and white-box output for all the candidate records for one run in the setting where adversary has access to 5\,000 records from the \dataset{Texas-100X} training distribution. The records with Hispanic ethnicity (depicted with red dots) are correlated with both imputation confidence and white-box output. However, there are a considerable number of Hispanic records for which the imputation has low confidence but the white-box attack has a high value (depicted as blue dots in \autoref{fig:scatter_texas_med_5000_wb}). These are the records that are most harmed by model release---from imputation by an adversary with just knowledge of the underlying distribution they would be predicted as being in the majority class, but an attribute inference attack using the released model is able to confidently (and correctly) predict them in the minority class. In particular, for \dataset{Texas-100X}, of the 76 records for which the imputation confidence is between 0.00 and 0.30 (imputing non-Hispanic) and the white-box output is greater than 0.90, 41 have attribute value Hispanic. Averaged across five runs, there are 38.60 $\pm$ 4.41 Hispanic records out of 75.80 $\pm$ 5.04 records in this region, with around 51\% PPV. The white-box attack successfully identifies these records as having the sensitive attribute value. As with the black-box case, \attack{IP} trained on 5,000 records does not have high confidence due to the limited number of Hispanic records in the imputation training set, which is why no candidate record has imputation confidence greater than 0.6 in \autoref{fig:scatter_texas_med_5000_wb}. If we train \attack{IP} on a larger set of 50,000 records, then we observe that the maximum imputation confidence value goes up to 1.0 as shown in \autoref{fig:scatter_texas_med_50000_wb}. As noted before, the absolute imputation confidence value does not affect the overall results. 
\autoref{tab:vulnerable_records} summarizes the statistics about such vulnerable records for both \dataset{Texas-100X} and \dataset{Census19}. We report on further experiments in \autoref{sec:defenses} on the impact of removing these vulnerable records from the training data. 

%% file: 8_defenses.tex
\section{Evaluating Possible Defenses}\label{sec:defenses}

The experiments reported on in the previous section show that white-box attribute inference attacks can pose serious privacy threats in situations where an adversary has limited prior knowledge of the training distribution. 
Here, we evaluate two potential defenses that aim to mitigate the privacy threat of these inference attacks. The first approach is to remove the subset of vulnerable candidate records from the training set and retrain the model. The second defense approach is to train the model with differential privacy. 
Our results show that these defenses do not substantially mitigate the privacy risk.

\begin{table*}[tb]
    \centering
    \begin{tabular}{clccccccccc}
        \toprule
        & & \multicolumn{3}{c}{$\cA$ knows train distribution $\cD$} & \multicolumn{3}{c}{$\cA$ knows different distribution $\cD_{HP}$} & \multicolumn{3}{c}{$\cA$ knows different distribution $\cD_{LP}$} \\ 
        & $|D_{\mathrm{aux}}|$ & 5\,000 & 500 & 50 & 5\,000 & 500 & 50 & 5\,000 & 500 & 50 \\ \midrule
        \parbox[t]{2mm}{\multirow{4}{*}{\rotatebox[origin=c]{90}{\dataset{Texas-100X}}}} & \attack{IP} & 0.62 $\pm$ 0.05 & 0.39 $\pm$ 0.03 & 0.24 $\pm$ 0.01 & 0.63 $\pm$ 0.03 & 0.39 $\pm$ 0.03 & 0.40 $\pm$ 0.03 & 0.44 $\pm$ 0.02 & 0.41 $\pm$ 0.05 & 0.37 $\pm$ 0.05 \\
        & \attack{WB} & 0.54 $\pm$ 0.02 & {\color{red} 0.54 $\pm$ 0.00} & {\color{red} 0.46 $\pm$ 0.07} & 0.57 $\pm$ 0.02 & {\color{red} 0.51 $\pm$ 0.02} & {\color{red} 0.48 $\pm$ 0.04} & {\color{red} 0.51 $\pm$ 0.02} & 0.48 $\pm$ 0.03 & {\color{red} 0.54 $\pm$ 0.05} \\
        & \attack{WB$\cdot$IP} & 0.61 $\pm$ 0.02 & {\color{red} 0.58 $\pm$ 0.05} & {\color{red} 0.45 $\pm$ 0.03} & 0.60 $\pm$ 0.03 & {\color{red} 0.53 $\pm$ 0.06} & {\color{red} 0.52 $\pm$ 0.05} & {\color{red} 0.51 $\pm$ 0.04} & 0.48 $\pm$ 0.04 & {\color{red} 0.54 $\pm$ 0.06} \\
        & \attack{WB$\diamondsuit$IP} & 0.64 $\pm$ 0.05 & {\color{red} 0.51 $\pm$ 0.05} & {\color{red} 0.46 $\pm$ 0.05} & 0.61 $\pm$ 0.06 & {\color{red} 0.55 $\pm$ 0.07} & {\color{red} 0.48 $\pm$ 0.02} & 0.48 $\pm$ 0.02 & 0.43 $\pm$ 0.06 & {\color{red} 0.52 $\pm$ 0.04} \\ \midrule
        
        \parbox[t]{2mm}{\multirow{4}{*}{\rotatebox[origin=c]{90}{\dataset{Census19}}}} & \attack{IP} & 0.91 $\pm$ 0.03 & 0.25 $\pm$ 0.02 & 0.55 $\pm$ 0.06 & 0.89 $\pm$ 0.02 & 0.11 $\pm$ 0.04 & 0.36 $\pm$ 0.06 & 0.90 $\pm$ 0.03 & 0.46 $\pm$ 0.04 & 0.42 $\pm$ 0.04 \\
        & \attack{WB} & 0.86 $\pm$ 0.03 & {\color{red} 0.85 $\pm$ 0.04} & {\color{red} 0.73 $\pm$ 0.06} & 0.86 $\pm$ 0.04 & {\color{red} 0.79 $\pm$ 0.12} & {\color{red} 0.62 $\pm$ 0.14} & 0.84 $\pm$ 0.03 & {\color{red} 0.83 $\pm$ 0.03} & {\color{red} 0.62 $\pm$ 0.12} \\
        & \attack{WB$\cdot$IP} & 0.89 $\pm$ 0.03 & {\color{red} 0.80 $\pm$ 0.07} & {\color{red} 0.74 $\pm$ 0.06} & 0.89 $\pm$ 0.00 & {\color{red} 0.74 $\pm$ 0.12} & {\color{red} 0.68 $\pm$ 0.11} & 0.88 $\pm$ 0.04 & {\color{red} 0.82 $\pm$ 0.01} & {\color{red} 0.72 $\pm$ 0.09} \\
        & \attack{WB$\diamondsuit$IP} & 0.90 $\pm$ 0.02 & {\color{red} 0.84 $\pm$ 0.03} & {\color{red} 0.79 $\pm$ 0.05} & 0.88 $\pm$ 0.02 & {\color{red} 0.86 $\pm$ 0.04} & {\color{red} 0.73 $\pm$ 0.06} & 0.83 $\pm$ 0.03 & {\color{red} 0.85 $\pm$ 0.02} & {\color{red} 0.75 $\pm$ 0.06} \\ \bottomrule
    \end{tabular}
    \caption{Comparing the PPV for predicting the sensitive value of top-100 records when model is re-trained after removing vulnerable records from training. \rm Reported results are for predicting \emph{Hispanic} ethnicity in \bfdataset{Texas-100X} and \emph{Asian} race in \bfdataset{Census19}. Cases in which the inference attack PPV (mean - std) is greater than the imputation PPV (mean + std) are highlighted in red.}
    \label{tab:ppv_top_100_banished}
\end{table*}

\begin{table}[tb]
    \centering
    \begin{tabular}{lcc}
        \toprule
        & \dataset{Texas-100X} & \dataset{Census19} \\ \midrule
        \# Total Records & 66.80 $\pm$ 9.47 & 4.00 $\pm$ 0.63 \\
        \# Old Vulnerable Records & 23.40 $\pm$ 4.67 & 2.20 $\pm$ 1.17 \\
        \# New Vulnerable Records & 10.80 $\pm$ 5.34 & 0.40 $\pm$ 0.49 \\
        Average PPV & 0.51 $\pm$ 0.04 & 0.64 $\pm$ 0.12 \\ \bottomrule
    \end{tabular}
    \caption{Impact of removing vulnerable records and retraining. \rm Candidate records in the vulnerable region that are labelled non-sensitive by \attack{IP} but considered sensitive by \attack{WB}, when the model is re-trained after removing the vulnerable records with sensitive value from training. Results are averaged over five runs.}
    \label{tab:vulnerable_records_after_retraining}
\end{table}

\subsection{Removing Vulnerable Training Records}\label{sec:removing_vulnerable}

In this straightforward defense method, the model trainer runs the attack to identify the most vulnerable training records, removes those from the training dataset, and re-trains the model. The intuition is that if the records are vulnerable due to some property of those records, then removing them from the training set would reduce their impact on the model and protect the privacy of those records. In theory, such a process could be repeated iteratively until there are no records remaining that are at high risk of exposure.
However, we find this defense method has limited effectiveness.

\autoref{tab:ppv_top_100_banished} shows the PPV of white-box attacks on predicting top-100 candidate records when the vulnerable records are removed from the model training for both Hispanic ethnicity in \dataset{Texas-100X} and Asian race on \bfdataset{Census19}. We conducted five separate executions for each data set, removing an average of 38.60 $\pm$ 4.41 records for \dataset{Texas-100X}, and 3.00 $\pm$ 1.67 records for \dataset{Census19}. We see small fluctuations in PPV of white-box attacks on \dataset{Texas-100X} when compared to not using the defense, but not in any clear direction or above the error margins. For instance, in the setting where adversary has access to 500 records from the training distribution, the PPV of \attack{WB} increases from 0.52 to 0.54 whereas it drops from 0.62 to 0.58 for \attack{WB$\cdot$IP}. The PPV of \attack{WB$\diamondsuit$IP} remains the same at 0.51 in this setting. However, all variation is within the noise. We observe similar fluctuations in PPV for \dataset{Census19}, though the variations are not outside the noise. For more detailed results across the varying threat models, see  \autoref{fig:ppv_texas_ethnicity_wb_banished} and \autoref{fig:ppv_census_race_wb_banished} in the appendix.  

\autoref{tab:vulnerable_records_after_retraining} shows the impact on the individually vulnerable candidate records after being removed from training set. For \dataset{Texas-100X}, we find that of the 38.60 $\pm$ 4.41 records that were most vulnerable, 23.40 $\pm$ 4.67 remain vulnerable in the retrained model even though they were removed from the training data. 
Furthermore, 10.80 $\pm$ 5.34 candidate records that were previously not vulnerable are now vulnerable. For \dataset{Census19}, the results are similar, but fewer records are vulnerable.
These results support our hypothesis that the model learns correlations from the training dataset, which reflect the underlying training distribution rather than leaking information about specific records in the training dataset. Hence, our observations are consistent with the hypothesis that the observed white-box attribute inference is really doing imputation from what is revealed about the training distribution by the model, not revealing specific information about training records.

\begin{table*}[tb]
    \centering
    \begin{tabular}{clccccccccc}
        \toprule
        & & \multicolumn{3}{c}{$\cA$ knows train distribution $\cD$} & \multicolumn{3}{c}{$\cA$ knows different distribution $\cD_{HP}$} & \multicolumn{3}{c}{$\cA$ knows different distribution $\cD_{LP}$} \\ 
        & $|D_{\mathrm{aux}}|$ & 5\,000 & 500 & 50 & 5\,000 & 500 & 50 & 5\,000 & 500 & 50 \\ \midrule
        \parbox[t]{2mm}{\multirow{4}{*}{\rotatebox[origin=c]{90}{\dataset{Texas-100X}}}} & \attack{IP} & 0.62 $\pm$ 0.05 & 0.39 $\pm$ 0.03 & 0.24 $\pm$ 0.01 & 0.63 $\pm$ 0.03 & 0.39 $\pm$ 0.03 & 0.40 $\pm$ 0.03 & 0.44 $\pm$ 0.02 & 0.41 $\pm$ 0.05 & 0.37 $\pm$ 0.05 \\
        & \attack{WB} & 0.49 $\pm$ 0.03 & {\color{red} 0.48 $\pm$ 0.05} & {\color{red} 0.46 $\pm$ 0.04} & 0.53 $\pm$ 0.08 & {\color{red} 0.54 $\pm$ 0.04} & {\color{red} 0.52 $\pm$ 0.02} & 0.51 $\pm$ 0.05 & {\color{red} 0.56 $\pm$ 0.06} & 0.45 $\pm$ 0.04 \\
        & \attack{WB$\cdot$IP} & 0.62 $\pm$ 0.04 & {\color{red} 0.60 $\pm$ 0.06} & {\color{red} 0.41 $\pm$ 0.04} & 0.57 $\pm$ 0.05 & {\color{red} 0.59 $\pm$ 0.10} & {\color{red} 0.50 $\pm$ 0.05} & {\color{red} 0.52 $\pm$ 0.04} & 0.50 $\pm$ 0.05 & {\color{red} 0.49 $\pm$ 0.02} \\
        & \attack{WB$\diamondsuit$IP} & 0.62 $\pm$ 0.06 & {\color{red} 0.57 $\pm$ 0.07} & {\color{red} 0.52 $\pm$ 0.06} & 0.60 $\pm$ 0.06 & {\color{red} 0.61 $\pm$ 0.06} & 0.48 $\pm$ 0.06 & 0.47 $\pm$ 0.02 & 0.43 $\pm$ 0.03 & 0.42 $\pm$ 0.05 \\ \midrule
        
        \parbox[t]{2mm}{\multirow{4}{*}{\rotatebox[origin=c]{90}{\dataset{Census19}}}} & \attack{IP} & 0.91 $\pm$ 0.03 & 0.25 $\pm$ 0.02 & 0.55 $\pm$ 0.06 & 0.89 $\pm$ 0.02 & 0.11 $\pm$ 0.04 & 0.36 $\pm$ 0.06 & 0.90 $\pm$ 0.03 & 0.46 $\pm$ 0.04 & 0.42 $\pm$ 0.04 \\
        & \attack{WB} & 0.88 $\pm$ 0.03 & {\color{red} 0.85 $\pm$ 0.04} & {\color{red} 0.76 $\pm$ 0.06} & 0.87 $\pm$ 0.05 & {\color{red} 0.88 $\pm$ 0.04} & {\color{red} 0.82 $\pm$ 0.07} & 0.86 $\pm$ 0.02 & {\color{red} 0.85 $\pm$ 0.04} & {\color{red} 0.81 $\pm$ 0.12} \\
        & \attack{WB$\cdot$IP} & 0.89 $\pm$ 0.04 & {\color{red} 0.78 $\pm$ 0.03} & {\color{red} 0.72 $\pm$ 0.02} & 0.88 $\pm$ 0.04 & {\color{red} 0.80 $\pm$ 0.04} & {\color{red} 0.77 $\pm$ 0.09} & 0.88 $\pm$ 0.04 & {\color{red} 0.83 $\pm$ 0.03} & {\color{red} 0.82 $\pm$ 0.05} \\
        & \attack{WB$\diamondsuit$IP} & 0.90 $\pm$ 0.03 & {\color{red} 0.86 $\pm$ 0.04} & {\color{red} 0.83 $\pm$ 0.03} & 0.89 $\pm$ 0.03 & {\color{red} 0.86 $\pm$ 0.03} & {\color{red} 0.84 $\pm$ 0.04} & 0.86 $\pm$ 0.04 & {\color{red} 0.87 $\pm$ 0.02} & {\color{red} 0.81 $\pm$ 0.06} \\ \bottomrule
    \end{tabular}
    \caption{Impact of training with differential privacy $(\epsilon = 1, \delta = 10^{-5})$. \rm Reported results are for predicting \emph{Hispanic} ethnicity in \bfdataset{Texas-100X} and \emph{Asian} race in \bfdataset{Census19}. Cases in which the inference attack PPV (mean - std) is greater than the imputation PPV (mean + std) are highlighted in red. The observed PPVs are similar to those in \autoref{tab:ppv_top_100}, where the model is trained without DP.}
    \label{tab:ppv_top_100_dp}
\end{table*}

\begin{table}[tb]
    \centering
    \begin{tabular}{lcccc}
        \toprule
        & \multicolumn{2}{c}{\bfdataset{Texas-100X}} & \multicolumn{2}{c}{\bfdataset{Census19}} \\ 
        & Train & Test & Train & Test \\ \midrule
        \attack{IP} & 0.62 $\pm$ 0.05 & 0.63 $\pm$ 0.02 & 0.91 $\pm$ 0.03 & 0.88 $\pm$ 0.01 \\
        \attack{WB} & 0.49 $\pm$ 0.03 & 0.48 $\pm$ 0.02 & 0.88 $\pm$ 0.03 & 0.89 $\pm$ 0.04 \\
        \attack{WB$\cdot$IP} & 0.62 $\pm$ 0.04 & 0.55 $\pm$ 0.02 & 0.89 $\pm$ 0.04 & 0.91 $\pm$ 0.02 \\
        \attack{WB$\diamondsuit$IP} & 0.62 $\pm$ 0.06 & 0.59 $\pm$ 0.02 & 0.90 $\pm$ 0.03 & 0.86 $\pm$ 0.01 \\ \bottomrule
    \end{tabular}
    \caption{Impact of differential privacy on train candidates vs test candidates $(\epsilon = 1, \delta = 10^{-5})$. \rm Reported results are for predicting \emph{Hispanic} ethnicity in \bfdataset{Texas-100X} and \emph{Asian} race in \bfdataset{Census19} when $\cA$ knows the train distribution and $|D_{\mathrm{aux}}|$ = 5\,000.}
    \label{tab:dp_train_vs_test}
\end{table}

\begin{comment}
\begin{table}[tb]
    \centering
    \begin{tabular}{lcc}
        \toprule
        & \dataset{Texas-100X} & \dataset{Census19} \\ \midrule
        \# Total Records & 56.80 $\pm$ 17.60 & 16.40 $\pm$ 5.20 \\
        \# Old Vulnerable Records & 19.20 $\pm$ 6.70 & 5.60 $\pm$ 2.00 \\
        \# New Vulnerable Records & 21.20 $\pm$ 9.00 & 6.20 $\pm$ 2.60 \\
        Average PPV & 0.72 $\pm$ 0.08 & 0.78 $\pm$ 0.20 \\ \bottomrule
    \end{tabular}
    \caption{Candidate records in the vulnerable region that are labelled non-sensitive by imputation but considered sensitive by the white-box attack, when the model is trained with Gaussian differential privacy ($\epsilon = 1$). Results are averaged over five runs.}
    \label{tab:vulnerable_records_after_dp}
\end{table}
\end{comment}

\subsection{Differential Privacy}\label{sec:differential_privacy}

Differential privacy~\cite{dwork2006calibrating} mechanisms typically limit the exposure of individual records by introducing noise in the training process. 
Differential privacy has been shown to both theoretically bound~\cite{yeom2018privacy,jayaraman2020revisiting} and experimentally mitigate membership inference risks~\cite{liu2021ml}. 
However, the theoretical guarantees provided by differential privacy are at the level of individual records and not at the level of statistical properties in a distribution.

To evaluate the impact of differential privacy mechanisms on attribute inference risks, we train private neural network models with Gaussian differential privacy~\cite{dong2019gaussian} for our experiments. Our implementation uses gradient perturbation~\cite{abadi2016deep} with a privacy loss budget $\epsilon = 1$ and $\delta = 10^{-5}$. The model achieves 30\% accuracy on both the training and testing sets for \bfdataset{Texas-100X}, and achieves 86\% accuracy on both training and testing sets for \bfdataset{Census19}.
\autoref{tab:ppv_top_100_dp} shows the PPV for the white-box attacks against models trained with differential privacy at different threat models. For the white-box attacks predicting top-100 records with Hispanic ethnicity in \dataset{Texas-100X}, we observe a slight fluctuation in PPV when we add differential privacy noise, but the difference is within the noise range. For predicting top-100 candidate records with Asian race in \bfdataset{Census19}, the PPV of \attack{WB} slightly increases from 0.85 to 0.88 in the setting where adversary has 5\,000 records from the training distribution. We observe a similar increase in PPV across all the other data size and distribution settings. 
This seems to be due to the increase in the correlation of neurons to the sensitive outcome caused by training with DP noise. The average Pearson correlation coefficient of the 10 most correlated neurons to Asian race for \dataset{Census19} increases from 0.29 $\pm$ 0.06 to 0.38 $\pm$ 0.07 when the privacy noise is added for the setting where adversary has 5\,000 records from the training distribution. 
We do not observe any significant change in the correlation values for \dataset{Texas-100X}. Detailed results are in \autoref{fig:ppv_texas_ethnicity_wb_1_eps} and \autoref{fig:ppv_census_race_wb_1_eps} in the appendix.

The limited (and apparently more negative than positive) privacy impact of differential privacy mechanisms should raise alarm, but does not contradict previous results that claimed attribute inference advantage (as defined by Yeom et al.\ as the predication accuracy difference between training and non-training records) is bounded by differential privacy. Differential privacy done at the level of individual records can provide a bound on the difference in inference accuracy between training and non-training records, but provides no assurances about inferences regarding the ability of an adversary to infer an attribute from distributional information leaked by the model. While the noise distorts the model parameters to reduce the impact of any single training record, it may inadvertently increase the correlation of a subset of neurons to the sensitive value. This is supported by the results in \autoref{tab:dp_train_vs_test}, where the white-box attacks obtain similar PPV values in predicting top-100 records from both training and non-training candidate sets. %This is in accordance with Cormode's earlier observations~\cite{cormode2011personal}. 
While attribute inference advantage according to Yeom et al.'s definition for \attack{WB} is 0.01 for \dataset{Texas-100X}, the advantage provided by the model (PPV gap between \attack{WB} and \attack{IP}) at $|D_{aux} = 50|$ is around 0.22 (from \autoref{tab:ppv_top_100_dp}). 
Differential privacy ensures that an individual training record cannot be distinguished from a non-training record, but provides no bound on inferences made based on statistical information revealed about the training distribution.

%% file: appendix.tex
\clearpage

\begin{figure*}
\section*{Appendix: Additional Results Across Different Threat Settings}
\vspace*{2ex}
     \centering
     \begin{subfigure}{0.32\textwidth}
         \centering
         \includegraphics[width=\textwidth]{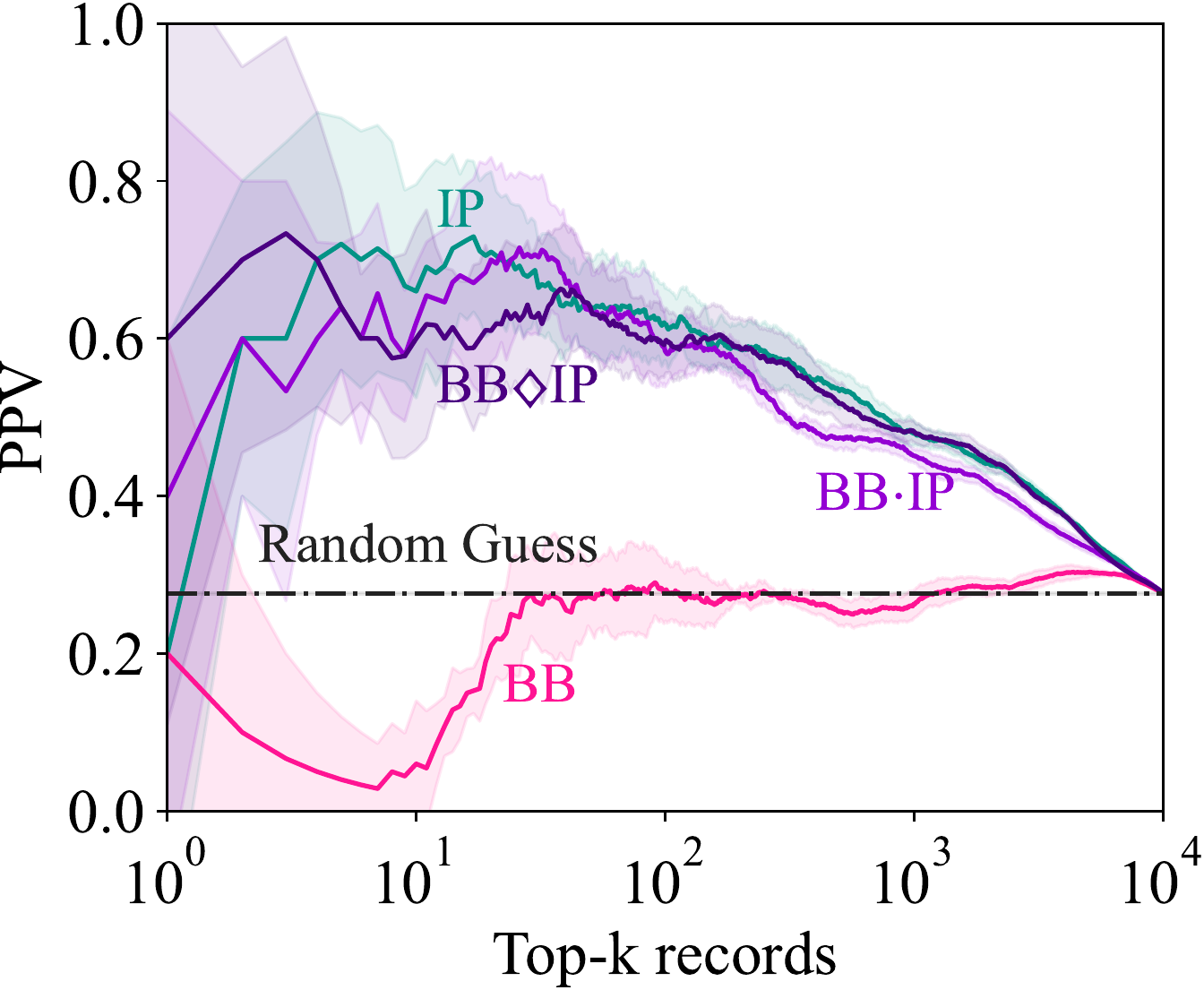}
         \caption{$|D_{aux}|$ = 5\,000}
         \label{fig:ppv_texas_ethnicity_low_bb}
     \end{subfigure}
     \begin{subfigure}{0.32\textwidth}
         \centering
         \includegraphics[width=\textwidth]{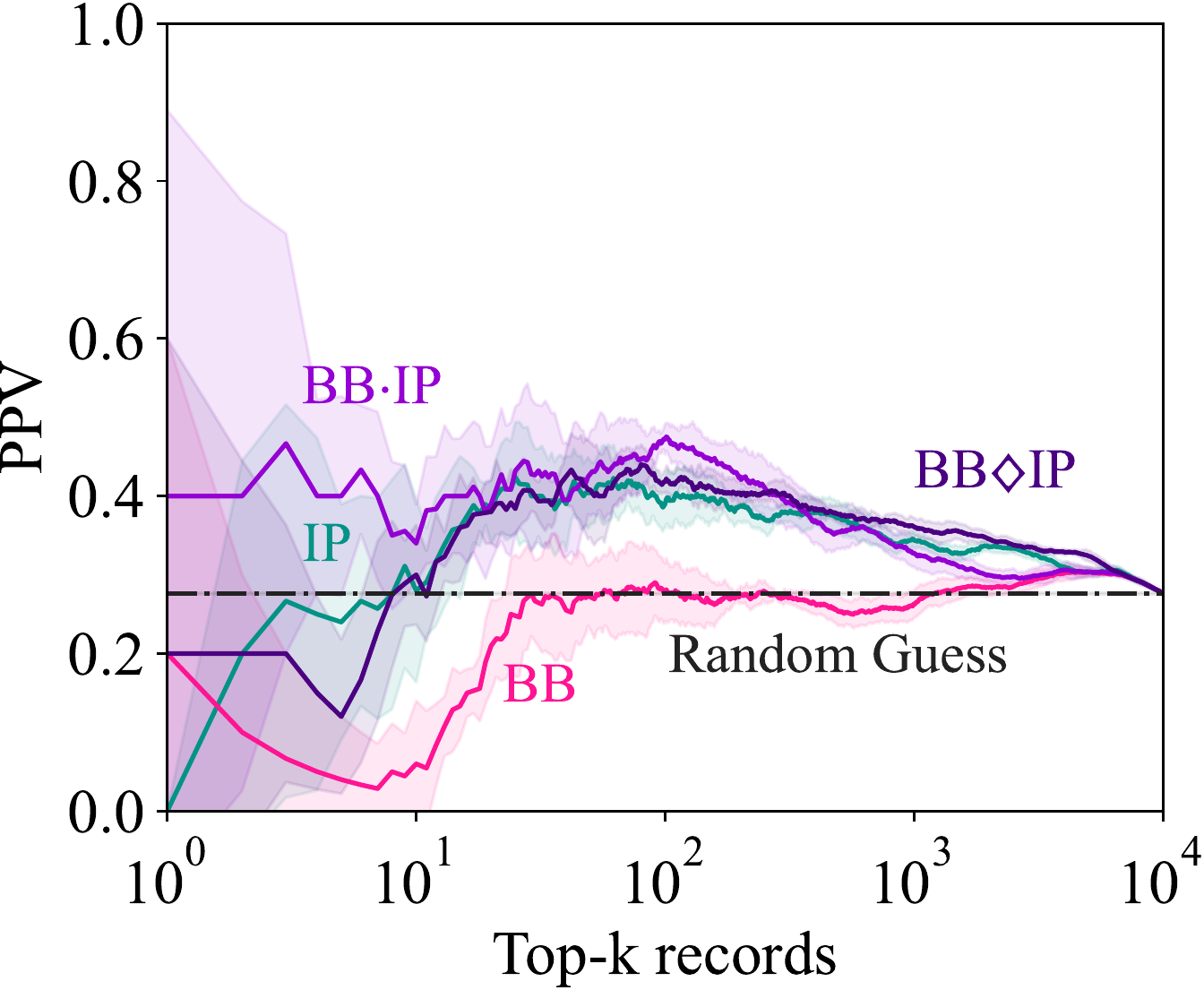}
         \caption{$|D_{aux}|$ = 500}
         \label{fig:ppv_texas_ethnicity_med_bb}
     \end{subfigure}
     \begin{subfigure}{0.32\textwidth}
         \centering
         \includegraphics[width=\textwidth]{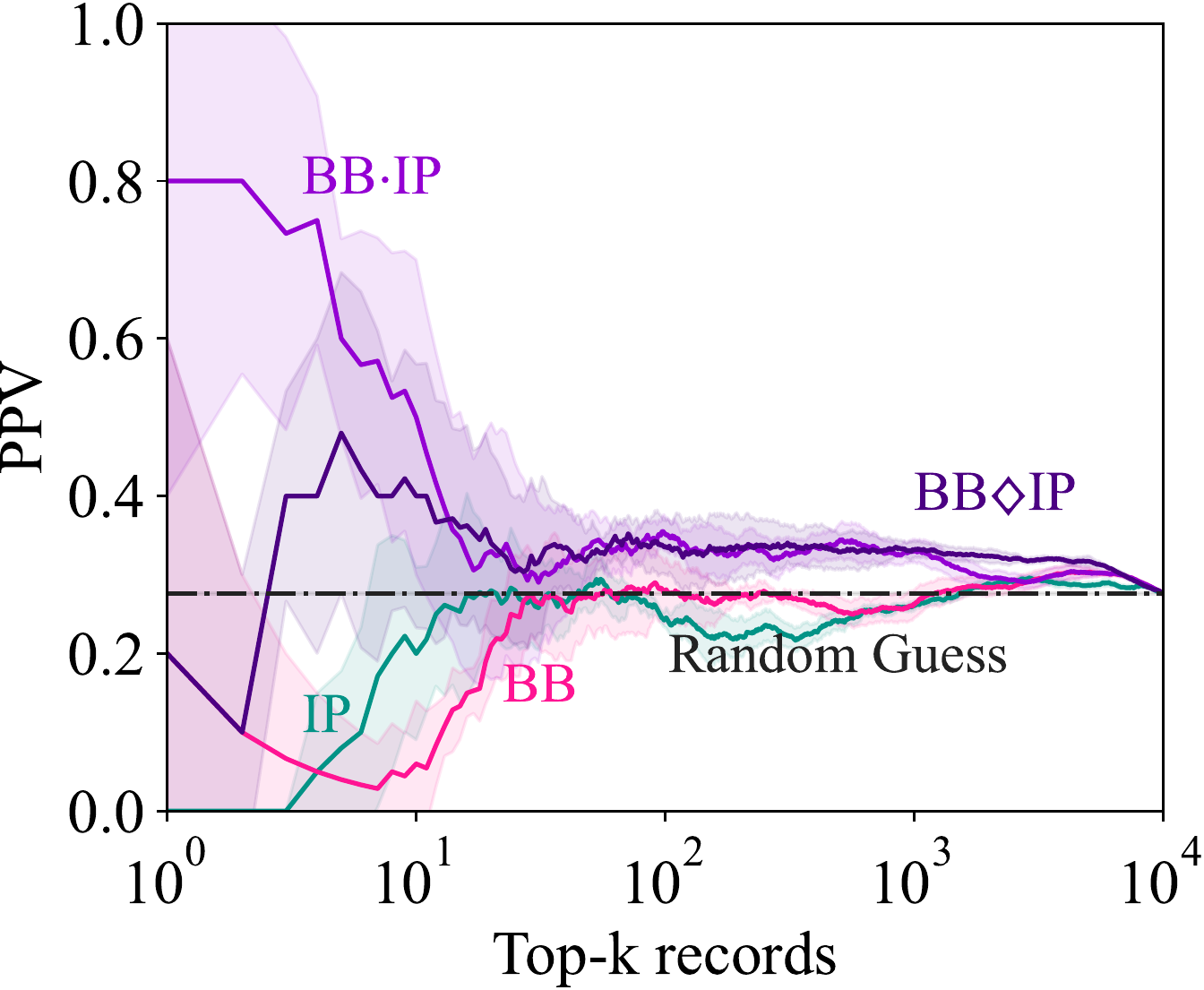}
         \caption{$|D_{aux}|$ = 50}
         \label{fig:ppv_texas_ethnicity_high_bb}
     \end{subfigure}
    \caption{Comparing the PPV of black-box attacks against imputation on predicting the Hispanic ethnicity among 10\,000 candidate training records in \bfdataset{Texas-100X} ($D_{aux} \sim \cD$). \rm Results are averaged over five runs.}
    \label{fig:ppv_texas_ethnicity_bb}
\end{figure*}

\begin{figure*}
     \centering
     \begin{subfigure}[b]{0.32\textwidth}
         \centering
         \includegraphics[width=\textwidth]{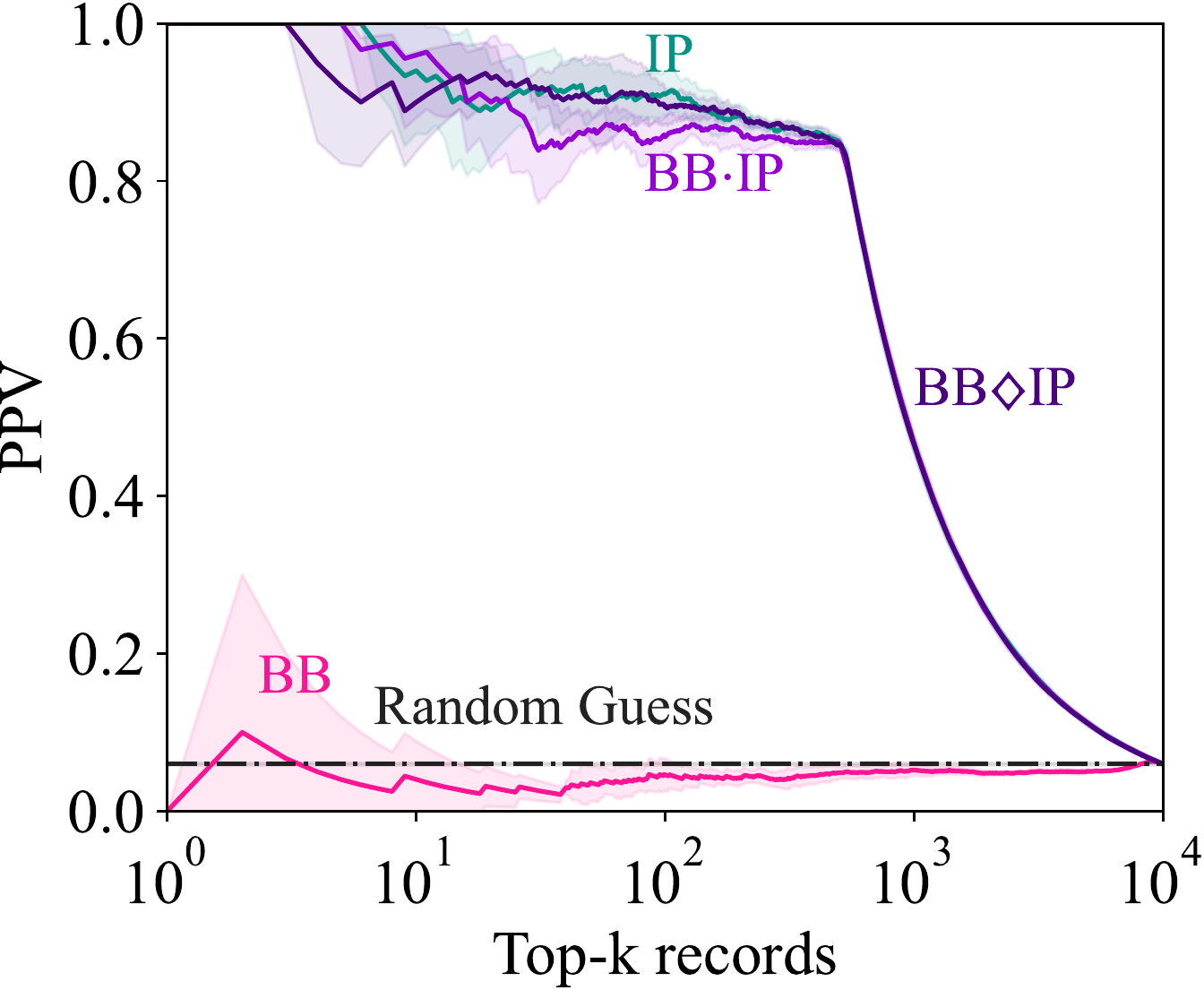}
         \caption{$|D_{aux}|$ = 5\,000}
         \label{fig:ppv_census_race_low_bb}
     \end{subfigure}
     \begin{subfigure}[b]{0.32\textwidth}
         \centering
         \includegraphics[width=\textwidth]{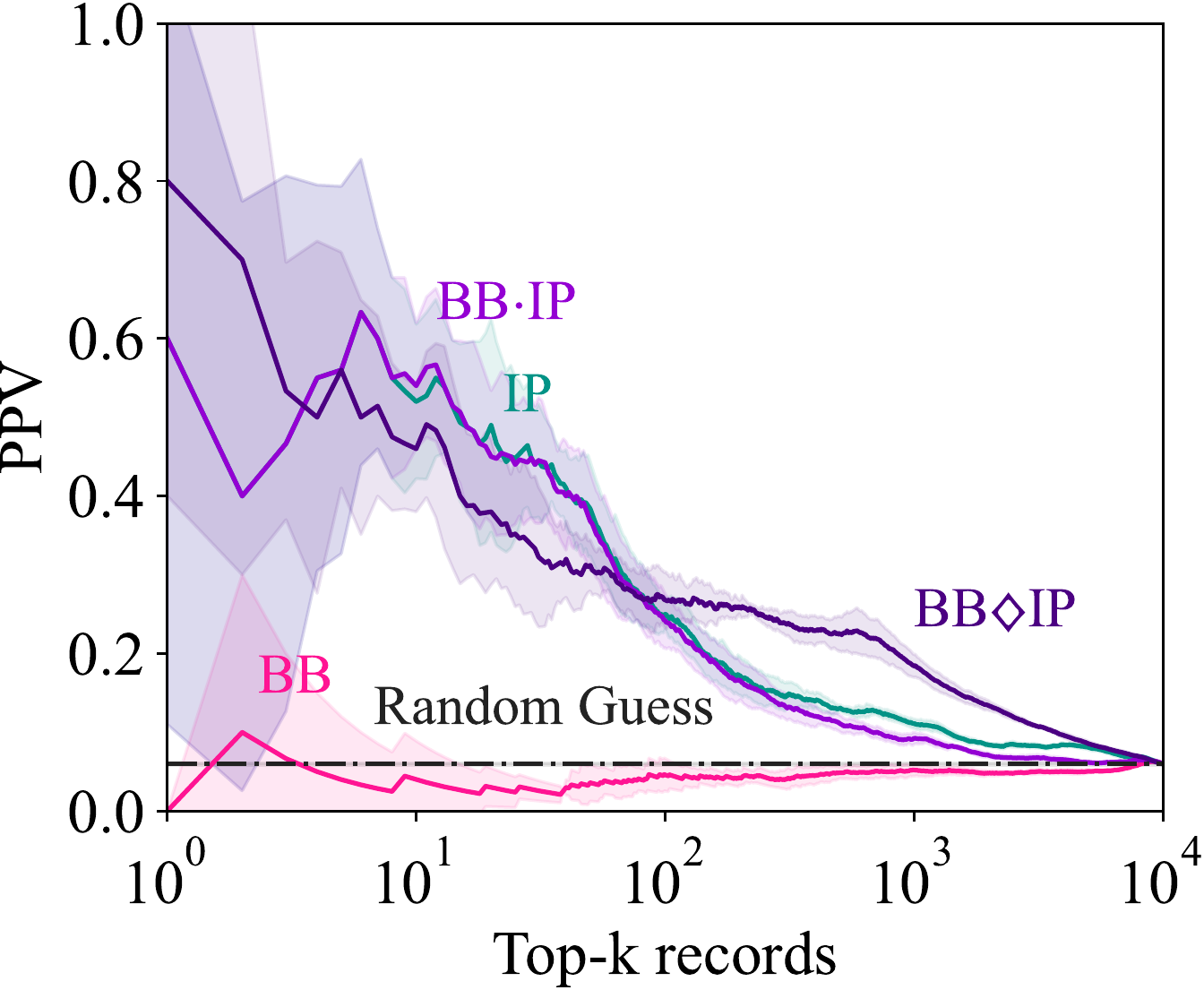}
         \caption{$|D_{aux}|$ = 500}
         \label{fig:ppv_census_race_med_bb}
     \end{subfigure}
     \begin{subfigure}[b]{0.32\textwidth}
         \centering
         \includegraphics[width=\textwidth]{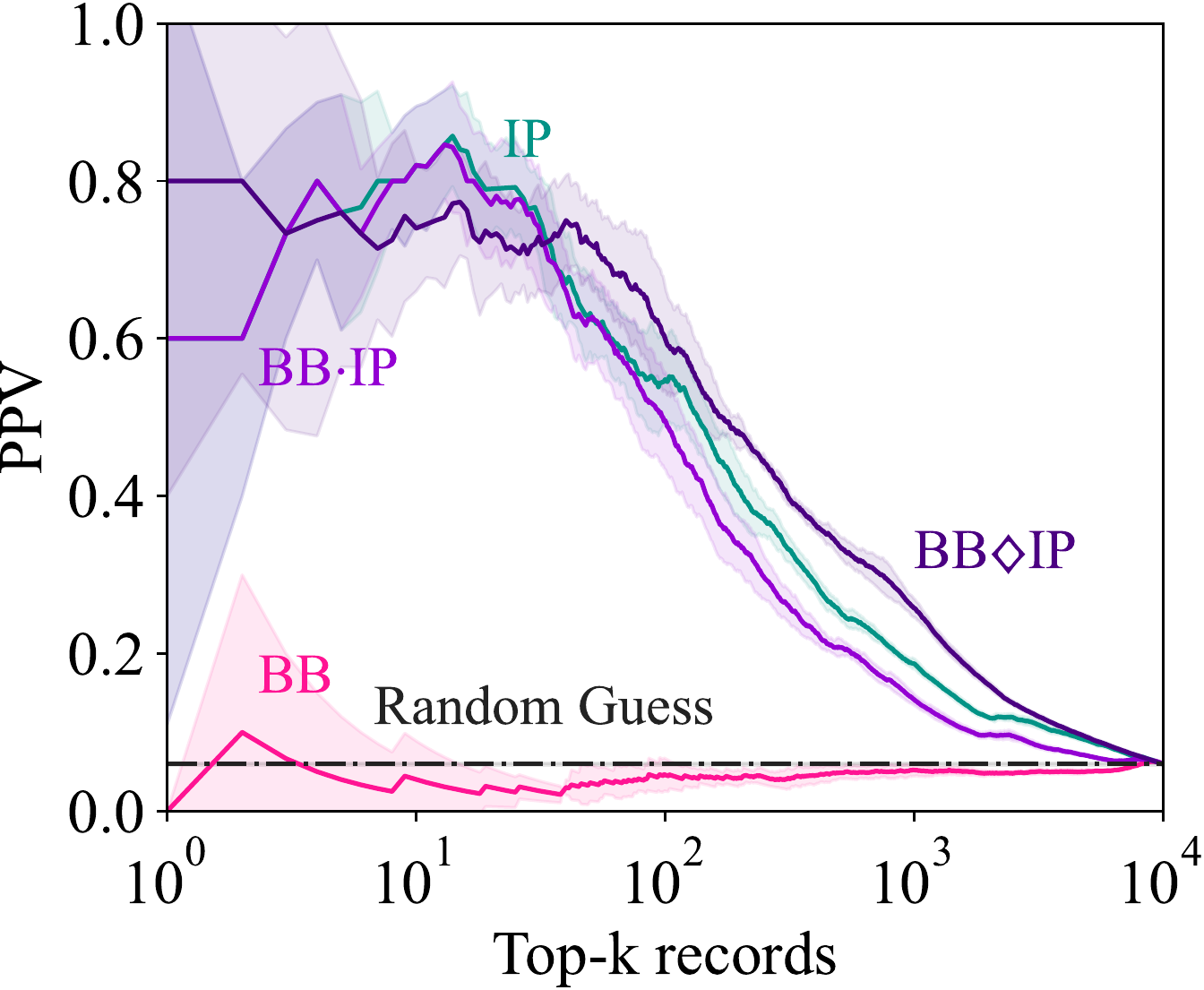}
         \caption{$|D_{aux}|$ = 50}
         \label{fig:ppv_census_race_high_bb}
     \end{subfigure}
    \caption{Comparing the PPV of black-box attacks against imputation on predicting the Asian race among 10\,000 candidate training records in \bfdataset{Census19} ($D_{aux} \sim \cD$). \rm Results are averaged over five runs.}
    \label{fig:ppv_census_race_bb}
\end{figure*}

\begin{figure*}[ptb]
     \centering
     \begin{subfigure}{0.32\textwidth}
         \centering
         \includegraphics[width=\textwidth]{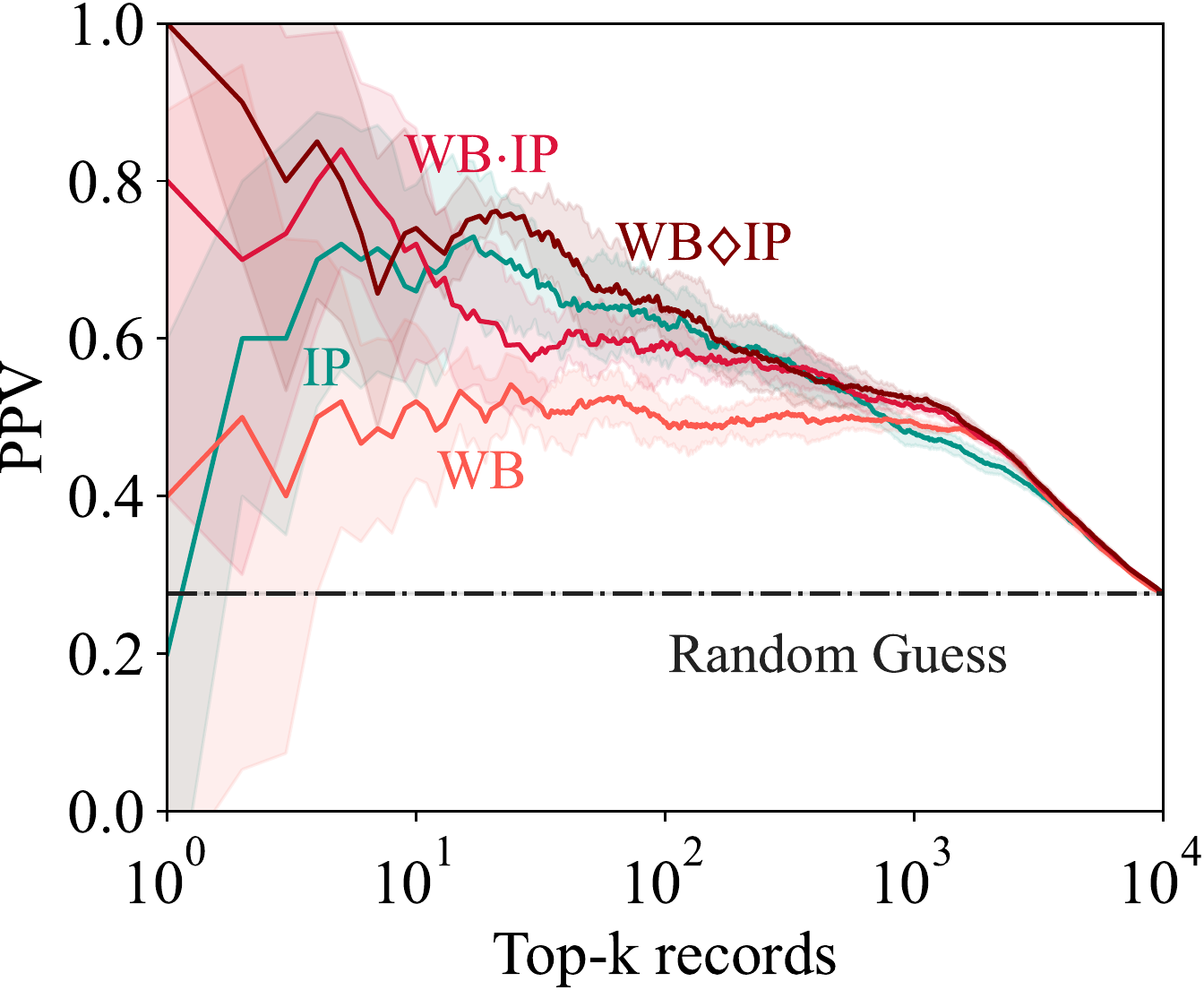}
         \caption{$|D_{aux}|$ = 5\,000}
         \label{fig:ppv_texas_ethnicity_low_wb}
     \end{subfigure}
     \begin{subfigure}{0.32\textwidth}
         \centering
         \includegraphics[width=\textwidth]{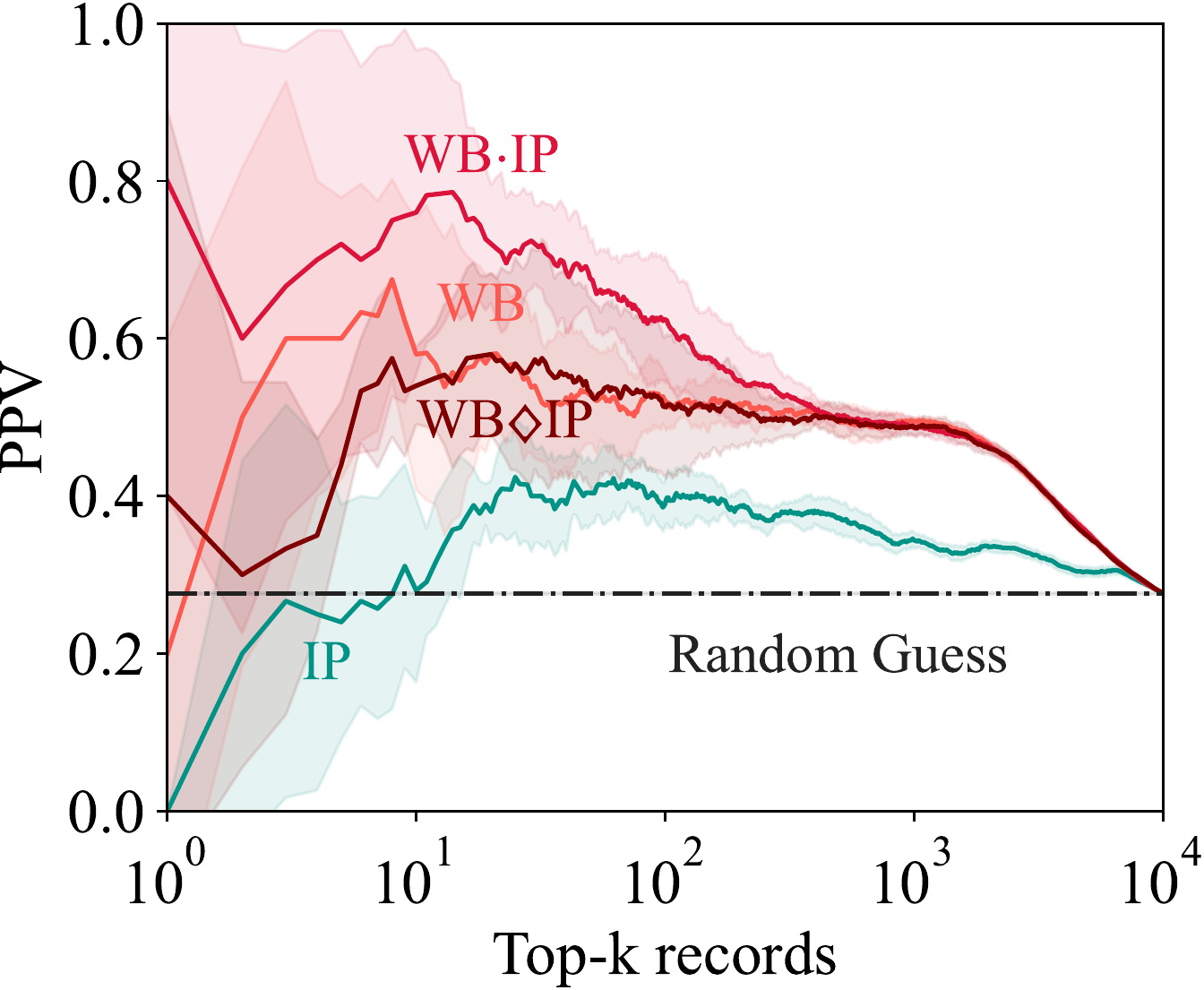}
         \caption{$|D_{aux}|$ = 500}
         \label{fig:ppv_texas_ethnicity_med_wb}
     \end{subfigure}
     \begin{subfigure}{0.32\textwidth}
         \centering
         \includegraphics[width=\textwidth]{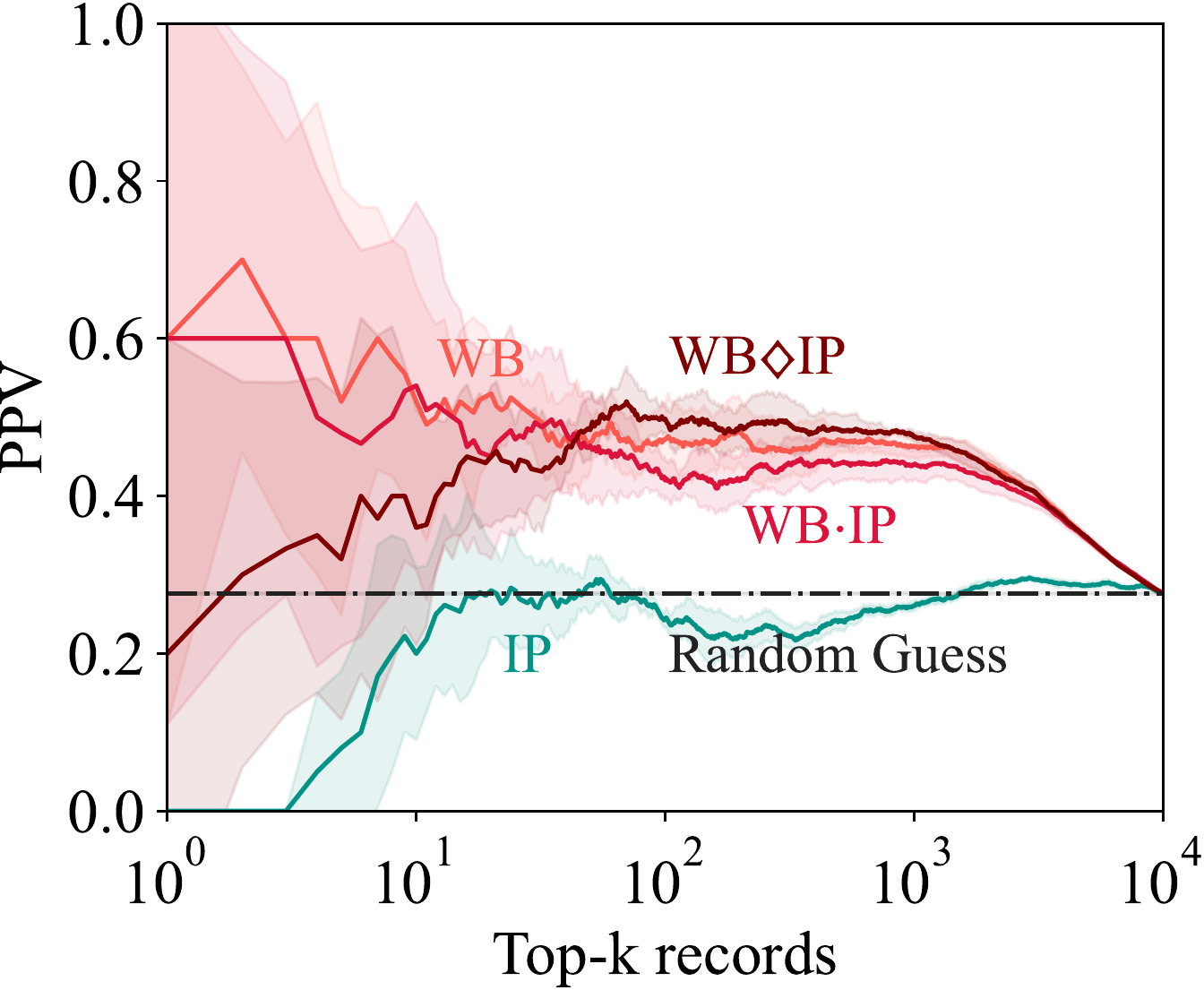}
         \caption{$|D_{aux}|$ = 50}
         \label{fig:ppv_texas_ethnicity_high_wb}
     \end{subfigure}
    \caption{Comparing the PPV of white-box attacks against imputation on predicting the Hispanic ethnicity among 10\,000 candidate training records in \bfdataset{Texas-100X} ($D_{aux} \sim \cD$). \rm Results are averaged over five runs.}
    \label{fig:ppv_texas_ethnicity_wb}
\end{figure*}

\begin{figure*}[ptb]
     \centering
     \begin{subfigure}{0.32\textwidth}
         \centering
         \includegraphics[width=\textwidth]{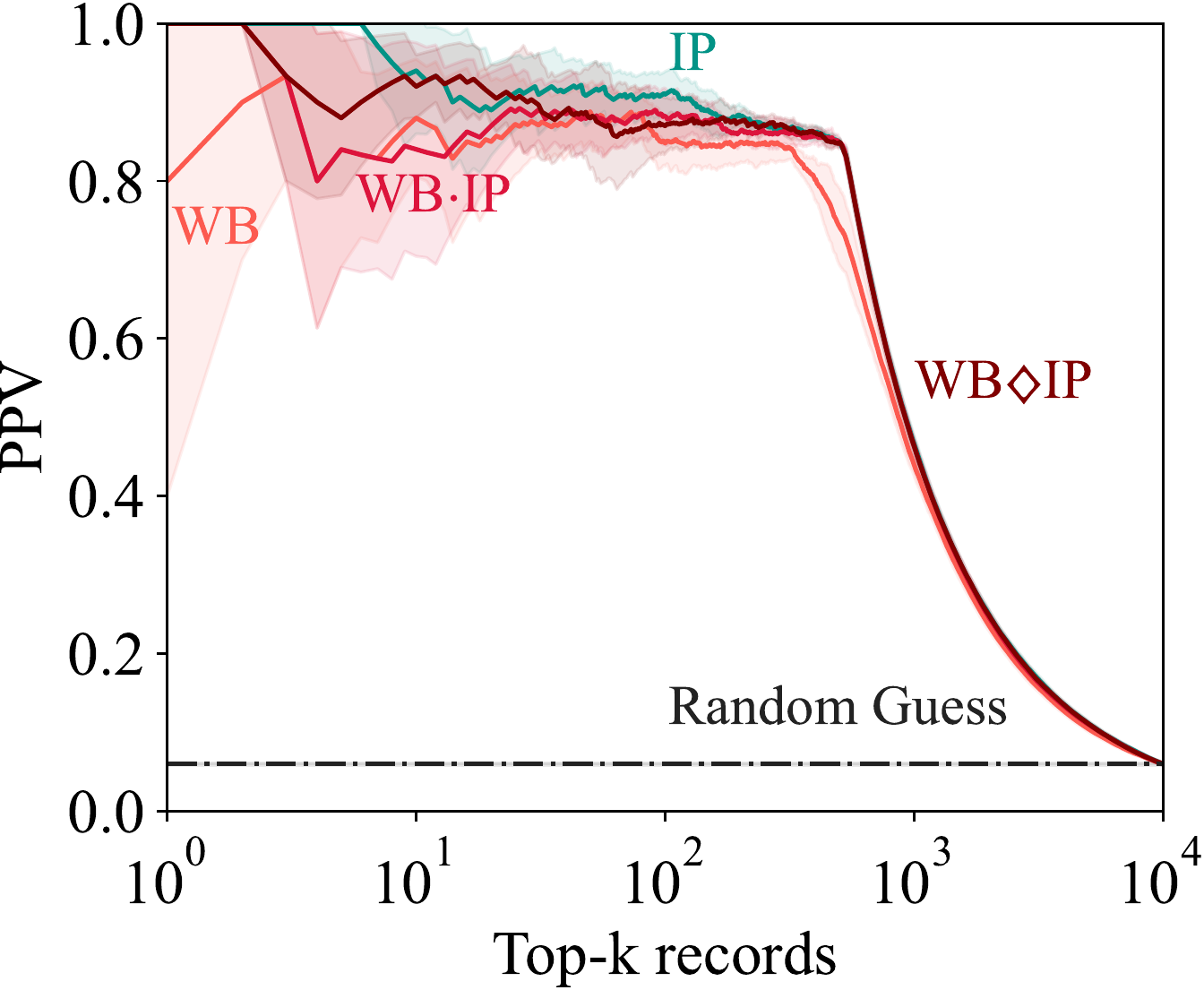}
         \caption{$|D_{aux}|$ = 5\,000}
         \label{fig:ppv_census_race_low_wb}
     \end{subfigure}
     \begin{subfigure}{0.32\textwidth}
         \centering
         \includegraphics[width=\textwidth]{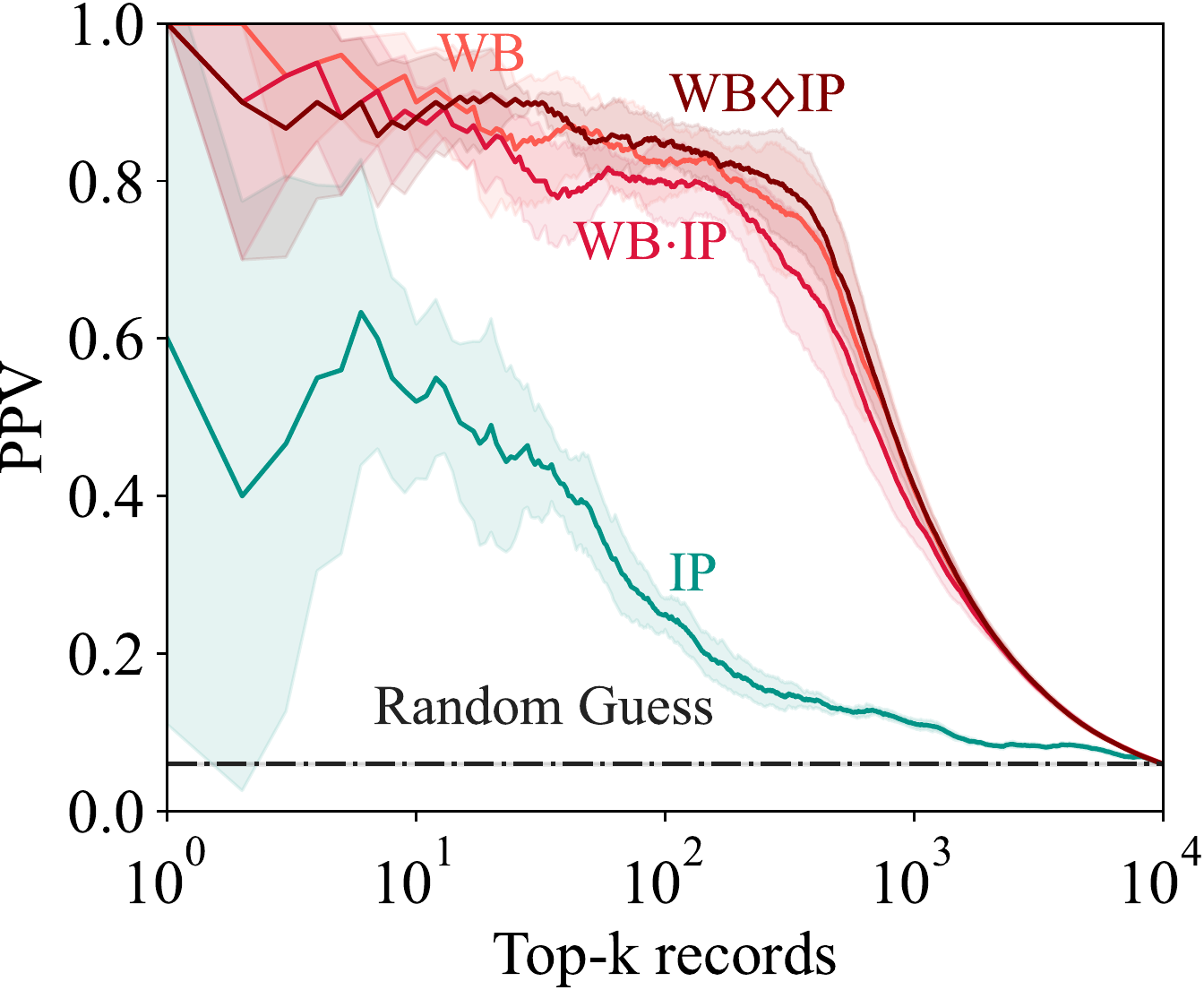}
         \caption{$|D_{aux}|$ = 500}
         \label{fig:ppv_census_race_med_wb}
     \end{subfigure}
     \begin{subfigure}{0.32\textwidth}
         \centering
         \includegraphics[width=\textwidth]{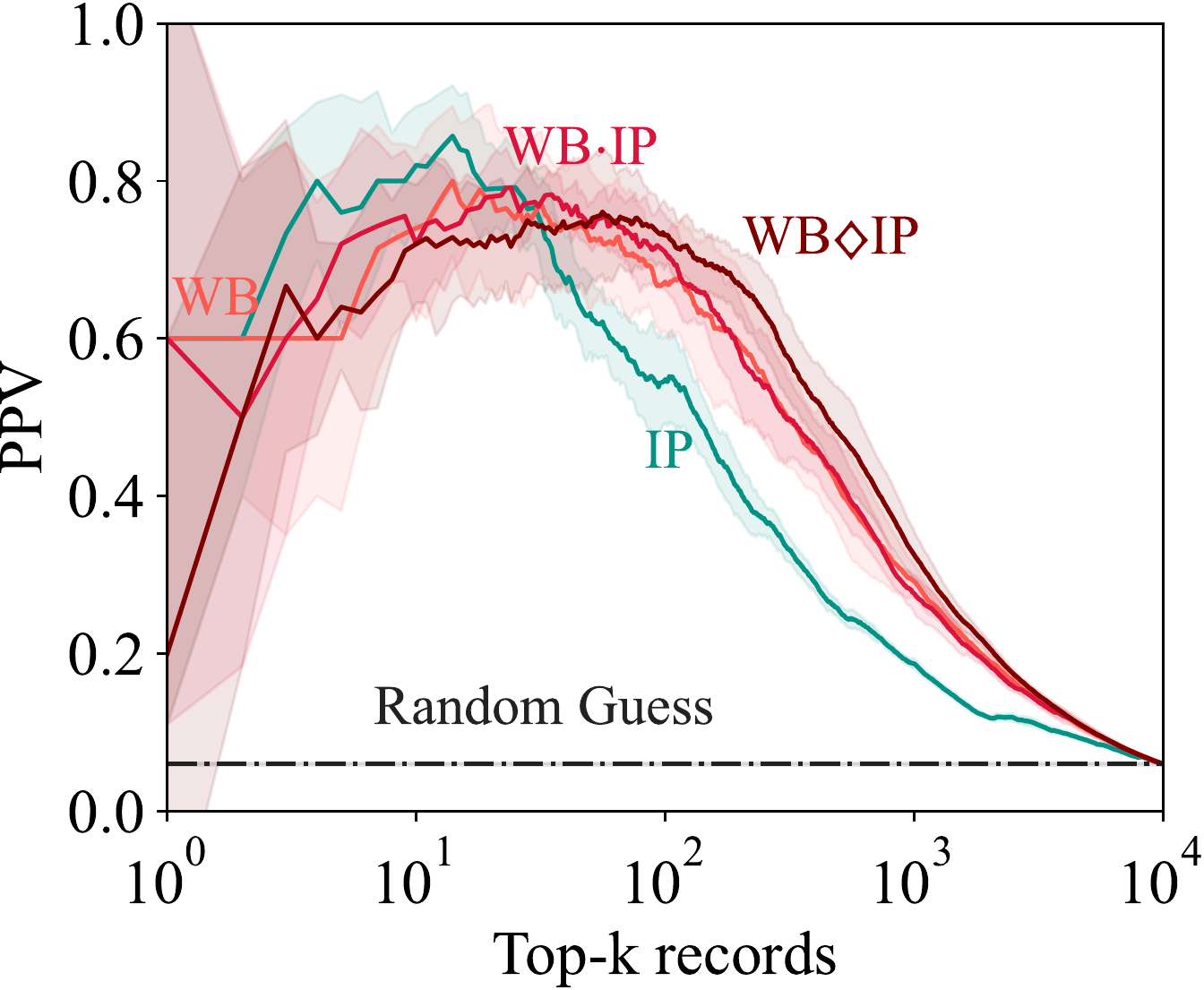}
         \caption{$|D_{aux}|$ = 50}
         \label{fig:ppv_census_race_high_wb}
     \end{subfigure}
    \caption{Comparing the PPV of white-box attacks against imputation on predicting the Asian race among 10\,000 candidate training records in \bfdataset{Census19} ($D_{aux} \sim \cD$). \rm Results are averaged over five runs.}
    \label{fig:ppv_census_race_wb}
\end{figure*}

\begin{figure*}[ptb]
     \centering
     \begin{subfigure}{0.32\textwidth}
         \centering
         \includegraphics[width=\textwidth]{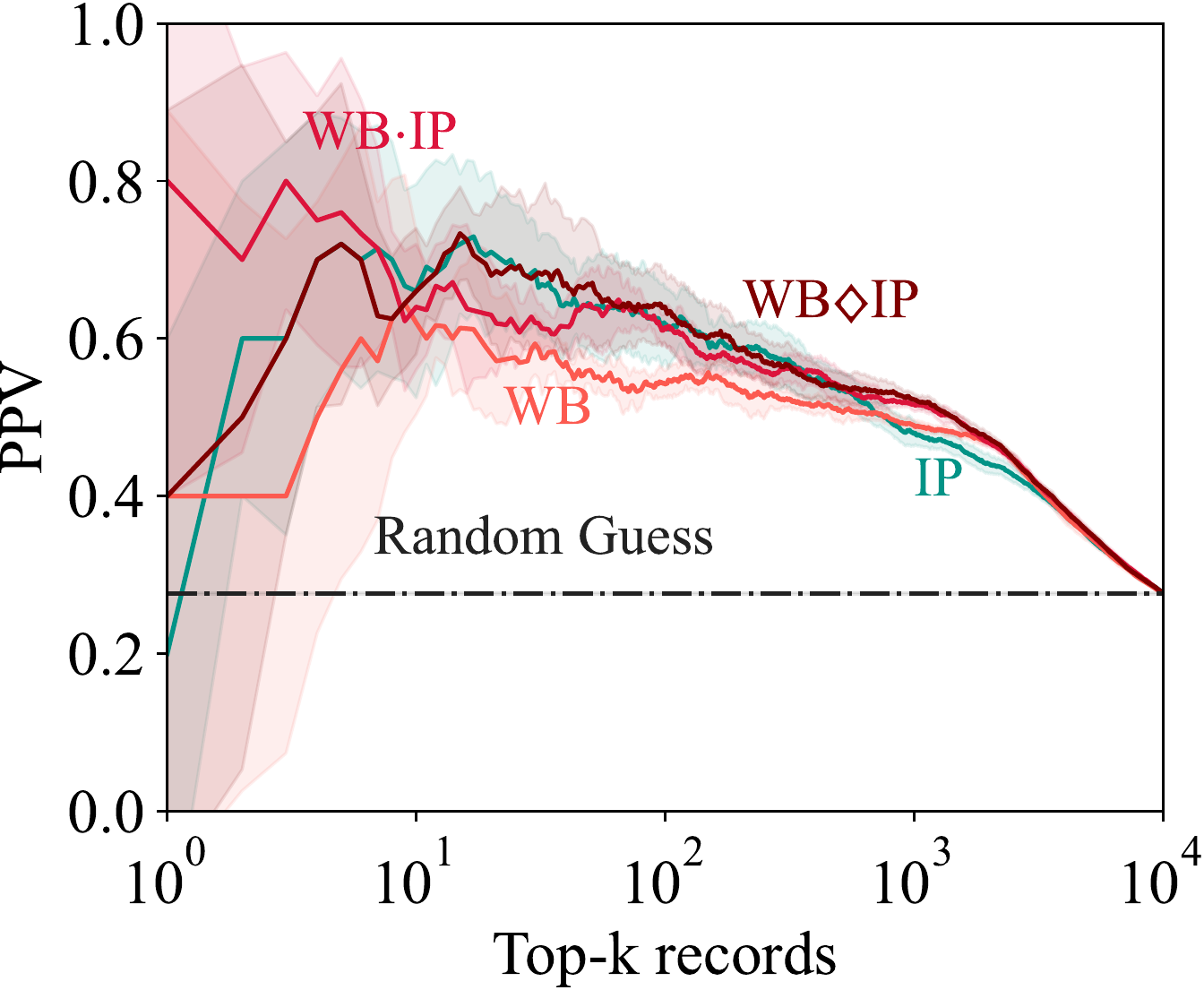}
         \caption{$|D_{aux}|$ = 5\,000}
         \label{fig:ppv_texas_ethnicity_low_wb_banished}
     \end{subfigure}
     \begin{subfigure}{0.32\textwidth}
         \centering
         \includegraphics[width=\textwidth]{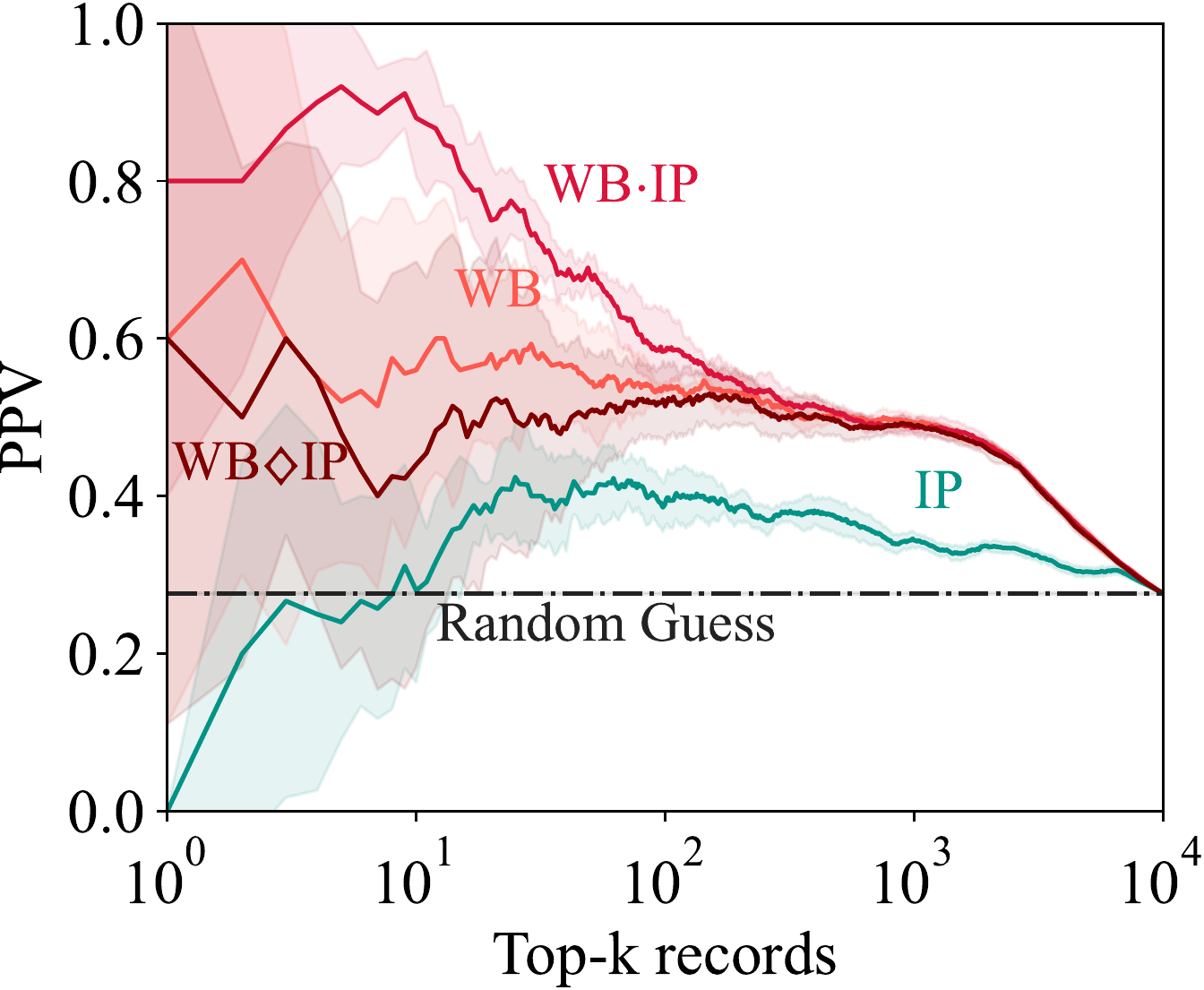}
         \caption{$|D_{aux}|$ = 500}
         \label{fig:ppv_texas_ethnicity_med_wb_banished}
     \end{subfigure}
     \begin{subfigure}{0.32\textwidth}
         \centering
         \includegraphics[width=\textwidth]{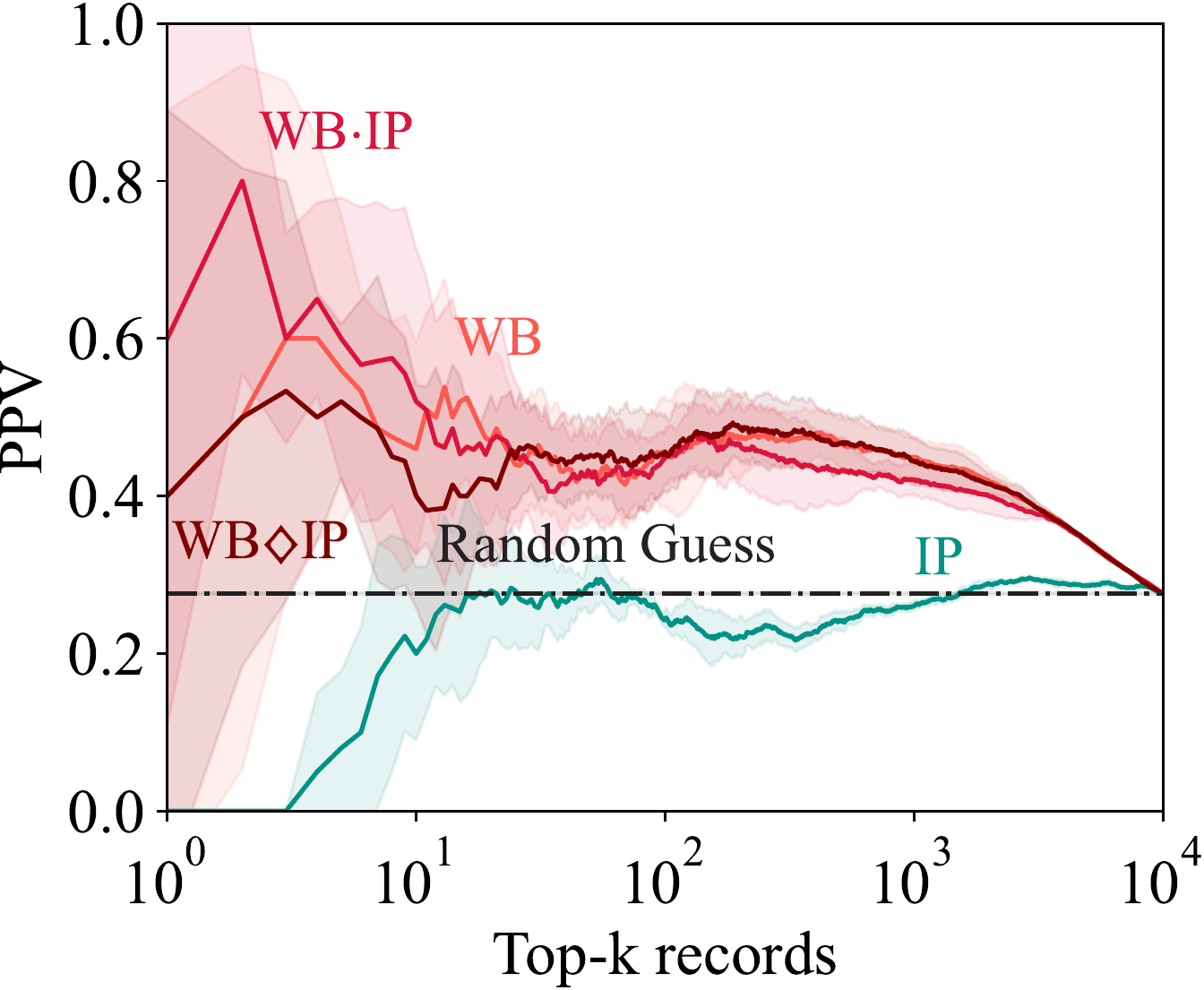}
         \caption{$|D_{aux}|$ = 50}
         \label{fig:ppv_texas_ethnicity_high_wb_banished}
     \end{subfigure}
    \caption{Comparing the PPV of white-box attacks on predicting the Hispanic ethnicity among 10\,000 candidate training records in \bfdataset{Texas-100X} ($D_{aux} \sim \cD$). \rm Model is retained after removing vulnerable records and the results are averaged over five runs.}
    \label{fig:ppv_texas_ethnicity_wb_banished}
\end{figure*}

\begin{figure*}[ptb]
     \centering
     \begin{subfigure}{0.32\textwidth}
         \centering
         \includegraphics[width=\textwidth]{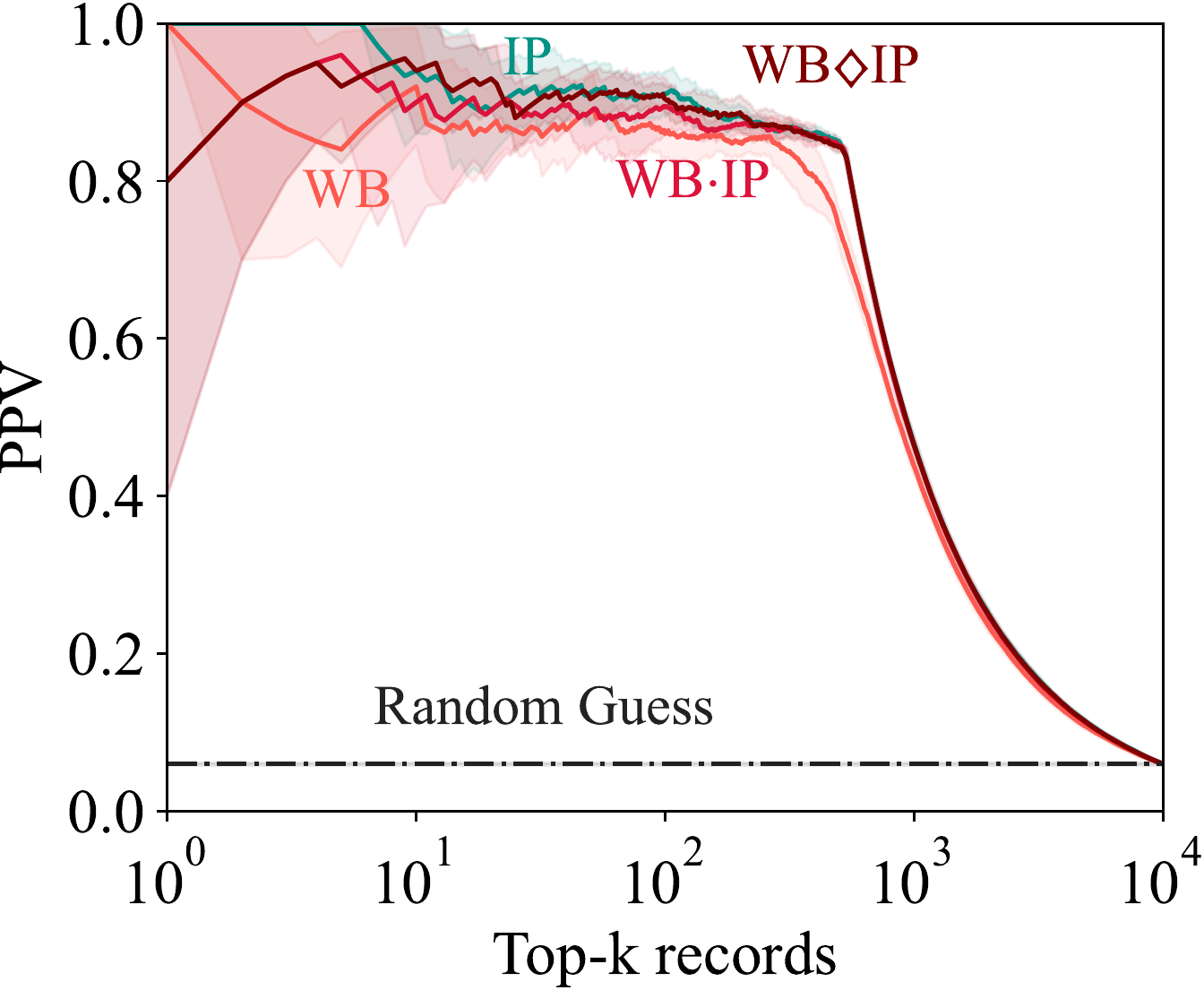}
         \caption{$|D_{aux}|$ = 5\,000}
         \label{fig:ppv_census_race_low_wb_banished}
     \end{subfigure}
     \begin{subfigure}{0.32\textwidth}
         \centering
         \includegraphics[width=\textwidth]{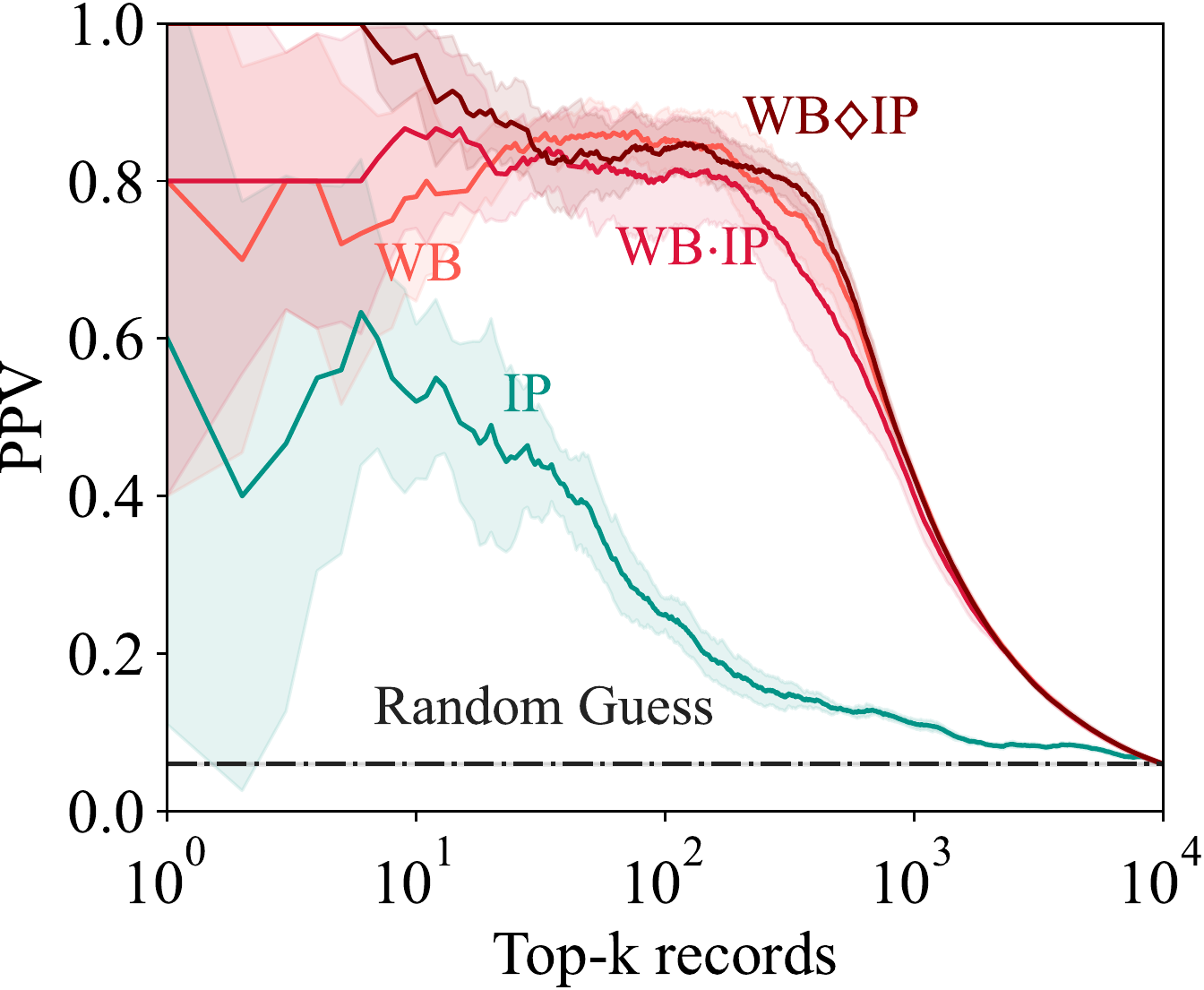}
         \caption{$|D_{aux}|$ = 500}
         \label{fig:ppv_census_race_med_wb_banished}
     \end{subfigure}
     \begin{subfigure}{0.32\textwidth}
         \centering
         \includegraphics[width=\textwidth]{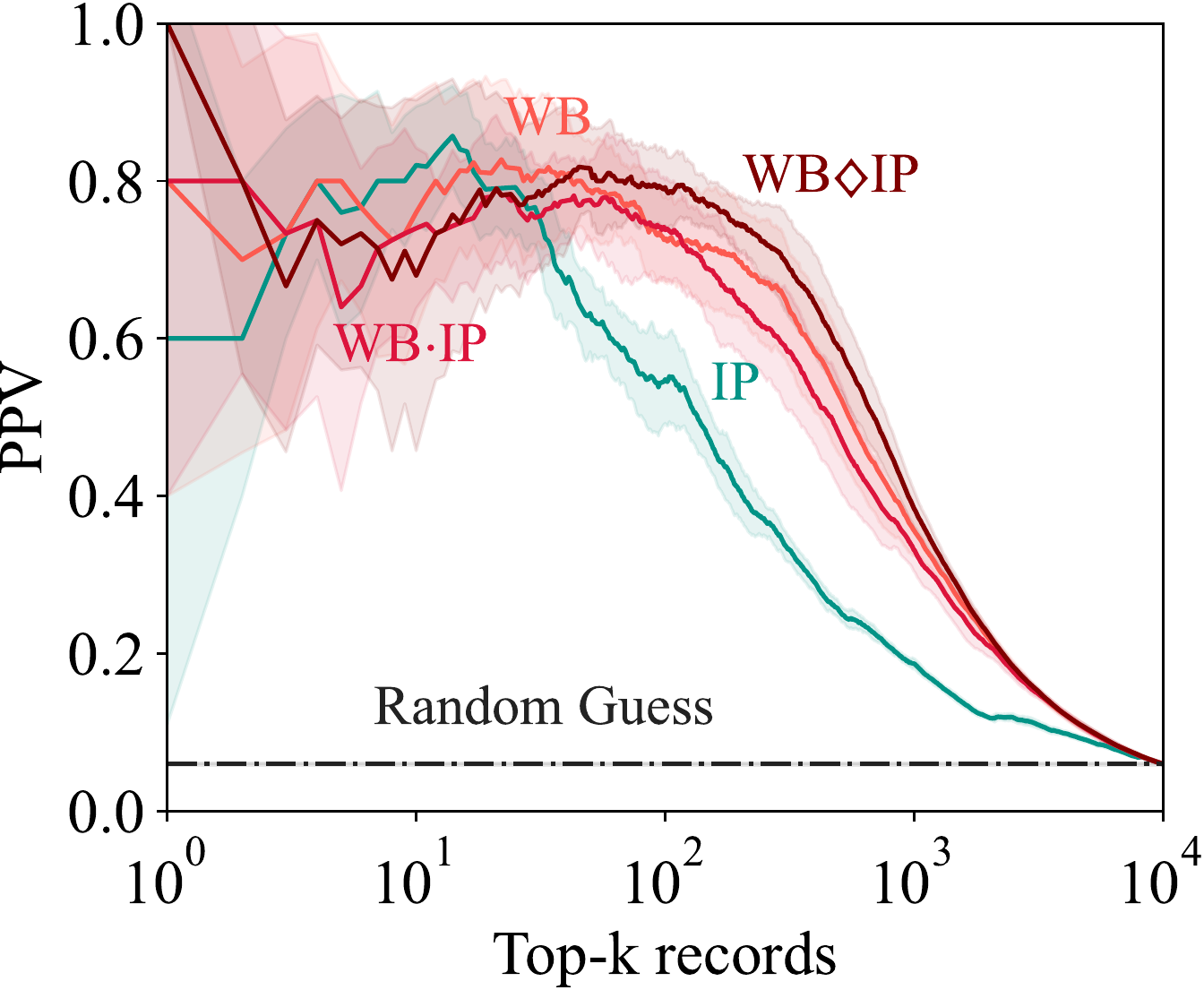}
         \caption{$|D_{aux}|$ = 50}
         \label{fig:ppv_census_race_high_wb_banished}
     \end{subfigure}
    \caption{Comparing the PPV of white-box attacks on predicting the Asian race among 10\,000 candidate training records in \bfdataset{Census19} ($D_{aux} \sim \cD$). \rm Model is retained after removing vulnerable records and the results are averaged over five runs.}
    \label{fig:ppv_census_race_wb_banished}
\end{figure*}

\begin{figure*}[ptb]
     \centering
     \begin{subfigure}{0.32\textwidth}
         \centering
         \includegraphics[width=\textwidth]{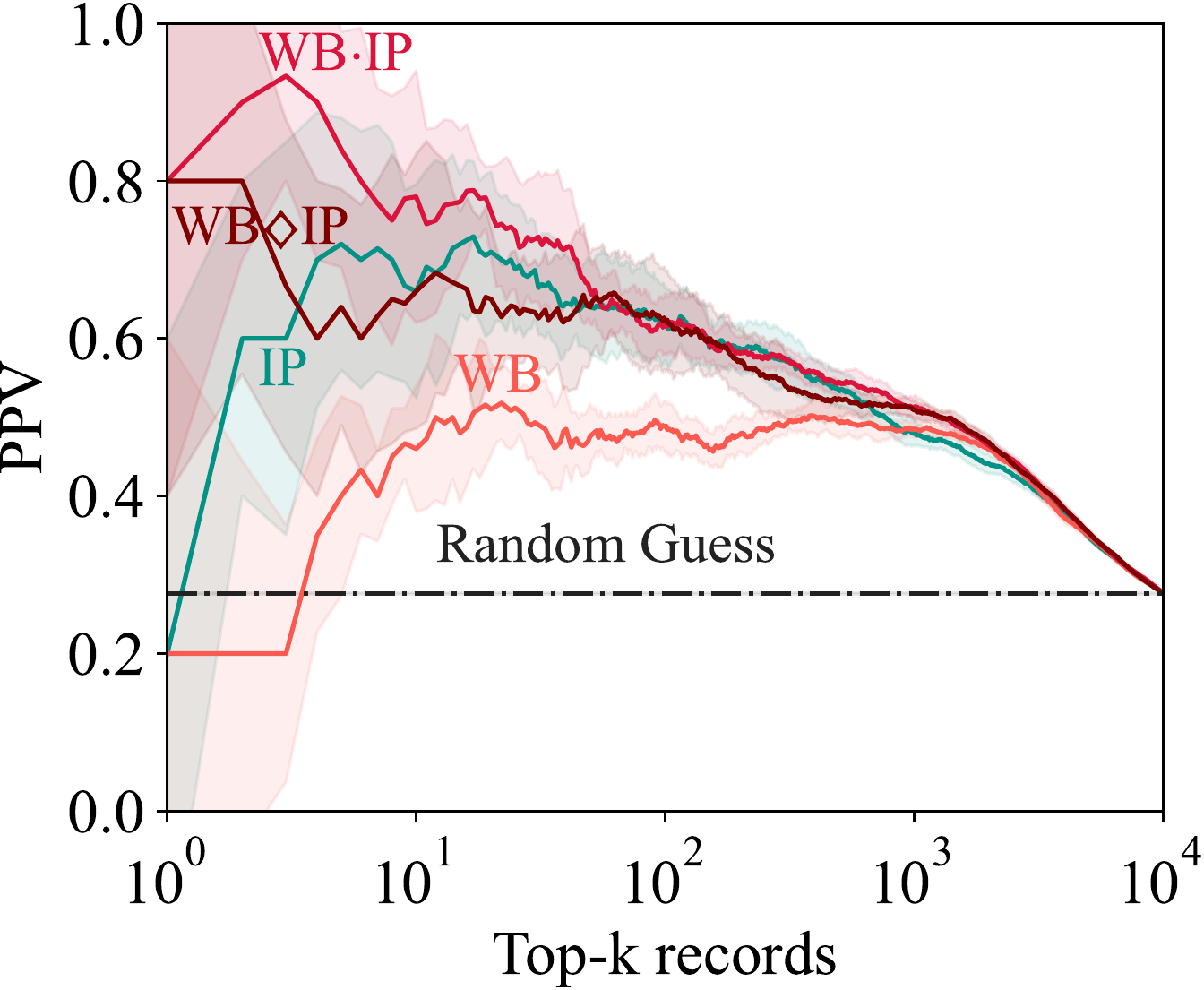}
         \caption{$|D_{aux}|$ = 5\,000}
         \label{fig:ppv_texas_ethnicity_low_wb_1_eps}
     \end{subfigure}
     \begin{subfigure}{0.32\textwidth}
         \centering
         \includegraphics[width=\textwidth]{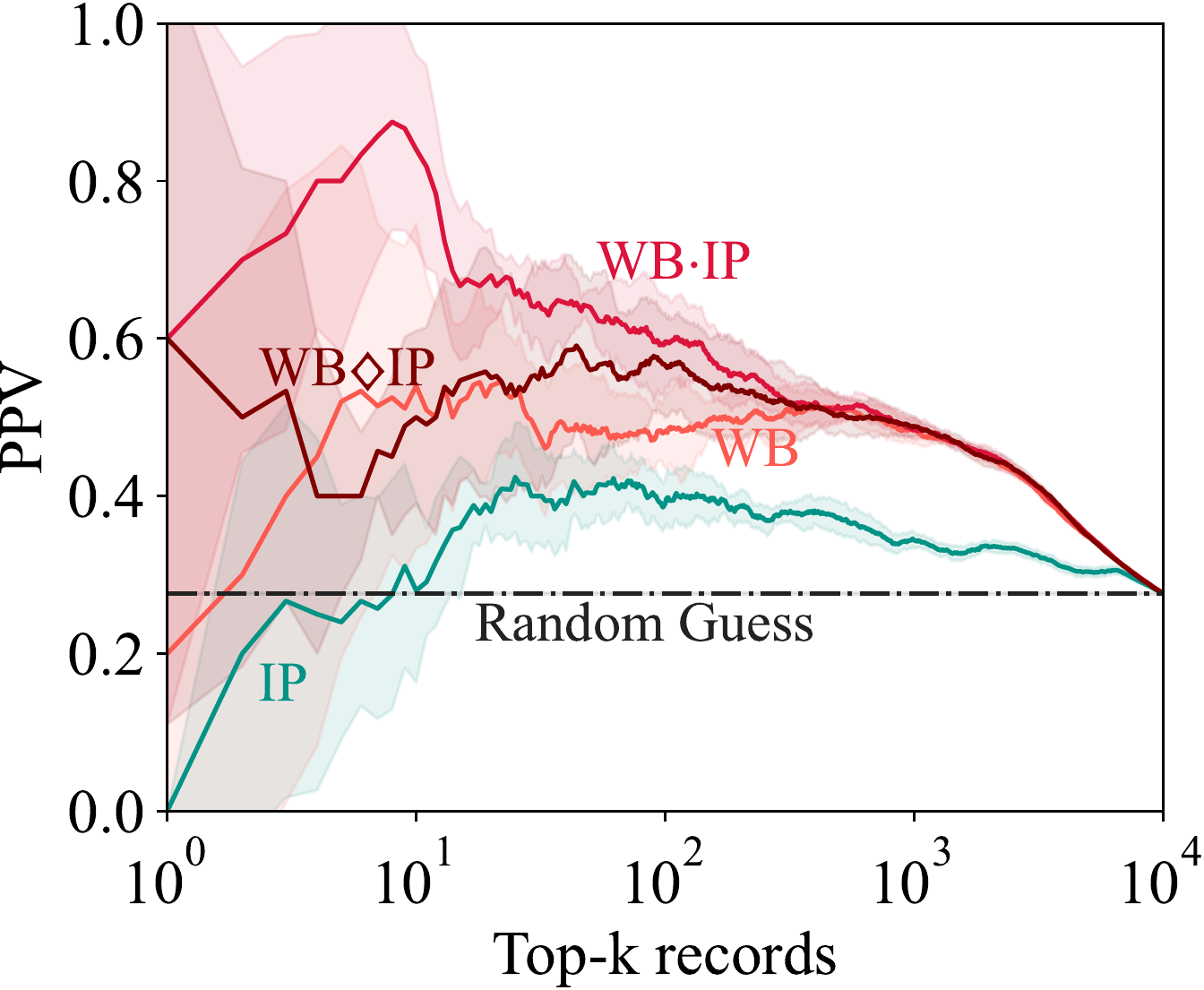}
         \caption{$|D_{aux}|$ = 500}
         \label{fig:ppv_texas_ethnicity_med_wb_1_eps}
     \end{subfigure}
     \begin{subfigure}{0.32\textwidth}
         \centering
         \includegraphics[width=\textwidth]{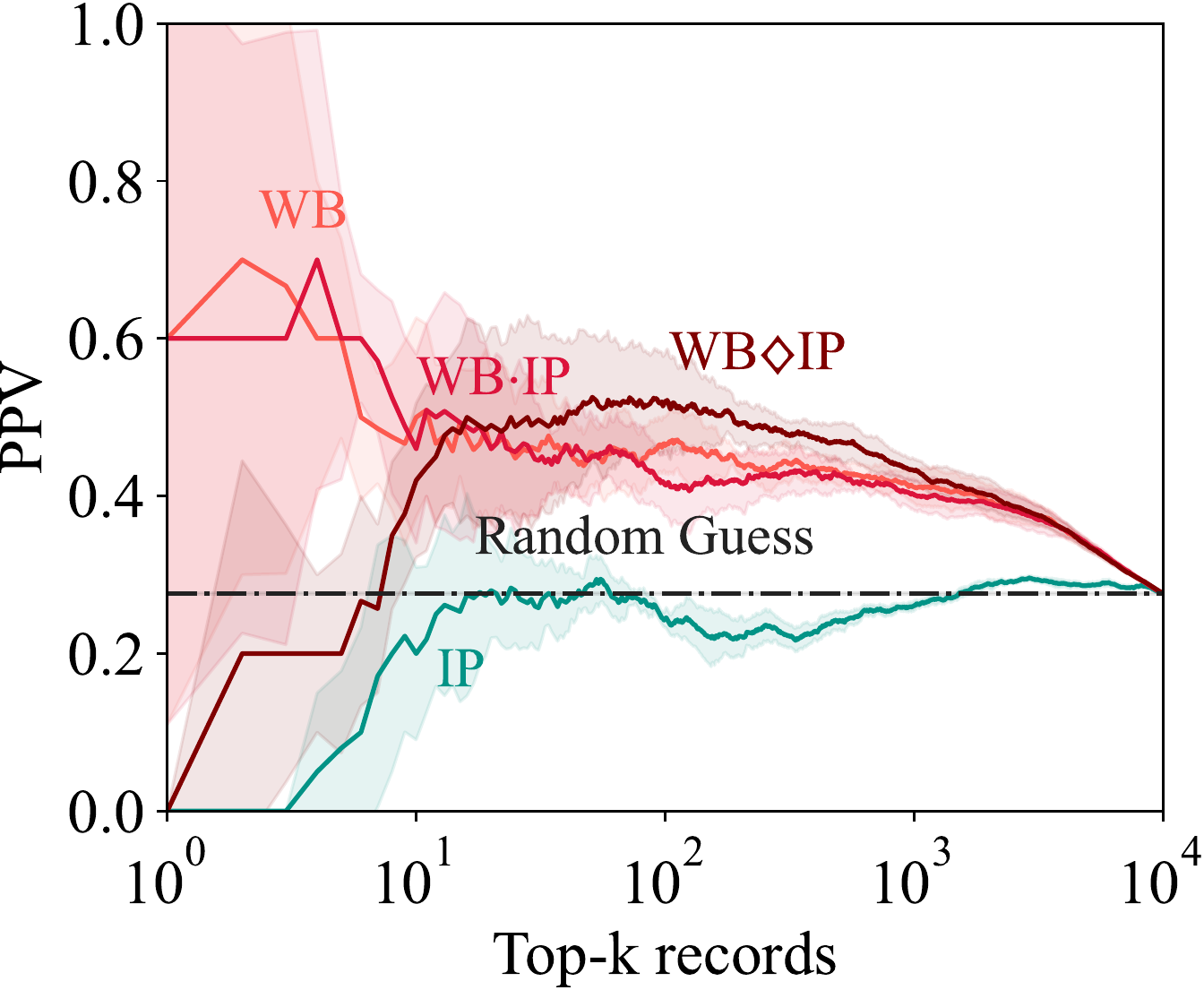}
         \caption{$|D_{aux}|$ = 50}
         \label{fig:ppv_texas_ethnicity_high_wb_1_eps}
     \end{subfigure}
    \caption{Comparing the PPV of white-box attacks on predicting the Hispanic ethnicity among 10\,000 candidate training records in \bfdataset{Texas-100X} ($D_{aux} \sim \cD$). \rm Model is trained with Gaussian differential privacy ($\epsilon = 1$) and results are averaged over five runs.}
    \label{fig:ppv_texas_ethnicity_wb_1_eps}
\end{figure*}

\begin{figure*}[ptb]
     \centering
     \begin{subfigure}{0.32\textwidth}
         \centering
         \includegraphics[width=\textwidth]{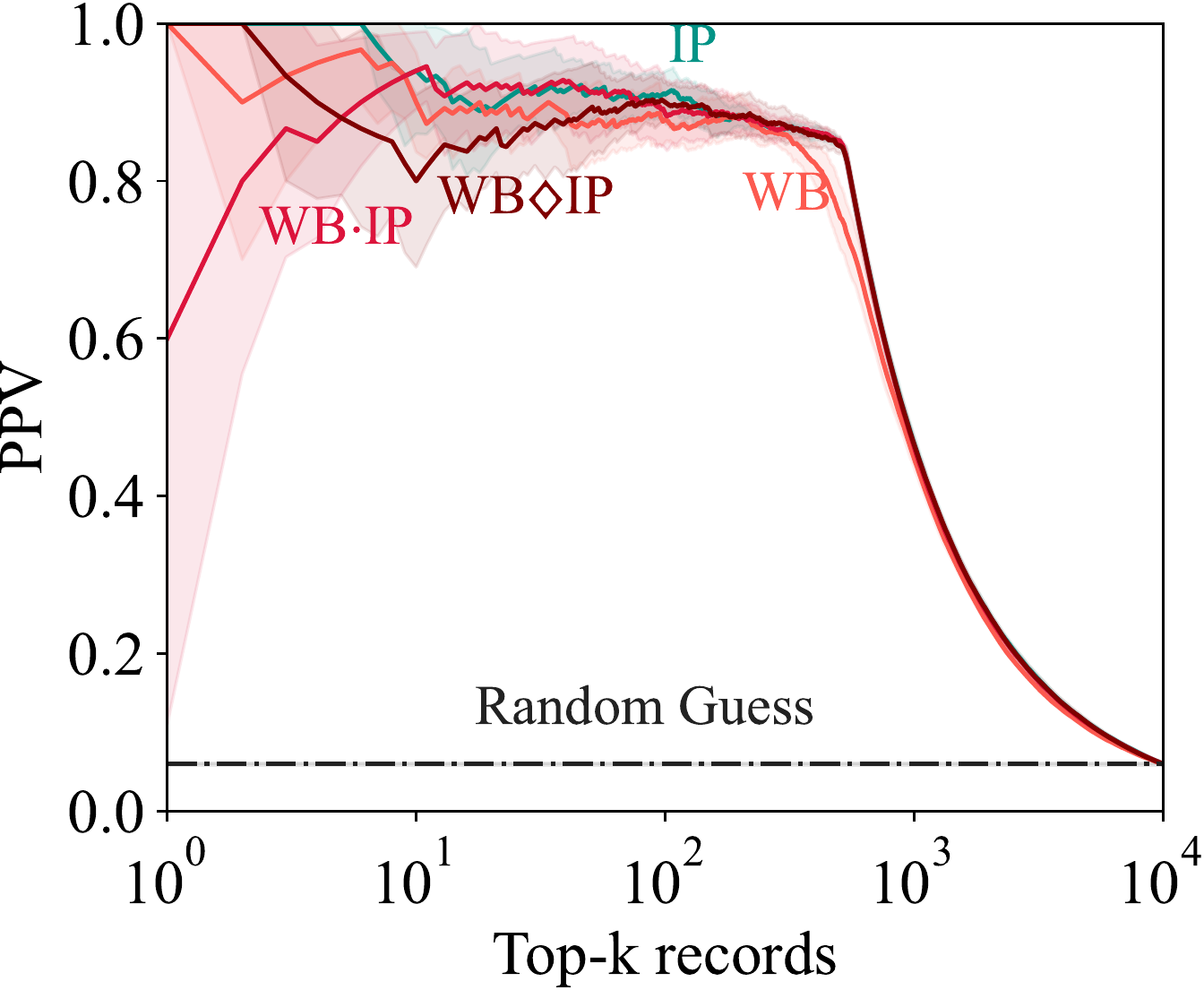}
         \caption{$|D_{aux}|$ = 5\,000}
         \label{fig:ppv_census_race_low_wb_1_eps}
     \end{subfigure}
     \begin{subfigure}{0.32\textwidth}
         \centering
         \includegraphics[width=\textwidth]{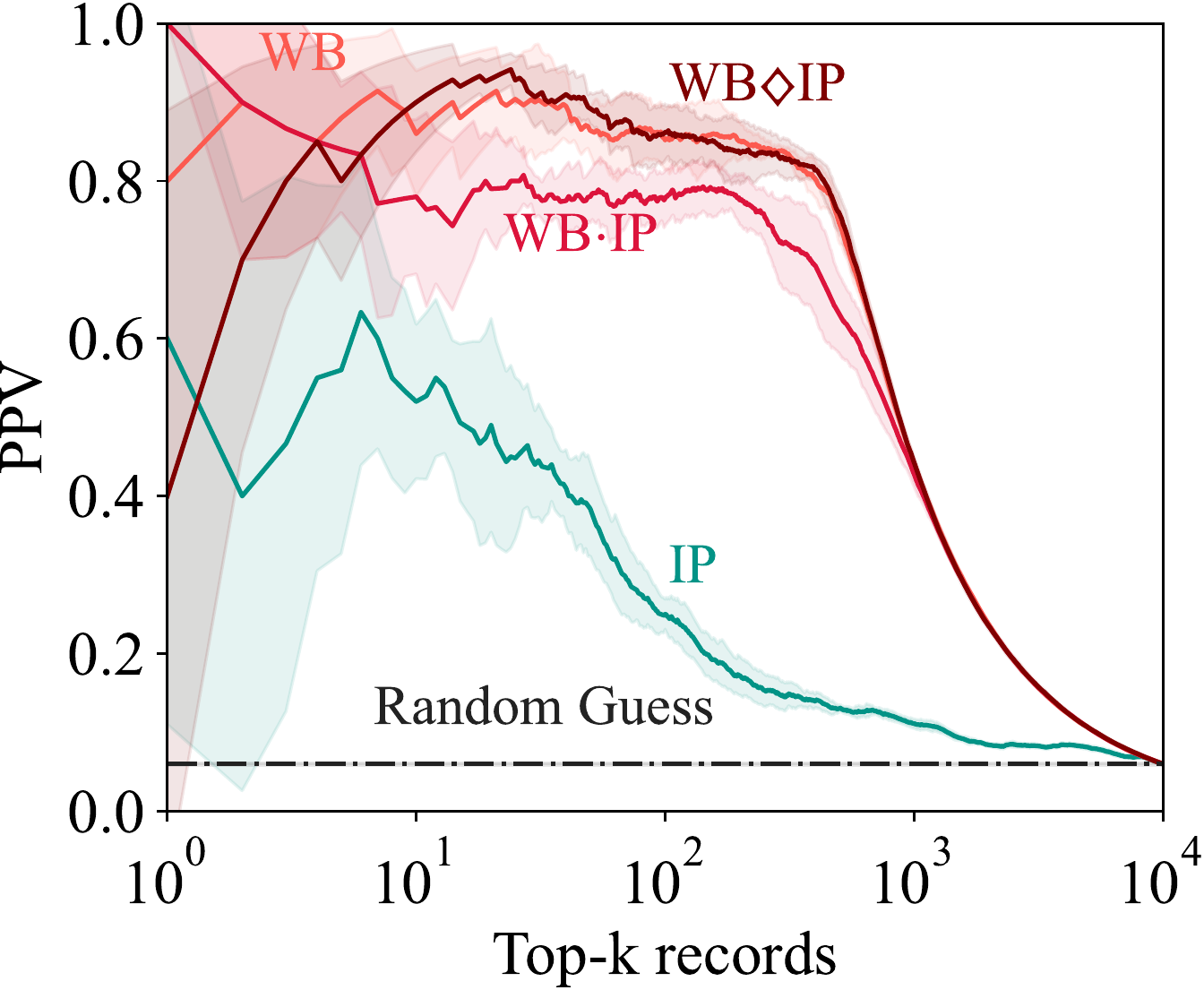}
         \caption{$|D_{aux}|$ = 500}
         \label{fig:ppv_census_race_med_wb_1_eps}
     \end{subfigure}
     \begin{subfigure}{0.32\textwidth}
         \centering
         \includegraphics[width=\textwidth]{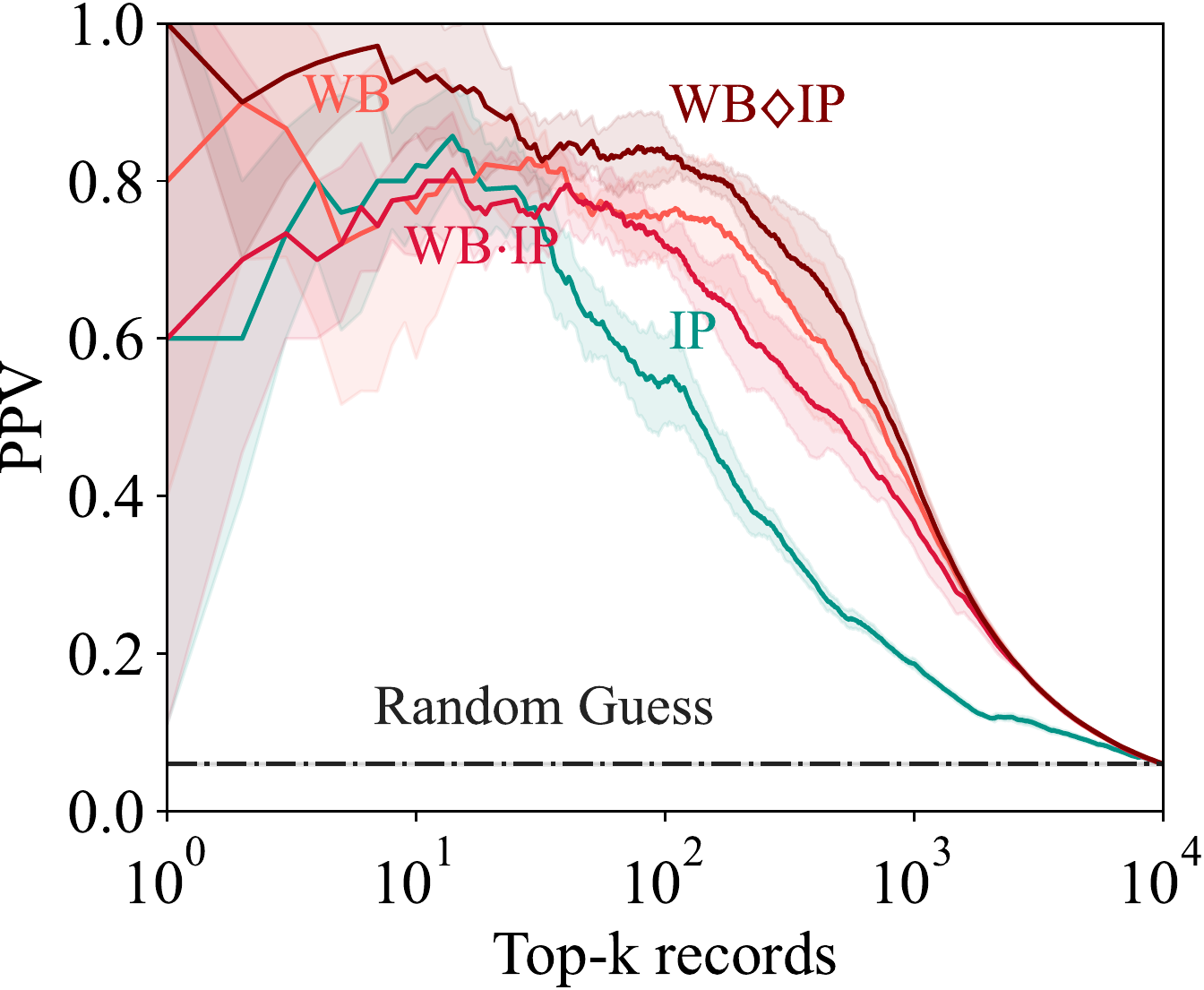}
         \caption{$|D_{aux}|$ = 50}
         \label{fig:ppv_census_race_high_wb_1_eps}
     \end{subfigure}
    \caption{Comparing the PPV of white-box attacks on predicting the Asian race among 10\,000 candidate training records in \bfdataset{Census19} ($D_{aux} \sim \cD$). \rm Model is trained with Gaussian differential privacy ($\epsilon = 1$) and results are averaged over five runs.}
    \label{fig:ppv_census_race_wb_1_eps}
\end{figure*}